\documentclass
	[
		10 pt
	]
	{article}

\usepackage
	[
		a4paper,
		left = 2.500 cm,
		right = 2.500 cm,
		top = 2.500 cm,
		bottom = 3.000 cm
	]
	{geometry}
\usepackage{tikz, pgf, pgfplots}
\usetikzlibrary
	{
		3d,
		arrows,											
		arrows.meta,
		automata,										
		backgrounds,
		bending,
		calc,
		chains,
		datavisualization.formats.functions,			
		decorations.pathmorphing,
		decorations.pathreplacing,
		decorations.text,
		fadings,
		fit,
		graphs,
		graphs.standard,
		matrix,
		positioning,									
		quantikz2,
		quotes,
		shadings,
		shadows,
		shadows.blur,
		shapes,
		shapes.geometric,
		trees
	}
\pgfplotsset{compat = newest}
\usepgflibrary
	{
		shapes.misc
	}
\usepgfplotslibrary
	{
		dateplot
	}
\usepackage{tikzpeople}
\usepackage{fontawesome5}
\usepackage{smartdiagram}

\allowdisplaybreaks
\usepackage{amsthm}
\usepackage{amssymb}
\usepackage[bb = boondox]{mathalfa}
\usepackage{dsfont}
\usepackage{newtxmath}
\usepackage{mathtools}
\usepackage{multicol}
\usepackage{physics2}
\usephysicsmodule{braket}
\usepackage{empheq}
\usepackage[ compat = newest ]{yquant}

\usepackage{sectsty}
\allsectionsfont{\color{WordBlue}}
\usepackage{xcolor}
\usepackage{varwidth}
\usepackage{graphicx}
\usepackage{wrapfig}
\usepackage{subcaption}
\usepackage{xurl}                                                                                  %
\usepackage{hyperref}
\hypersetup
	{
		colorlinks,
		linkcolor = {RedPurple!50!black},
		citecolor = {WordBlueDarker25},
		urlcolor = {WordAquaDarker50}
	}
\usepackage{tabularray}
\usepackage{colortbl}
\usepackage{float}
\usepackage{nicematrix}
\usepackage{cellspace}
\usepackage{makecell}
\usepackage{multirow}
\usepackage{caption}
\usepackage[ linesnumbered, tworuled, vlined ] {algorithm2e}
\usepackage{appendix}
\usepackage{enumitem}
\usepackage{siunitx}
\usepackage{xintexpr}
\usepackage{spreadtab}
\usepackage{framed}
\usepackage{svg}
\usepackage{lipsum}
\usepackage{listings}
\definecolor{MyLightRed}{RGB}{244, 213, 245}
\definecolor{Purple}{HTML}{911146}
\definecolor{PurpleDark}{RGB}{102, 0, 102}
\definecolor{RedDarkLight}{HTML}{ea005f}
\definecolor{RedDarkLightest}{HTML}{ff0088}
\definecolor{RedPurple}{HTML}{AA007F}
\definecolor{WordRed}{RGB}{255, 0, 102}
\definecolor{WordRedAccent5Lighter60}{HTML}{F5B5A7}
\definecolor{WordRedAccent5Darker25}{HTML}{B23214}
\definecolor{GreenDark}{HTML}{225522}
\definecolor{GreenLighter1}{HTML}{00B383}
\definecolor{GreenLighter2}{HTML}{00AA7F}
\definecolor{GreenLightest}{HTML}{00FFA0}
\definecolor{GreenTeal}{HTML}{008080}
\definecolor{WordLightGreen}{RGB}{140, 214, 192}
\definecolor{WordGreen}{RGB}{0, 176, 80}
\definecolor{BlueVeryDark}{HTML}{222255}
\definecolor{MyBlue}{RGB}{0, 64, 128}
\definecolor{MyDarkBlue}{RGB}{0, 51, 102}
\definecolor{MyVeryLightBlue}{RGB}{211, 245, 247}
\definecolor{WordBlue}{RGB}{19, 65, 99}
\definecolor{WordBlueDark}{RGB}{46, 116, 181}
\definecolor{WordBlueDarker}{RGB}{31, 78, 121}
\definecolor{WordBlueDarker25}{RGB}{54, 96, 146}
\definecolor{WordBlueDarker50}{RGB}{36, 64, 98}
\definecolor{WordBlueDarkest}{RGB}{0, 32, 96}
\definecolor{WordBlueLight}{RGB}{0, 112, 192}
\definecolor{WordBlueVeryLight}{HTML}{00B0F0}
\definecolor{WordIceBlue}{RGB}{223, 227, 229}
\definecolor{MagentaDark}{RGB}{106, 65, 152}
\definecolor{MagentaLight}{RGB}{128, 100, 162}
\definecolor{MagentaLighter}{RGB}{161, 106, 221}
\definecolor{MagentaVeryDark}{RGB}{97, 75, 128}
\definecolor{MagentaVeryLight}{RGB}{178, 162, 201}
\definecolor{WordAquaAccent2Darker25}{HTML}{398E98}
\definecolor{WordAquaDarker25}{HTML}{31869B}
\definecolor{WordAquaDarker50}{HTML}{215967}
\definecolor{WordAquaLighter40}{HTML}{92CDDC}
\definecolor{WordAquaLighter60}{HTML}{B7DEE8}
\definecolor{WordAquaLighter80}{HTML}{DAEEF3}
\definecolor{WordDarkerTeal}{RGB}{48, 82, 80}
\definecolor{WordDarkTeal}{RGB}{72, 123, 119}
\definecolor{WordDarkTealLighter80}{RGB}{207, 223, 234}
\definecolor{WordLightTeal}{RGB}{160, 199, 197}
\definecolor{WordVeryLightTeal}{RGB}{223, 236, 235}
\definecolor{WordTurquoiseLighter80}{RGB}{209, 238, 249}
\definecolor{Brown}{HTML}{666633}
\definecolor{WordGoldAccent1Darker25}{HTML}{C49A00}
\definecolor{WordGoldAccent1Lighter40}{HTML}{FFDF6A}
\definecolor{WordOrangeAccent2Lighter60}{HTML}{FCD3A4}
\definecolor{WordOrangeAccent4Lighter60}{HTML}{F7C5A1}
\definecolor{LavenderBlush}{RGB}{255, 240, 245}
\definecolor{MediumTurquoise}{RGB}{72, 209, 204}
\definecolor{PowderBlue}{RGB}{176, 224, 230}
\definecolor{SkyBlue}{RGB}{135, 206, 235}
\definecolor{Azure2}{RGB}{224, 238, 238}
\definecolor{Azure3}{RGB}{193, 205, 205}
\definecolor{CadetBlue4}{RGB}{83, 134, 139}
\definecolor{DarkSeaGreen1}{RGB}{193, 255, 193}
\definecolor{DeepPink4}{RGB}{139, 10, 80}
\definecolor{Honeydew2}{RGB}{224, 238, 224}
\definecolor{LightSkyBlue1}{RGB}{176, 226, 255}
\definecolor{LightSkyBlue3}{RGB}{141, 182, 205}
\definecolor{LightSkyBlue4}{RGB}{96, 123, 139}
\definecolor{LightSteelBlue1}{RGB}{202, 225, 255}
\definecolor{LightSteelBlue4}{RGB}{110, 123, 139}
\definecolor{MediumPurple1}{RGB}{171, 130, 255}
\definecolor{PaleTurquoise3}{RGB}{150, 205, 205}
\definecolor{PaleVioletRed3}{RGB}{205, 104, 137}
\definecolor{Purple1}{RGB}{155, 48, 255}
\definecolor{SeaGreen1}{RGB}{84, 255, 159}
\definecolor{SeaGreen2}{RGB}{78, 238, 148}
\definecolor{SeaGreen3}{RGB}{67, 205, 128}
\definecolor{SkyBlue1}{HTML}{87CEFF}
\definecolor{SkyBlue4}{RGB}{74, 112, 139}
\definecolor{SteelBlue1}{RGB}{99, 184, 255}
\definecolor{Thistle3}{RGB}{205, 181, 205}
\definecolor{Turquoise4}{RGB}{0, 134, 139}
\definecolor{VioletRed1}{RGB}{255, 62, 150}
\definecolor{VioletRed2}{RGB}{208, 32, 144}
\definecolor{VioletRed3}{RGB}{199, 21, 133}
\definecolor{VioletRed4}{RGB}{139, 10, 80}
\usepackage
	[
		skins,
		breakable,
		listings,
		raster,
		theorems
	]
	{tcolorbox}
\NewTcbTheorem
	[
		number within = section
	]
	{definition}
	{Definition}
	{
		breakable,
		skin = enhanced,
		enhanced jigsaw,			
		attach boxed title to top left = { xshift = -5.000 mm, yshift = 0.000 mm },
		boxed title style = { boxrule = 0.000 pt, sharp corners = all },
		varwidth boxed title,
		fonttitle = \bfseries,
		coltitle = SkyBlue!30!black,
		colbacktitle = SkyBlue!70!white,
		colframe = SkyBlue,
		colback = SkyBlue!20!white,
		sharp corners = all,
		toprule = 0.000 mm,
		bottomrule = 0.500 mm,
		leftrule = 0.500 mm,
		rightrule = 0.000 mm,
	}
	{def}
\NewTcbTheorem
	[
		number within = section
	]
	{example}
	{Example}
	{
		breakable,
		skin = enhanced,
		enhanced jigsaw,			
		attach boxed title to top left = { xshift = -5.000 mm, yshift = 0.000 mm },
		boxed title style = { boxrule = 0.000 pt, sharp corners = all },
		varwidth boxed title,
		fonttitle = \bfseries,
		coltitle = PowderBlue!30!black,
		colbacktitle = PowderBlue!70!white,
		colframe = PowderBlue,
		colback = white,
		sharp corners = all,
		toprule = 0.000 mm,
		bottomrule = 0.500 mm,
		leftrule = 0.500 mm,
		rightrule = 0.000 mm,
	}
	{xmp}
\NewTcbTheorem
	[
		number within = section
	]
	{theorem}
	{Theorem}
	{
		breakable,
		skin = enhanced,
		enhanced jigsaw,			
		attach boxed title to top left = { xshift = -5.000 mm, yshift = 0.000 mm },
		boxed title style = { boxrule = 0.000 pt, sharp corners = all },
		varwidth boxed title,
		fonttitle = \bfseries,
		coltitle = VioletRed3!20!black,
		colbacktitle = VioletRed2!30!white,
		colframe = VioletRed4,
		colback = VioletRed1!03,
		sharp corners = all,
		toprule = 0.000 mm,
		bottomrule = 0.500 mm,
		leftrule = 0.500 mm,
		rightrule = 0.000 mm,
	}
	{thr}
\NewTcbTheorem
	[
		number within = section
	]
	{proposition}
	{Proposition}
	{
		breakable,
		skin = enhanced,
		enhanced jigsaw,			
		attach boxed title to top left = { xshift = -5.000 mm, yshift = 0.000 mm },
		boxed title style = { boxrule = 0.000 pt, sharp corners = all },
		varwidth boxed title,
		fonttitle = \bfseries,
		coltitle = cyan5!30!black,
		colbacktitle = cyan7!50!white,
		colframe = cyan5,
		colback = cyan9!07,
		sharp corners = all,
		toprule = 0.000 mm,
		bottomrule = 0.500 mm,
		leftrule = 0.500 mm,
		rightrule = 0.000 mm,
	}
	{prp}
\NewTcbTheorem
	[
		number within = section
	]
	{corollary}
	{Corollary}
	{
		breakable,
		skin = enhanced,
		enhanced jigsaw,			
		attach boxed title to top left = { xshift = -5.000 mm, yshift = 0.000 mm },
		boxed title style = { boxrule = 0.000 pt, sharp corners = all },
		varwidth boxed title,
		fonttitle = \bfseries,
		coltitle = cyan5!30!black,
		colbacktitle = cyan7!50!white,
		colframe = cyan5,
		colback = cyan9!07,
		sharp corners = all,
		toprule = 0.000 mm,
		bottomrule = 0.500 mm,
		leftrule = 0.500 mm,
		rightrule = 0.000 mm,
	}
	{crl}

\newcounter{mathseed}
\pgfmathsetseed{\arabic{mathseed}}
\pgfdeclarelayer{background}
\pgfsetlayers{background, main}
\pgfdeclaredecoration{irregular fractal line}{init}
{
	\state{init}[width=\pgfdecoratedinputsegmentremainingdistance]
	{
		\pgfpathlineto{\pgfpoint{random*\pgfdecoratedinputsegmentremainingdistance}{(random*\pgfdecorationsegmentamplitude-0.02)*\pgfdecoratedinputsegmentremainingdistance}}
		\pgfpathlineto{\pgfpoint{\pgfdecoratedinputsegmentremainingdistance}{0pt}}
	}
}
\def\tornpaper#1{%
	\ifthenelse{\isodd{\value{mathseed}}}
	{%
		\tikz
		{
			\node[inner sep = 1em] (A) {#1};		
			\begin{pgfonlayer}{background}			
				\fill[paper]						
				\pgfextra{\pgfmathsetseed{\arabic{mathseed}}\addtocounter{mathseed}{1}}%
				{decorate[irregular cloudy border]{decorate{decorate{decorate{decorate[ragged border]{
										(A.north west) -- (A.north east)
				}}}}}}
				-- (A.south east)
				\pgfextra{\pgfmathsetseed{\arabic{mathseed}}}%
				{decorate[irregular spiky border]{decorate{decorate{decorate{decorate[ragged border]{
										-- (A.south west)
				}}}}}}
				-- (A.north west);
			\end{pgfonlayer}
		}
	}
	{%
		\tikz{
			\node[inner sep=1em] (A) {#1};  
			\begin{pgfonlayer}{background}  
				\fill[paper] 
				\pgfextra{\pgfmathsetseed{\arabic{mathseed}}\addtocounter{mathseed}{1}}%
				{decorate[irregular spiky border]{decorate{decorate{decorate{decorate[ragged border]{
										(A.north east) -- (A.north west)
				}}}}}}
				-- (A.south west)
				\pgfextra{\pgfmathsetseed{\arabic{mathseed}}}%
				{decorate[irregular cloudy border]{decorate{decorate{decorate{decorate[ragged border]{
										-- (A.south east)
				}}}}}}
				-- (A.north east);
		\end{pgfonlayer}}
	}
}
\title
	{
		Probabilistic Links Between Quantum Classification of Patterns of Boolean Functions and Hamming Distance
	}

\usepackage{orcidlink}
\newcommand{\orcidicon}[1]{\href{https://orcid.org/#1}{\includegraphics[height=\fontcharht\font`\B]{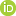}}}

\author
{
	Theodore Andronikos$^1$\orcidicon{0000-0002-3741-1271},
	Constantinos Bitsakos$^2$\orcidicon{0009-0003-3669-0453},
	Konstantinos Nikas$^2$\orcidicon{0000-0003-4424-9951},
	\\
	Georgios I. Goumas$^2$\orcidicon{0000-0001-7811-4831}
	and
	Nectarios Koziris$^2$\orcidicon{0000-0002-4890-8427}
	\\
	$^1$ \
	Department of Informatics, Ionian University, \\
	7 Tsirigoti Square, 49100 Corfu, Greece; \\
	andronikos@ionio.gr
	\\
	$^2$ \ 
	Computing Systems Laboratory, \\
	National Technical University of Athens, Greece; \\
	\{kbitsak, knikas, goumas, nkoziris\}@cslab.ece.ntua.gr
}
\begin{document}

\maketitle

\begin{abstract}
	This article investigates the probabilistic relationship between quantum classification of Boolean functions and their Hamming distance. By integrating concepts from quantum computing, information theory, and combinatorics, we explore how Hamming distance serves as a metric for analyzing deviations in function classification. Our extensive experimental results confirm that the Hamming distance is a pivotal metric for validating nearest neighbors in the process of classifying random functions. One of the significant conclusions we arrived is that the successful classification probability decreases monotonically with the Hamming distance. However, key exceptions were found in specific classes, revealing intra-class heterogeneity. We have established that these deviations are not random but are systemic and predictable. Furthermore, we were able to quantify these irregularities, turning potential errors into manageable phenomena. The most important novelty of this work is the demarcation, for the first time to the best of our knowledge, of precise Hamming distance intervals for the classification probability. These intervals bound the possible values the probability can assume, and provide a new foundational tool for probabilistic assessment in quantum classification. Practitioners can now endorse classification results with high certainty or dismiss them with confidence. This framework can significantly enhance any quantum classification algorithm's reliability and decision-making capability.
\end{abstract}
\section{Introduction} \label{sec: Introduction}

Today Quantum computing is advancing rapidly, with recent breakthroughs suggesting that transformative quantum systems may soon emerge, even though current quantum computers are not yet capable of fully disrupting classical paradigms. Major industry players have made significant strides.
IBM introduced the 1,121-qubit Condor and high-performance R2 Heron, building on the 127-qubit Eagle \cite{IBMEagle2021} and 433-qubit Osprey \cite{IBMOsprey2022} \cite{IBMCondor2023, IBMHeron2024}.
Google demonstrated quantum computers outperforming advanced supercomputers \cite{GoogleWillow2024, NatureGoogle2024}.
Microsoft unveiled the Majorana 1 quantum chip \cite{Aasen2025, Aghaee2025, MicrosoftMajorana2025}
D-Wave’s quantum annealer solved a scientifically significant problem faster than classical computers \cite{King2025, NatureD-Wave2025}.
China’s 105-qubit Zuchongzhi 3.0 processor marked a technological milestone \cite{Gao2025, APSNewsZuchongzhi3.0-2025}.
Other innovations include novel design concepts \cite{Cacciapuoti2024, Illiano2024}, hardware advancements \cite{Photonic2024, NuQuantum2024}, and distributed quantum computing, where two processors were linked via a photonic network to function as a single system \cite{Main2025, OxfordNewsEvents2025}. These developments underscore the growing feasibility of distributed quantum systems and their potential for applications like quantum classification of Boolean functions.

This study explores quantum classification of Boolean functions from a novel perspective, focusing on probabilistic correlations with Hamming distance. Quantum algorithms can perfectly classify functions with specific properties (probability 1.0), but our goal is to understand classification behavior when functions deviate from these properties, using Hamming distance as a metric. The Deutsch-Jozsa algorithm \cite{Deutsch1992} pioneered this field, followed by extensions such as multidimensional variants \cite{Cleve1998}, analyses of balanced functions \cite{Chi2001, Holmes2003}, and generalizations \cite{Ballhysa2004, Qiu2018, OssorioCastillo2023}. Practical applications \cite{Qiu2020, Zhengwei2021} and distributed versions \cite{Tanasescu2019, Li2025} further advanced the field, while \cite{Nagata2020} extended Deutsch’s algorithm to binary Boolean functions. Quantum learning literature often focuses on distinguishing constant and balanced functions \cite{Bshouty1998, Farhi1999, Servedio2004, Hunziker2009, Yoo2014, Cross2015}, but a recent study \cite{Andronikos2025a} targets imbalanced functions, proposing the Boolean Function Pattern Quantum Classifier (BFPQC) for classifying functions with specific behavioral patterns.

We present our exposition as a quantum game featuring Alice and Bob, using the engaging nature of games to clarify complex concepts. Quantum games, popularized since 1999 \cite{Meyer1999, Eisert1999}, often outperform classical strategies \cite{Andronikos2018,Andronikos2021,Andronikos2022a}, as exemplified by the Prisoners’ Dilemma \cite{Eisert1999} and other abstract games \cite{Koh2024}. Beyond entertainment, quantum games have addressed serious problems like cryptographic protocols \cite{Bennett1984,Ampatzis2021, Ampatzis2022,Ampatzis2023,Andronikos2023,Andronikos2023a,Andronikos2023b,Karananou2024, Andronikos2024,Andronikos2024a,Andronikos2024b,Andronikos2025,Andronikos2025c}, and quantum classification of Boolean functions \cite{Andronikos2025a,Andronikos2025b}. Our game-based framework makes the probabilistic links between quantum classification and Hamming distance more accessible, offering a novel tool for advancing quantum algorithm design.
Furthermore, many classical systems can be transformed into quantum versions, including political frameworks, as demonstrated in recent studies \cite{Andronikos2022}. Games make be envisioned in unconventional situations, such as those featuring biological systems, which have actually attracted a lot of attention during the last decade \cite{Theocharopoulou2019,Kastampolidou2020a,Kostadimas2021}. It may even be the case to discover biosystems that employ biostrategies that perform better than conventional strategies—even in the well-known Prisoners' Dilemma game (see for example \cite{Kastampolidou2020,Kastampolidou2021,Papalitsas2021,Kastampolidou2023,Adam2023}).
This perspective reinforces our belief that the game-theoretic framework may facilitate a deeper understanding of quantum classification algorithms.

\textbf{Contribution}. This work capitalizes upon recent prior studies, such as \cite{Andronikos2025a, Andronikos2025b}, to investigate the usefulness of the Hamming distance as metric between Boolean functions, that can yield meaningful insights when the quantum classification algorithm encounters an unknown function outside its scope. This is a challenging benchmark that stresses the robustness and versatility limits of any quantum classification algorithm. The main strengths of this work are summarized below.

\begin{itemize}
	\item
	Our extensive experimental results robustly validate the Hamming distance as a critical and effective metric for verifying or disqualifying potential nearest neighbor candidates in the classification process.
	\item
	As anticipated, we confirm that the classification threshold is a monotonically decreasing function of the Hamming distance. This foundational trend, however, is punctuated by the notable exceptions identified for specific classes in subsection \ref{subsec: Interpretation Of The Experimental Results}. These exceptions are not mere statistical artifacts; they provide crucial insight into intra-class heterogeneity, demonstrating how individual functions can cause significant deviations from the global behavioral pattern.
	\item
	Crucially, our analysis reveals that even these deviations are systemic. The irregularities observed in special classes are not random but can be rigorously quantified and predicted. This transforms a potential source of error into a manageable, well-understood aspect of the classification landscape.
	\item
	To the best of our knowledge, this work is the first to establish precise intervals for Hamming distances that bound the possible values of the classification threshold. By quantifying this relationship, we provide a foundational tool for probabilistic reasoning within quantum classification paradigms.
	\item
	This capability is of paramount practical importance. It equips practitioners with a powerful framework to either:
	\begin{enumerate}
		\item	confidently endorse a classification result with high certainty when it is likely a true nearest neighbor, or
		\item	decisively dismiss a result when the probability of it being a genuine nearest neighbor is vanishingly small.
	\end{enumerate}
	This directly enhances the reliability and efficiency of the classification system.
\end{itemize}

\subsection*{Organization} \label{subsec: Organization}

This article is structured in the following way. Section \ref{sec: Introduction} introduces the topic and provides references to relevant literature. Section \ref{sec: Preliminary Concepts} contains a brief overview of key concepts and technical details. Section \ref{sec: Perfectly Classifiable Boolean Functions} defined the notion of perfectly classifiable Boolean function, while the following Section \ref{sec: What About Nonclassifiable Boolean Functions?} introduces the theoretical machinery that establishes the framework for classifying functions outside of the promised class. Section \ref{sec: What Do Experiments Show?} presents and analyzes the entire suite of experiments we have undertaken in this work. Finally, the paper wraps up with a summary and a discussion of the Nearest Basis Ket Game in Section \ref{sec: Discussion and Conclusions}.

\section{Preliminary concepts} \label{sec: Preliminary Concepts}

This section introduces the notation and terminology that will be used throughout this work.

\begin{itemize}
	\item	
	$\mathbb{ B }$ denotes the binary set $\{ 0, 1 \}$.
	\item	
	A \emph{bit vector} $\mathbf{ b }$ is a sequence of $n$ bits: $b_{ n - 1 } \dots b_{ 0 }$. For succinctness it is helpful to consider it as an element of $\mathbb{ B }^{ n }$ and write $\mathbf{ b } \in \mathbb{ B }^{ n }$. The special bit vectors $\mathbf{ 0 } = 0 \dots 0$ and $\mathbf{ 1 } = 1 \dots 1$ represent vectors with all bits set to zero and one, respectively.
	\item	
	Bit vectors $\mathbf{ b } \in \mathbb{ B }^{ n }$ are written in boldface to distinguish them from scalar quantities. Throughout this work, $\mathbf{ b }$ will also be interpreted as the binary representation of an integer $b$.
	\item	
	Each bit vector $\mathbf{ b } \in \mathbb{ B }^{ n }$ corresponds to one of the $2^{ n }$ basis kets forming the computational basis of a $2^{ n }$-dimensional Hilbert space.
\end{itemize}

The Hamming distance is a metric used to quantify the difference between two strings of equal length over a fixed alphabet. In our setting, which involves bit vectors, if we are given two bit vectors $\mathbf{ a }, \mathbf{ b } \in \mathbb{ B }^{ n }$, their Hamming distance $d_H ( \mathbf{ a }, \mathbf{ b } )$ is defined as the number of positions at which the corresponding bits differ.

\begin{definition} {Hamming Distance} { Hamming Distance}
	Let $\mathbf{ a } = a_{ n - 1 } \dots a_{ 0 }$ and $\mathbf{ b } = b_{ n - 1 } \dots b_{ 0 }$ be two binary strings in $\mathbb{ B }^{ n }$, $n \geq 1$. Their Hamming distance, denoted by $d_{ H } ( \mathbf{ a }, \mathbf{ b } )$, is given by:
	\begin{align}
		\label{eq: Hamming Distance}
		d_{ H } ( \mathbf{ a }, \mathbf{ b } )
		\coloneq
		\sum_{ i = 0 }^{ n - 1 }
		a_{ i }
		\oplus
		b_{ i }
		\ ,
	\end{align}
	where $\oplus$ stands for the XOR operation, since $a_{ i } \oplus b_{ i } = 1$ if and only if $a_{ i } \neq b_{ i }$.
\end{definition}

\begin{definition} {Boolean Function} { Boolean Function}
	A Boolean function $f \colon \mathbb{ B }^{ n } \to \mathbb{ B }$, $n \geq 1$, maps bit vectors to a single bit.
\end{definition}

Oracles are a cornerstone of quantum computing, acting as black-box components in many quantum algorithms. They encapsulate a function or specific information within a quantum circuit, enabling efficient problem-solving compared to classical methods in certain cases. Oracles evaluate a function or verify conditions without revealing their internal workings, often providing insights into a function’s properties or identifying solutions. For this article, we adopt the following definition.

\begin{definition} {Oracle \& Unitary Transform} { Oracle & Unitary Transform}
	An \emph{oracle} is a black box that implements a Boolean function $f$. As a black box, its internal mechanics are opaque, but it is assumed to operate correctly. The oracle enables the construction of a unitary transform $U_{ f }$ that encodes the behavior of $f$ according to the standard schema:
	\begin{align}
		\label{eq: Generic Unitary Transform U_f}
		U_{ f }
		\colon
		\ket{ y }
		\
		\ket{ \mathbf{ x } }
		\rightarrow
		\ket{ y \oplus f( \mathbf{ x } ) }
		\
		\ket{ \mathbf{ x } }
		\ .
	\end{align}
\end{definition}

This oracle is commonly referred to as a Deutsch-Jozsa oracle. Other oracle types, such as the Grover oracle, are used for different purposes, such as solution identification. In this work, we assume all oracles and their unitary transforms follow Equation \eqref{eq: Generic Unitary Transform U_f} and are used to infer properties of a function based on its behavior. The complexity of oracle-based algorithms is typically measured by query complexity, defined as the number of oracle queries made by the algorithm.

For reference, we define the states $\ket{+}$ and $\ket{-}$ as follows:

\begin{tcolorbox}
	[
		enhanced,
		breakable,
		center title,
		fonttitle = \bfseries,
		grow to left by = 0.000 cm,
		grow to right by = 0.000 cm,
		colback = white,			
		enhanced jigsaw,			
		sharp corners,
		toprule = 0.001 pt,
		bottomrule = 0.001 pt,
		leftrule = 0.001 pt,
		rightrule = 0.001 pt,
	]
	\begin{minipage} [ b ] { 0.475 \textwidth }
		\begin{align}
			\label{eq: Ket +}
			\ket{ + }
			=
			H
			\ket{ 0 }
			=
			\frac
			{ \ket{ 0 } + \ket{ 1 } }
			{ \sqrt{ 2 } }
		\end{align}
	\end{minipage}
	\hfill	
	\begin{minipage} [ b ] { 0.475 \textwidth }
		\begin{align}
			\label{eq: Ket -}
			\ket{ - }
			=
			H
			\ket{ 1 }
			=
			\frac
			{ \ket{ 0 } - \ket{ 1 } }
			{ \sqrt{ 2 } }
		\end{align}
	\end{minipage}
\end{tcolorbox}

To extract useful information from the schema in equation \eqref{eq: Generic Unitary Transform U_f}, we set $\ket{ y } = \ket{-}$, yielding the following form:

\begin{align}
	\label{eq: Unitary Transform U_f}
	U_{ f }
	\colon
	\ket{ - }
	\
	\ket{ \mathbf{ x } }
	\rightarrow
	( - 1 )^{ f ( \mathbf{ x } ) }
	\
	\ket{ - }
	\
	\ket{ \mathbf{ x } }
	\ .
\end{align}

Figures \ref{fig: The Quantum Circuit for the Generic Unitary Transform U_f} and \ref{fig: The Quantum Circuit for the Unitary Transform U_f} illustrate the unitary transforms $U_{ f }$ implementing schemata \eqref{eq: Generic Unitary Transform U_f} and \eqref{eq: Unitary Transform U_f}, respectively.

\begin{tcolorbox}
	[
		enhanced,
		breakable,
		center title,
		fonttitle = \bfseries,
		colbacktitle = SkyBlue4,
		grow to left by = 0.000 cm,
		grow to right by = 0.000 cm,
		colback = SkyBlue1!08,
		enhanced jigsaw,			
		frame hidden,
		sharp corners,
	]
	\begin{minipage} [ b ] { 0.450 \textwidth }
		\begin{figure}[H]
			\centering
			\begin{tikzpicture} [ scale = 0.800, transform shape ]
				\begin{yquant}
					qubits { $IR \colon \ket{ \mathbf{ x } }$ } IR;
					qubit { $OR \colon \ket{ y }$ } OR;
					[ name = Input, WordBlueDarker, line width = 0.250 mm, label = { [ label distance = 0.600 cm ] north: Input } ]
					barrier ( - ) ;
					hspace { 0.100 cm } IR;
					[ draw = SkyBlue4, fill = SkyBlue4, x radius = 0.900 cm, y radius = 0.700 cm ] box { \color{white} \Large \sf{U}$_{ f }$} (-);
					[ name = Output, WordBlueDarker, line width = 0.250 mm, label = { [ label distance = 0.600 cm ] north: Output } ]
					barrier ( - ) ;
					output { $\ket{ \mathbf{ x } }$ } IR;
					output { $\ket{ \mathbf{ y \oplus f ( \mathbf{ x } ) } }$ } OR;
					\node [ below = 1.250 cm ] at (Input) { $\ket{ \psi_{ i } }$ };
					\node [ below = 1.250 cm ] at (Output) { $\ket{ \psi_{ o } }$ };
				\end{yquant}
				\node [ anchor = center, below = 1.250 cm of Input ] (PhantomNode) { };
			\end{tikzpicture}
			\caption{This figure shows the unitary transform $U_{ f }$, which is based on the oracle for the function $f$ and implements the standard schema \eqref{eq: Generic Unitary Transform U_f}.}
			\label{fig: The Quantum Circuit for the Generic Unitary Transform U_f}
		\end{figure}
	\end{minipage}
	\hspace{ 0.750 cm }
	\begin{minipage} [ b ] { 0.450 \textwidth }
		\begin{figure}[H]
			\centering
			\begin{tikzpicture} [ scale = 0.800, transform shape ]
				\begin{yquant}
					qubits { $IR \colon \ket{ \mathbf{ x } }$ } IR;
					qubit { $OR \colon \ket{ - }$ } OR;
					[ name = Input, WordBlueDarker, line width = 0.250 mm, label = { [ label distance = 0.600 cm ] north: Input } ]
					barrier ( - ) ;
					hspace { 0.100 cm } IR;
					[ draw = SkyBlue4, fill = SkyBlue4, x radius = 0.900 cm, y radius = 0.700 cm ] box { \color{white} \Large \sf{U}$_{ f }$} (-);
					[ name = Output, WordBlueDarker, line width = 0.250 mm, label = { [ label distance = 0.600 cm ] north: Output } ]
					barrier ( - ) ;
					output { $\ket{ \mathbf{ x } }$ } IR;
					output { $( - 1 )^{ f ( \mathbf{ x } ) } \ket{ - }$ } OR;
					\node [ below = 1.250 cm ] at (Input) { $\ket{ \psi_{ i } }$ };
					\node [ below = 1.250 cm ] at (Output) { $\ket{ \psi_{ o } }$ };
				\end{yquant}
				\node [ anchor = center, below = 1.250 cm of Input ] (PhantomNode) { };
			\end{tikzpicture}
			\caption{This figure shows the unitary transform $U_{ f }$, again based on the oracle for the function $f$, but now implementing the schema \eqref{eq: Unitary Transform U_f}.}
			\label{fig: The Quantum Circuit for the Unitary Transform U_f}
		\end{figure}
	\end{minipage}
\end{tcolorbox}

All quantum circuits in this work, including those in Figures \ref{fig: The Quantum Circuit for the Generic Unitary Transform U_f} and \ref{fig: The Quantum Circuit for the Unitary Transform U_f}, adhere to the following conventions:

\begin{itemize}
	\item	
	Qubits are arranged according to the Qiskit convention \cite{Qiskit2025}, using little-endian qubit indexing, where the least significant qubit appears at the top and the most significant at the bottom.
	\item	
	$IR$ denotes the quantum input register, comprising a specified number of qubits.
	\item	
	$OR$ is the single-qubit output register, initialized to an arbitrary state $\ket{y}$ in Figure \ref{fig: The Quantum Circuit for the Generic Unitary Transform U_f} and to $\ket{-}$ in Figure \ref{fig: The Quantum Circuit for the Unitary Transform U_f}.
	\item	
	$U_{ f }$ represents the unitary transform, with its exact form depending on $f$ and remaining unspecified, but satisfying equation \eqref{eq: Generic Unitary Transform U_f} in Figure \ref{fig: The Quantum Circuit for the Generic Unitary Transform U_f} and equation \eqref{eq: Unitary Transform U_f} in Figure \ref{fig: The Quantum Circuit for the Unitary Transform U_f}.
\end{itemize}

In the literature, the term ``promise'' refers to a guaranteed property of the Boolean function $f$, meaning $f$ satisfies the property with certainty (probability $1.0$). A classic example is the Deutsch–Jozsa algorithm, which assumes $f$ is either \emph{balanced} or \emph{constant}.

We frequently refer to the following state, termed the \emph{perfect superposition state}:

\begin{align}
	\label{eq: The Perfect SuperPosition State}
	\ket{ \varphi }
	&=
	2^{ - \frac { n } { 2 } }
	\sum_{ \mathbf{ x } \in \mathbb{ B }^{ n } }
	\ket{ \mathbf{ x } }
	\ ,
\end{align}

which serves as the standard starting point for most quantum algorithms. It is typically generated by applying the $n$-fold Hadamard transform to the initial state $\ket{0}^{\otimes n}$.

Bitwise addition modulo $2$ extends the XOR operation to bit vectors, providing a natural generalization.

\begin{definition} {Bitwise Addition Modulo $2$} { Bitwise Addition Modulo $2$}
	For two bit vectors $\mathbf{ x }, \mathbf{ y } \in \mathbb{ B }^{ n }$, with $\mathbf{ x } = x_{ n - 1 } \dots x_{ 0 }$ and $\mathbf{ y } = y_{ n - 1 } \dots y_{ 0 }$, their bitwise sum modulo $2$, denoted by $\mathbf{ x } \oplus \mathbf{ y }$, is defined as:
	\begin{align}
		\label{eq: Bitwise Addition Modulo $2$}
		\mathbf{ x }
		\oplus
		\mathbf{ y }
		\coloneq
		( x_{ n - 1 } \oplus y_{ n - 1 } )
		\dots
		( x_{ 0 } \oplus y_{ 0 } )
		\ .
	\end{align}
\end{definition}

We use the symbol $\oplus$ for both single-bit addition modulo $2$ and bitwise addition modulo $2$, as the context clarifies the intended operation.

\section{Perfectly classifiable Boolean functions} \label{sec: Perfectly Classifiable Boolean Functions}

Quantum Boolean classification algorithms typically enable the classification of Boolean functions that satisfy specific properties with a probability of $1.0$. In the literature, these functions are often described as satisfying a ``promise,'' meaning they are guaranteed to possess certain characteristics, giving rise to the term ``promise problem.'' For example, in the quantum classification algorithm introduced in \cite{Andronikos2025a}, the promise is that the behavior of the classifiable Boolean functions expresses ``imbalance'' in the number of zero and one values they assume. 

To facilitate the analysis of Boolean functions and their classification, we shall introduce the concept of \emph{pattern bit vectors}, which provide a compact and computationally convenient representation of Boolean functions. This abstraction enables the use of the Hamming distance between two pattern bit vectors as a measure of the dissimilarity between their corresponding Boolean functions.

\begin{definition} {Pattern Bit Vector} { Pattern Vector}
	Let $f \colon \mathbb{ B }^{ n } \to \mathbb{ B }$, $n \geq 1$, be a Boolean function. We define the \emph{pattern bit vector} of $f$ as a unique bit vector that encodes the function’s output values across all possible inputs. The formal construction is as follows:
	\begin{itemize}
		\item	
		The pattern bit vector $\mathbf{ p }_{ f } = p_{ 2^{ n } - 1} \dots p_{ 1 } p_{ 0 } \in \mathbb{ B }^{ 2^{ n } }$ is defined such that $p_{ i } = f ( \mathbf { i } )$, where $\mathbf{ i } \in \mathbb{ B }^{ n }$ is the binary bit vector representing the integer $i$, for $0 \leq i \leq 2^{ n } - 1$. Thus, $\mathbf{ p }_{ f }$ systematically records the values of $f ( \mathbf{ i } )$ as $\mathbf{ i }$ ranges over all possible inputs in $\mathbb{ B }^{ n }$.
		\begin{tblr}
			{
				colspec =
				{
					Q [ r, m, 4.000 cm ]
					Q [ c, m, 1.400 cm ]
					Q [ c, m, 0.350 cm ]
					Q [ c, m, 1.400 cm ]
					Q [ c, m, 0.000 cm ]
					Q [ c, m, 1.400 cm ]
				},
				rowspec =
				{
					| [ 6.000 pt, SkyBlue!20!white ]
					Q
					Q
					Q
					Q
					Q
					Q
					Q
					| [ 6.000 pt, SkyBlue!20!white ]
				}
			}
			\emph{Position:}
			&
			\SetCell { bg = RedPurple!75, fg = black } $2^{ n } - 1$
			&
			\dots
			&
			\SetCell { bg = RedPurple!50, fg = black } $1$
			&
			&
			\SetCell { bg = RedPurple!25, fg = black } $0$
			\\
			&
			$\downarrow$
			&
			\dots
			&
			$\downarrow$
			&
			&
			$\downarrow$
			\\
			\emph{Position in binary:}
			&
			\SetCell { bg = GreenLighter2!75, fg = black } $\mathbf{ 1 \dots 1 1 }$
			&
			\dots
			&
			\SetCell { bg = GreenLighter2!50, fg = black } $\mathbf{ 0 \dots 0 1 }$
			&
			&
			\SetCell { bg = GreenLighter2!25, fg = black } $\mathbf{ 0 \dots 0 0 }$
			\\
			&
			$\downarrow$
			&
			\dots
			&
			$\downarrow$
			&
			&
			$\downarrow$
			\\
			\emph{Function value:}
			&
			\SetCell { bg = WordBlueVeryLight!75, fg = black } $f ( \mathbf{ 1 \dots 1 1 } )$
			&
			\dots
			&
			\SetCell { bg = WordBlueVeryLight!50, fg = black } $f ( \mathbf{ 0 \dots 0 1 } )$
			&
			&
			\SetCell { bg = WordBlueVeryLight!25, fg = black } $f ( \mathbf{ 0 \dots 0 0 } )$
			\\
			&
			$\downarrow$
			&
			\dots
			&
			$\downarrow$
			&
			&
			$\downarrow$
			\\
			\emph{Pattern bit:}
			&
			\SetCell { bg = cyan7, fg = black } $p_{ 2^{ n } - 1 }$
			&
			\dots
			&
			\SetCell { bg = cyan8, fg = black } $p_{ 1 }$
			&
			&
			\SetCell { bg = cyan9, fg = black } $p_{ 0 }$
		\end{tblr}
		\item	
		The \emph{negation} of the pattern bit vector, denoted $\overline{ \mathbf{ p }_{ f } } = \overline{ p_{ 2^{ n } - 1 } } \dots \overline{ p_{ 1 } } \ \overline{p_{ 0 } }$, is the pattern bit vector corresponding to the negated function $\overline{ f }$, where $\overline{ f } ( \mathbf{ i } ) = \neg f ( \mathbf{ i } )$ for all $\mathbf{ i } \in \mathbb{ B }^{ n }$.
	\end{itemize}
\end{definition}

\begin{definition} {Function Realizing Pattern Vector} { Function Realizing Pattern Vector}
	Conversely, given any pattern bit vector $\mathbf{ p }$ $=$ $p_{ 2^{ n } - 1 }$ $\dots$ $p_{ 1 }$ $p_{ 0 }$ $\in$ $\mathbb{ B }^{ 2^{ n } }$, we define the Boolean function $f_{ \mathbf{ p } } \colon \mathbb{ B }^{ n } \to \mathbb{ B }$ such that $f_{ \mathbf{ p } } ( \mathbf{ i } ) = p_{ i }$, where $\mathbf{ i } \in \mathbb{ B }^{ n }$ is the binary bit vector representing the integer $i$, for $0 \leq i \leq 2^{ n } - 1$. This establishes a unique Boolean function $f_{ \mathbf{ p } }$ associated with the pattern bit vector $\mathbf{ p }$, termed the \emph{realization} of $\mathbf{ p }$.
\end{definition}

The preceding definitions establish a bijective correspondence between Boolean functions $f \colon \mathbb{ B }^{ n } \to \mathbb{ B }$ and their pattern bit vectors in $\mathbb{ B }^{ 2^{ n } }$. In essence, a Boolean function and its pattern bit vector are two representations of the same mathematical object, akin to dual perspectives of a single entity. Knowing the behavior of a Boolean function allows the construction of its pattern bit vector, while the pattern bit vector fully encodes the information required to reconstruct the function. This duality, visualized in Figure \ref{fig: Boolean Function - Pattern Vector Equivalence}, is particularly advantageous, as it enables simple and efficient manipulation of functions via their bit vector representations.

\begin{tcolorbox}
	[
		enhanced,
		breakable,
		grow to left by = 0.000 cm,
		grow to right by = 0.000 cm,
		colback = SkyBlue1!08,
		enhanced jigsaw,			
		frame hidden,
		sharp corners,
	]
	\begin{figure}[H]
		\centering
		\begin{tikzpicture} 
			[
			scale = 1.000,
			node distance = 1.000 cm,
			concept/.style =
			{
				circle,						
				thick,						
				inner sep = 1.000 pt,		
				text width = 2.250 cm,		
				align = center,				
				font = \sffamily,			
				text = black				
			}
			]
			\node [ concept, shade, outer color = RedPurple!75, inner color = white ] (A) at ( 0, 0 ) { \textbf{Boolean function} };
			\node [ concept, shade, outer color = WordBlueVeryLight, inner color = white ] (B) at ( 8.000, 0 ) { \textbf{Pattern Bit Vector} };
			\draw [ Stealth-Stealth, line width = 7.750 pt, GreenLighter2 ] (A) -- (B);
		\end{tikzpicture}
		\caption{The duality between Boolean functions and their pattern bit vectors.}
		\label{fig: Boolean Function - Pattern Vector Equivalence}
	\end{figure}
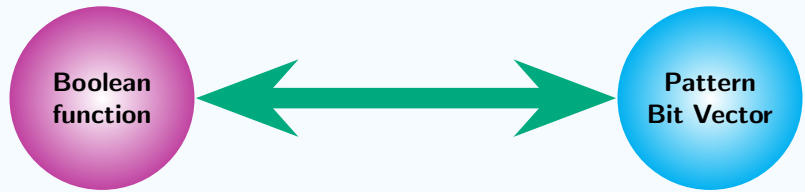
\end{tcolorbox}

\begin{example} {Functions \& Patterns} { Functions & Patterns}

For example, consider a Boolean function $f \colon \mathbb{ B }^{ 2 } \to \mathbb{ B }$ defined by $f ( 00 ) = 0$, $f ( 01 ) = 1$, $f ( 10 ) = 0$, and $f ( 11 ) = 1$. Its pattern bit vector is $\mathbf{ p }_{ f } = p_{ 3 } p_{ 2 } p_{ 1 } p_{ 0 } = f ( 11 ) f ( 10 ) f ( 01 ) f ( 00 ) = 1010$. Conversely, given the pattern bit vector $\mathbf{ p }_{ f } = 1010$, we can reconstruct the function $f_{ \mathbf{ p } }$ with the same input-output mappings. This equivalence is visually illustrated in Figure \ref{fig: Boolean Function - Pattern Vector Equivalence}, which highlights the one-to-one correspondence between Boolean functions and their pattern bit vectors. The concept of pattern bit vector provides a succinct way to represent the behavior of more complex Boolean functions, such as function $g \colon \mathbb{ B }^{ 4 } \to \mathbb{ B }$, which takes the values shown in Table \ref{tbl: Truth Values of the Functions g} and is described by the pattern vector $\mathbf{ p }_{ g } = 0000 \ 0001 \ 0001 \ 1110$.

\begin{tcolorbox}
	[
		enhanced,
		breakable,
		grow to left by = 0.500 cm,
		grow to right by = 2.000 cm,
		colback = white,
		enhanced jigsaw,			
		frame hidden,
		sharp corners,
	]
	\begin{table}[H]
		{\small
			\caption{Truth values of the function $g \colon \mathbb{ B }^{ 4 } \to \mathbb{ B }$.}
			\label{tbl: Truth Values of the Functions g}
			\centering
			\SetTblrInner { rowsep = 1.000 mm }
			\begin{tblr}
				{
					colspec =
					{
						Q [ c, m, 0.500 cm ]
						| [ 1.000 pt, WordAquaDarker25 ]
						| [ 1.000 pt, WordAquaDarker25 ]
						Q [ c, m, 0.450 cm ]
						| [ 0.500 pt, WordAquaDarker25 ]
						Q [ c, m, 0.450 cm ]
						| [ 0.500 pt, WordAquaDarker25 ]
						Q [ c, m, 0.450 cm ]
						| [ 0.500 pt, WordAquaDarker25 ]
						Q [ c, m, 0.450 cm ]
						| [ 0.500 pt, WordAquaDarker25 ]
						Q [ c, m, 0.450 cm ]
						| [ 0.500 pt, WordAquaDarker25 ]
						Q [ c, m, 0.450 cm ]
						| [ 0.500 pt, WordAquaDarker25 ]
						Q [ c, m, 0.450 cm ]
						| [ 0.500 pt, WordAquaDarker25 ]
						Q [ c, m, 0.450 cm ]
						| [ 0.500 pt, WordAquaDarker25 ]
						Q [ c, m, 0.450 cm ]
						| [ 0.500 pt, WordAquaDarker25 ]
						Q [ c, m, 0.450 cm ]
						| [ 0.500 pt, WordAquaDarker25 ]
						Q [ c, m, 0.450 cm ]
						| [ 0.500 pt, WordAquaDarker25 ]
						Q [ c, m, 0.450 cm ]
						| [ 0.500 pt, WordAquaDarker25 ]
						Q [ c, m, 0.450 cm ]
						| [ 0.500 pt, WordAquaDarker25 ]
						Q [ c, m, 0.450 cm ]
						| [ 0.500 pt, WordAquaDarker25 ]
						Q [ c, m, 0.450 cm ]
						| [ 0.500 pt, WordAquaDarker25 ]
						Q [ c, m, 0.450 cm ]
					},
					rowspec =
					{
						|
						[ 3.500 pt, WordAquaDarker25 ]
						|
						[ 0.750 pt, WordAquaDarker25 ]
						|
						[ 0.250 pt, white ]
						Q
						|
						[ 0.500 pt, WordAquaDarker25 ]
						Q
						|
						[ 3.500 pt, WordAquaDarker50 ]
					}
				}
				&
				\SetCell { bg = WordAquaLighter40, fg = black } $\mathbf{ 0000 }$
				&
				\SetCell { bg = WordAquaLighter40, fg = black } $\mathbf{ 0001 }$
				&
				\SetCell { bg = WordAquaLighter40, fg = black } $\mathbf{ 0010 }$
				&
				\SetCell { bg = WordAquaLighter40, fg = black } $\mathbf{ 0011 }$
				&
				\SetCell { bg = WordAquaLighter40, fg = black, font = \bfseries } $\mathbf{ 0100 }$
				&
				\SetCell { bg = WordAquaLighter40, fg = black, font = \bfseries } $\mathbf{ 0101 }$
				&
				\SetCell { bg = WordAquaLighter40, fg = black, font = \bfseries } $\mathbf{ 0110 }$
				&
				\SetCell { bg = WordAquaLighter40, fg = black, font = \bfseries } $\mathbf{ 0111 }$
				&
				\SetCell { bg = WordAquaLighter40, fg = black, font = \bfseries } $\mathbf{ 1000 }$
				&
				\SetCell { bg = WordAquaLighter40, fg = black, font = \bfseries } $\mathbf{ 1001 }$
				&
				\SetCell { bg = WordAquaLighter40, fg = black, font = \bfseries } $\mathbf{ 1010 }$
				&
				\SetCell { bg = WordAquaLighter40, fg = black, font = \bfseries } $\mathbf{ 1011 }$
				&
				\SetCell { bg = WordAquaLighter40, fg = black, font = \bfseries } $\mathbf{ 1100 }$
				&
				\SetCell { bg = WordAquaLighter40, fg = black, font = \bfseries } $\mathbf{ 1101 }$
				&
				\SetCell { bg = WordAquaLighter40, fg = black, font = \bfseries } $\mathbf{ 1110 }$
				&
				\SetCell { bg = WordAquaLighter40, fg = black, font = \bfseries } $\mathbf{ 1111 }$
				\\
				$g$
				&
				$0$
				&
				$1$
				&
				$1$
				&
				$1$
				&
				$1$
				&
				$0$
				&
				$0$
				&
				$0$
				&
				$1$
				&
				$0$
				&
				$0$
				&
				$0$
				&
				$0$
				&
				$0$
				&
				$0$
				&
				$0$
				\\
			\end{tblr}
		}
	\end{table}
\end{tcolorbox}

\end{example}

\begin{definition} {Pattern Basis} { Pattern Basis}
	A \emph{pattern basis} of rank $n$, $n \geq 1$, denoted by $P_{ n }$, is a list of $2^{ n }$ pattern bit vectors $( \mathbf{ p }_{ 0 }, \mathbf{ p }_{ 1 }, \dots, \mathbf{ p }_{ 2^{ n } - 1 } )$, each of length $2^{ n }$, that satisfy the pairwise \emph{orthogonality} property:
	\begin{align}
		\label{eq: Orthogonality Property}
		\mathbf{ p }_{ i }
		\oplus
		\mathbf{ p }_{ j }
		\ \text{contains} \ 2^{ n - 1 } \ 0s \ \text{and} \ 2^{ n - 1 } \ 1s
		\ , \ 0 \leq i \neq j \leq n - 1
		\ .
	\end{align}
\end{definition}

\begin{example} {Pattern Bases} { Pattern Bases}

Let $B_{ 1 }$ and $Q_{ 2 }$ be the following lists of pattern vectors:

\begin{align}
	\label{eq: Pattern Basis $B_{ 1 }$}
	B_{ 1 }
	&\coloneq
	(
	\mathbf{ e }_{ 0 } = 00,
	\mathbf{ e }_{ 1 } = 01
	)
	\ ,
	\quad
	\text{and}
	\\
	\label{eq: Pattern Basis $Q_{ 2 }$}
	Q_{ 2 }
	&\coloneq
	(
	\mathbf{ g }_{ 0 } = 0001,
	\mathbf{ g }_{ 1 } = 0010,
	\mathbf{ g }_{ 2 } = 0100,
	\mathbf{ g }_{ 3 } = 1000
	)
	\ .
\end{align}

Every $\mathbf{ e }_{ i }$ has length $2^{ 1 } = 2$ and is orthogonal to every other $\mathbf{ e }_{ j }$, $i \neq j$, and, similarly, every $\mathbf{ g }_{ i }$ has length $2^{ 2 } = 4$ and is orthogonal to every other $\mathbf{ g }_{ j }$, $i \neq j$. Furthermore, $B_{ 1 }$ consists of $2^{ 1 } = 2$ pattern bit vectors and $Q_{ 2 }$ consists of $2^{ 2 } = 4$ pattern bit vectors. Therefore, by Definition \ref{def: Pattern Basis}, $B_{ 1 }$ and $Q_{ 2 }$ are two pattern bases of rank $1$ and $2$, respectively.

\end{example}

\begin{definition} {Function Class Realizing Pattern Basis} { Function Class Realizing Pattern Basis}
	Let $P_{ n }$ $=$ $( \mathbf{ p }_{ 0 }, \mathbf{ p }_{ 1 },$ \dots $, \mathbf{ p }_{ 2^{ n } - 1 } )$ be a pattern basis of rank $n$. To $P_{ n }$ we associate the class $F_{ P_{ n } }$ $=$ $( f_{ \mathbf{ p }_{ 0 } }, f_{ \mathbf{ p }_{ 1 } },$ \dots $, f_{ \mathbf{ p }_{ 2^{ n } - 1 } } )$ of Boolean functions from $\mathbb{ B }^{ n }$ to $\mathbb{ B }$ such that $f_{ \mathbf{ p }_{ i } }$, $0 \leq i \leq 2^{ n } - 1$, is the function realizing the pattern bit vector $\mathbf{ p }_{ i }$. We say that the class of Boolean functions $F_{ P_{ n } }$ realizes the pattern basis $P_{ n }$.
\end{definition}

\begin{definition} {The Pattern Basis Classification Scheme} { The Pattern Basis Classification Scheme}
	Let $P_{ n }$ be a pattern basis of rank $n$ and let $F_{ P_{ n } }$ be the class of Boolean functions realizing $P_{ n }$. We say that $P_{ n }$ admits a \emph{classification scheme}, if there exists an abstract quantum circuit, such as the one depicted in Figure \ref{fig: The Pattern Basis Classification Scheme}, satisfying the \emph{pattern basis classification} property. This property asserts that if $U_{ f_{ \mathbf{ p }_{ i } } }$ is an oracle for the function $f_{ \mathbf{ p }_{ i } } \in F_{ P_{ n } }$ that realizes the pattern bit vector $\mathbf{ p }_{ i } \in P_{ n }$, then, with probability $1.0$, the outcome of the final measurement is the basis ket $\ket{ \mathbf { i } }$, $\mathbf { i }$ being the binary representation of the index $i$.
	
	We clarify the symbolism employed in Figure \ref{fig: The Pattern Basis Classification Scheme}:
	
	\begin{itemize}
		\item	
		$IR$ denotes the quantum input register, containing $n$ qubits.
		\item	
		$OR$ is the single-qubit output register, initialized to $\ket{-}$.
		\item	
		$U_{ f_{ \mathbf{ p }_{ i } } }$ designates the unitary transform that implements the oracle for the function $f_{ \mathbf{ p }_{ i } }$. Its exact form depends on $f_{ \mathbf{ p }_{ i } }$ and satisfies equation \eqref{eq: Unitary Transform U_f}.
		\item	
		$\mathbf{ G }$ is the $n \times n$ the unitary transform that acts on $IR$ and implements the pattern basis classification scheme.
	\end{itemize}
	
	The unitary transform $\mathbf{ G }$ is, henceforth, referred to as the \emph{classifier} for $P_{ n }$ and $F_{ P_{ n } }$. The outcome $\ket{ \mathbf { i } }$ is called the \emph{classification measurement} for $f_{ \mathbf{ p }_{ i } }$ and is denoted by $CM ( f_{ \mathbf{ p }_{ i } } )$. The elements of $P_{ n }$ and $F_{ P_{ n } }$ are termed the \emph{perfectly classifiable} pattern bit vectors and Boolean functions, respectively. Functions that do not belong to $F_{ P_{ n } }$ are termed \emph{nonclassifiable} with respect to $\mathbf{ G }$.
\end{definition}

\begin{tcolorbox}
	[
	enhanced,
	breakable,
	grow to left by = 1.000 cm,
	grow to right by = 0.000 cm,
	colback = SkyBlue1!08,
	enhanced jigsaw,			
	frame hidden,
	sharp corners,
	]
	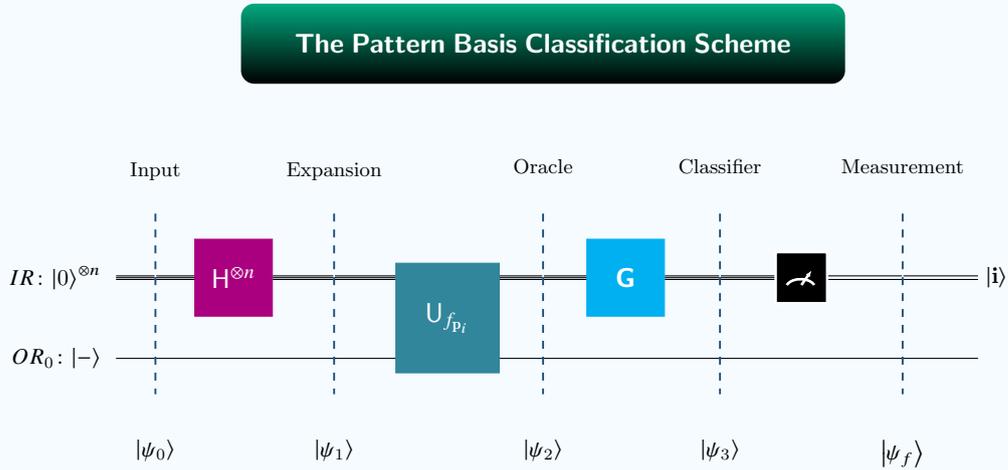
\begin{figure}[H]
		\centering
		\begin{tikzpicture} [ scale = 0.850, transform shape ]
			\begin{yquant}
				nobit AUX_IB_0_0;
				[ name = Charlie ] qubits { $IR \colon \ket{ 0 }^{ \otimes n }$ } IR;
				nobit AUX_IB_0_2;
				qubit { $OR_{ 0 } \colon \ket{ - }$ } OR;
				nobit AUX_IB_0_4;
				[
				name = Input,
				WordBlueDarker,
				line width = 0.250 mm,
				label = { [ label distance = 0.500 cm ] north: {\small Input} }
				]
				barrier ( - ) ;
				[ draw = RedPurple, fill = RedPurple, radius = 0.600 cm ] box {\color{white} \large \sf{H}$^{ \otimes n }$} IR;
				[
				name = Expansion,
				WordBlueDarker,
				line width = 0.250 mm,
				label = { [ label distance = 0.500 cm ] north: {\small Expansion} }
				]
				barrier ( - ) ;
				[ draw = WordAquaDarker25, fill = WordAquaDarker25, x radius = 0.800 cm, y radius = 0.450 cm ] box { \color{white} \large \sf{U}$_{ f_{ \mathbf{ p }_{ i } } }$} ( IR - OR );
				[
				name = Oracle,
				WordBlueDarker,
				line width = 0.250 mm,
				label = { [ label distance = 0.600 cm ] north: {\small Oracle} }
				]
				barrier ( - ) ;
				[ draw = WordBlueVeryLight, fill = WordBlueVeryLight, x radius = 0.600 cm, y radius = 0.600 cm ] box { \color{white} \large \sf{\textbf{G}}} IR;
				[
				name = Classifier,
				WordBlueDarker,
				line width = 0.250 mm,
				label = { [ label distance = 0.600 cm ] north: {\small Classifier} }
				]
				barrier ( - ) ;
				[ line width = .350 mm, draw = white, fill = black, radius = 0.400 cm ] measure IR [ 0 ];
				[
				name = Measurement,
				WordBlueDarker,
				line width = 0.250 mm,
				label = { [ label distance = 0.600 cm ] north: {\small Measurement} }
				]
				barrier ( - ) ;
				output { $\ket{ \mathbf { i } }$ } IR;
				\node [ below = 2.000 cm ] at (Input) { $\ket{ \psi_{ 0 } }$ };
				\node [ below = 2.000 cm ] at (Expansion) { $\ket{ \psi_{ 1 } }$ };
				\node [ below = 2.000 cm ] at (Oracle) { $\ket{ \psi_{ 2 } }$ };
				\node [ below = 2.000 cm ] at (Classifier) { $\ket{ \psi_{ 3 } }$ };
				\node [ below = 2.000 cm ] at (Measurement) { $\ket{ \psi_{ f } }$ };
			\end{yquant}
			\node
			[
			above = 2.750 cm of Oracle,
			rectangle,							
			rounded corners = 5.000 pt,			
			minimum width = 8.000 cm,
			minimum height = 1.250 cm,
			thick,								
			inner sep = 5.000 pt,				
			text width = 9.000 cm,				
			align = center,						
			anchor = center,
			font = \sffamily,					
			text = white,						
			shade,
			top color = GreenLighter2, bottom color = black,
			]
			(Label)
			{
				\large \textbf{The Pattern Basis Classification Scheme}
			};
		\end{tikzpicture}
		\caption{The above quantum circuit implements the pattern basis classification scheme.}
		\label{fig: The Pattern Basis Classification Scheme}
	\end{figure}
\end{tcolorbox}

\begin{example} {Classifiers for $B_{ 1 }$ and $Q_{ 2 }$} { Classifiers for $B_{ 1 }$ and $Q_{ 2 }$}

For the pattern basis $B_{ 1 }$ of Example \ref{xmp: Pattern Bases}, we designate by $F_{ B_{ 1 } } = ( f_{ \mathbf{ e }_{ 0 } }, f_{ \mathbf{ e }_{ 1 } } )$ the class realizing $B_{ 1 }$. The function $f_{ \mathbf{ e }_{ 0 } }$ captures the ``constancy'' property in the sense that it always takes the value $0$, while the function $f_{ \mathbf{ e }_{ 1 } }$ captures the ``balance'' property in the sense that $f_{ \mathbf{ e }_{ 1 } } ( 0 ) = 0$ and $f_{ \mathbf{ e }_{ 1 } } ( 1 ) = 1$. It is very easy to verify that the Hadamard transform $H$ is the classifier for $B_{ 1 }$ and $F_{ B_{ 1 } }$.

For the pattern basis $Q_{ 2 }$ of Example \ref{xmp: Pattern Bases}, we denote by $F_{ Q_{ 2 } } = ( f_{ \mathbf{ g }_{ 0 } }, f_{ \mathbf{ g }_{ 1 } }, f_{ \mathbf{ g }_{ 2 } }, f_{ \mathbf{ g }_{ 3 } } )$ the class realizing $Q_{ 2 }$. For completeness, we repeat the pattern bit vectors $\mathbf{ g }_{ 0 }, \mathbf{ g }_{ 1 }, \mathbf{ g }_{ 2 }, \mathbf{ g }_{ 3 }$, together with the truth values of the functions $f_{ \mathbf{ g }_{ 0 } }, f_{ \mathbf{ g }_{ 1 } }, f_{ \mathbf{ g }_{ 2 } }, f_{ \mathbf{ g }_{ 3 } }$ they realize in Table \ref{tbl: Pattern Bit Vectors & Truth Values of the Functions in $F_{ I_{ 2 } }$}.

\begin{tcolorbox}
	[
		enhanced,
		breakable,
		grow to left by = 0.000 cm,
		grow to right by = 0.000 cm,
		colback = SkyBlue1!08,
		enhanced jigsaw,			
		frame hidden,
		sharp corners,
	]
	\begin{table}[H]
		{\small
			\caption{The pattern bit vectors and the truth values of the functions $f_{ \mathbf{ g }_{ 0 } }, f_{ \mathbf{ g }_{ 1 } }, f_{ \mathbf{ g }_{ 2 } }, f_{ \mathbf{ g }_{ 3 } }$.}
			\label{tbl: Pattern Bit Vectors & Truth Values of the Functions in $F_{ I_{ 2 } }$}
			\centering
			\SetTblrInner { rowsep = 1.000 mm }
			\begin{tblr}
				{
					colspec =
					{
						Q [ c, m, 0.500 cm ]
						| [ 1.000 pt, WordAquaDarker25 ]
						| [ 1.000 pt, WordAquaDarker25 ]
						Q [ c, m, 4.000 cm ]
						| [ 0.500 pt, WordAquaDarker25 ]
						Q [ c, m, 1.500 cm ]
						| [ 0.500 pt, WordAquaDarker25 ]
						Q [ c, m, 1.500 cm ]
						| [ 0.500 pt, WordAquaDarker25 ]
						Q [ c, m, 1.500 cm ]
						| [ 0.500 pt, WordAquaDarker25 ]
						Q [ c, m, 1.500 cm ]
					},
					rowspec =
					{
						|
						[ 3.500 pt, WordAquaDarker25 ]
						|
						[ 0.750 pt, WordAquaDarker25 ]
						|
						[ 0.250 pt, white ]
						Q
						|
						[ 0.500 pt, WordAquaDarker25 ]
						Q
						|
						[ 0.500 pt, WordAquaDarker25 ]
						Q
						|
						[ 0.500 pt, WordAquaDarker25 ]
						Q
						|
						[ 0.500 pt, WordAquaDarker25 ]
						Q
						|
						[ 3.500 pt, WordAquaDarker50 ]
					}
				}
				&
				\SetCell { bg = WordAquaDarker50, fg = white } Pattern Bit Vector
				&
				\SetCell { bg = WordAquaLighter40, fg = black } $\mathbf{ 00 }$
				&
				\SetCell { bg = WordAquaLighter40, fg = black } $\mathbf{ 01 }$
				&
				\SetCell { bg = WordAquaLighter40, fg = black } $\mathbf{ 10 }$
				&
				\SetCell { bg = WordAquaLighter40, fg = black } $\mathbf{ 11 }$
				\\
				$f_{ \mathbf{ g }_{ 0 } }$
				&
				0001
				&
				$1$
				&
				$0$
				&
				$0$
				&
				$0$
				\\
				$f_{ \mathbf{ g }_{ 1 } }$
				&
				0010
				&
				$0$
				&
				$1$
				&
				$0$
				&
				$0$
				\\
				$f_{ \mathbf{ g }_{ 2 } }$
				&
				0100
				&
				$0$
				&
				$0$
				&
				$1$
				&
				$0$
				\\
				$f_{ \mathbf{ g }_{ 3 } }$
				&
				1000
				&
				$0$
				&
				$0$
				&
				$0$
				&
				$1$
				\\
			\end{tblr}
		}
	\end{table}
\end{tcolorbox}

The four functions $f_{ \mathbf{ g }_{ 0 } }, f_{ \mathbf{ g }_{ 1 } }, f_{ \mathbf{ g }_{ 2 } }, f_{ \mathbf{ g }_{ 3 } }$ exhibit a common pattern: for precisely one element $\mathbf{ x } \in \mathbb{ B }^{ 2 }$ their value is $1$, while for the remaining three elements their value is $0$. This shows that there is an imbalance between the number of inputs that take value 1 and the number of inputs that take value 0. The important thing is that this imbalance exhibits constant ratio; the number of inputs corresponding to value 1 over the total number of inputs is $\frac { 1 } { 4 }$. Therefore, these function capture the ``constant ratio imbalance'' property and, specifically, the $\frac { 1 } { 4 }$ imbalance ratio. In \cite{Andronikos2025a} it was shown that $C_{ 2 }$, defined by the equation \eqref{eq: Unitary Transform C_2}, is the classifier for $Q_{ 2 }$ and $F_{ Q_{ 2 } }$. The matrix representation of $C_{ 2 }$ is given by equation \eqref{eq: Unitary Transform C_2 - Explicit Matrix Form}. It is quite easy to construct $C_{ 2 }$ using standard quantum gates readily available in contemporary quantum computers, such as Hadamard, $Z$ and controlled-$Z$ gates, as demonstrated in Figure \ref{fig: Classifier Kernel C2}.

\begin{align}
	\label{eq: Unitary Transform C_2}
	C_{ 2 }
	=
	( H \otimes H )
	\
	CZ
	\
	( Z \otimes Z )
	\
	( H \otimes H )
\end{align}

\begin{tcolorbox}
	[
		enhanced,
		breakable,
		grow to left by = 0.000 cm,
		grow to right by = 0.000 cm,
		colback = white,
		enhanced jigsaw,			
		frame hidden,
		sharp corners,
	]
	\begin{minipage} [ t ] { 0.450 \textwidth }
		\begin{align}
			\label{eq: Unitary Transform C_2 - Explicit Matrix Form}
			\NiceMatrixOptions{ cell-space-limits = 1.500 pt }
			C_{ 2 }
			=
			\begin{bNiceMatrix}[ margin ] 
				- \frac { 1 } { 2 } & \phantom{-} \frac { 1 } { 2 } & \phantom{-} \frac { 1 } { 2 } & \phantom{-} \frac { 1 } { 2 } \\
				\phantom{-} \frac { 1 } { 2 } & - \frac { 1 } { 2 } & \phantom{-} \frac { 1 } { 2 } & \phantom{-} \frac { 1 } { 2 } \\
				\phantom{-} \frac { 1 } { 2 } & \phantom{-} \frac { 1 } { 2 } & - \frac { 1 } { 2 } & \phantom{-} \frac { 1 } { 2 } \\
				\phantom{-} \frac { 1 } { 2 } & \phantom{-} \frac { 1 } { 2 } & \phantom{-} \frac { 1 } { 2 } & - \frac { 1 } { 2 } \\
			\end{bNiceMatrix}
		\end{align}
	\end{minipage}
	\hspace{ 0.500 cm }
	\begin{minipage} [ t ] { 0.450 \textwidth }
		\begin{figure}[H]
			\centering
			\includegraphics [ scale = 0.700, trim = {0.000cm 0.750cm 0.000cm 0.500cm}, clip ] {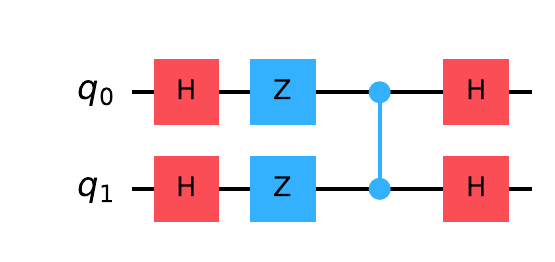}
			\caption{This figure shows the quantum circuit that implements the $C_{ 2 }$ classifier.}
			\label{fig: Classifier Kernel C2}
		\end{figure}
	\end{minipage}
\end{tcolorbox}

\end{example}

We introduce a mechanism for creating new pattern bases from existing pattern bases, by utilizing a concept of product between pattern bit vectors and pattern bases, as outlined below.

\begin{definition} {New Pattern Bases From Old} { New Pattern Bases From Old} \

	If $\mathbf{ p }$ $=$ $p_{ 2^{ n } - 1 }$ $\dots$ $p_{ 1 }$ $p_{ 0 }$ and $\mathbf{ q }$ $=$ $q_{ 2^{ m } - 1 }$ $\dots$ $q_{ 1 }$ $q_{ 0 }$ are two pattern bit vectors of length $2^{ n }$ and $2^{ m }$, respectively, their product $\mathbf{ p } \odot \mathbf{ q }$ is pattern bit vector of length $2^{ n + m }$ defined as
	\begin{align}
		\label{eq: Pattern Bit Vectors Product}
		\mathbf{ p }
		\odot
		\mathbf{ q }
		=
		\mathbf{ s }_{ 2^{ n } - 1 } \dots \mathbf{ s }_{ 1 } \mathbf{ s }_{ 0 }
		\ ,
		\quad
		\text{where}
		\quad
		\mathbf{ s }_{ i }
		=
		\begin{cases}
			\ \mathbf{ q }						& \text{if } p_{ i } = 0 
			\\[ 1.000 ex ]
			\ \overline{ \mathbf{ q } }			& \text{if } p_{ i } = 1 
		\end{cases}
		\ ,
		\quad
		0 \leq i \leq 2^{ n } - 1
		\ .
	\end{align}

	If $P_{ n }$ $=$ $( \mathbf{ p }_{ 0 }, \mathbf{ p }_{ 1 },$ \dots $, \mathbf{ p }_{ 2^{ n } - 1 } )$ and $Q_{ m }$ $=$ $( \mathbf{ q }_{ 0 },$ $\mathbf{ q }_{ 1 },$ $\dots $, $\mathbf{ q }_{ 2^{ m } - 1 } )$ are two pattern bases of rank $n$ and $m$, respectively, their product $P_{ n }$ $\odot$ $Q_{ m }$ is a pattern basis containing $2^{ n + m }$ pattern bit vectors, each of length $2^{ n + m }$, which is constructed as
	\begin{align}
		\label{eq: Pattern Bases Product}
		P_{ n }
		\odot
		Q_{ m }
		\coloneq
		(
		\
		&\mathbf{ p }_{ 0 } \odot \mathbf{ q }_{ 0 },
		\mathbf{ p }_{ 0 } \odot \mathbf{ q }_{ 1 },
		\dots ,
		\mathbf{ p }_{ 0 } \odot \mathbf{ q }_{ 2^{ m } - 1 },
		\nonumber
		\\
		&\mathbf{ p }_{ 1 } \odot \mathbf{ q }_{ 0 },
		\mathbf{ p }_{ 1 } \odot \mathbf{ q }_{ 1 },
		\dots ,
		\mathbf{ p }_{ 1 } \odot \mathbf{ q }_{ 2^{ m } - 1 },
		\nonumber
		\\
		&\dots ,
		\nonumber
		\\
		&\mathbf{ p }_{ 2^{ n } - 1 } \odot \mathbf{ q }_{ 0 },
		\mathbf{ p }_{ 2^{ n } - 1 } \odot \mathbf{ q }_{ 1 },
		\dots ,
		\mathbf{ p }_{ 2^{ n } - 1 } \odot \mathbf{ q }_{ 2^{ m } - 1 }
		\
		)
		\ .
	\end{align}
\end{definition}

The use of same symbol $\odot$ for the product operation between two pattern bit vectors, and between two pattern bases should cause no confusion because the context always makes clear the intended operation.

\begin{example} {Example of Pattern Bases Product} { Example Of Pattern Bases Product}

This example is meant to illustrate how the product between two pattern bases is constructed. For simplicity, we compute the product of $Q_{ 2 }$, defined by equation \eqref{eq: Pattern Basis $Q_{ 2 }$}, with itself.

\begin{align}
	\label{eq: Pattern Basis $Q_{ 4 }$}
	Q_{ 4 }
	\coloneq
	Q_{ 2 }
	\odot
	Q_{ 2 }
	\ .
\end{align}

To explain the inner workings of Definition \ref{def: New Pattern Bases From Old} works, we enumerate the elements of $Q_{ 4 }$ in Table \ref{tbl: Pattern Basis $Q_{ 4 }$}, indicating their derivation from $Q_{ 2 }$.

\begin{tcolorbox}
	[
		enhanced,
		breakable,
		grow to left by = 0.500 cm,
		grow to right by = 1.500 cm,
		colback = white,
		enhanced jigsaw,			
		frame hidden,
		sharp corners,
	]
	\begin{table}[H]
		\caption{This table contains the pattern bit vectors of $Q_{ 4 }$.}
		\label{tbl: Pattern Basis $Q_{ 4 }$}
		\centering
		\SetTblrInner { rowsep = 1.200 mm }
		\begin{tblr}
			{
				colspec =
				{
					Q [ c, m, 2.000 cm ]
					| [ 0.750 pt, WordAquaDarker25 ]
					| [ 0.750 pt, WordAquaDarker25 ]
					Q [ c, m, 2.850 cm ]
					| [ 0.500 pt, WordAquaDarker25 ]
					Q [ c, m, 2.850 cm ]
					| [ 0.500 pt, WordAquaDarker25 ]
					Q [ c, m, 2.850 cm ]
					| [ 0.500 pt, WordAquaDarker25 ]
					Q [ c, m, 2.850 cm ]
				},
				rowspec =
				{
					| [ 3.500 pt, WordAquaDarker25 ]
					| [ 0.750 pt, WordAquaDarker25 ]
					| [ 0.250 pt, white ]
					Q
					|
					Q
					| [ 0.150 pt, WordAquaDarker25 ]
					Q
					| [ 0.150 pt, WordAquaDarker25 ]
					Q
					| [ 0.150 pt, WordAquaDarker25 ]
					Q
					| [ 3.500 pt, WordAquaDarker50 ]
				}
			}
			\SetCell { bg = WordAquaDarker50, fg = white, font = \bfseries \small } $\mathbf{ Q }_{ 2 }$ Pattern \quad Bit Vectors
			&
			\SetCell { bg = WordAquaLighter40, fg = black, font = \bfseries \small } $\mathbf{ g }_{ i } \odot \mathbf{ g }_{ 0 }, \ 0 \leq i \leq 3$
			&
			\SetCell { bg = WordAquaLighter40, fg = black, font = \bfseries \small } $\mathbf{ g }_{ i } \odot \mathbf{ g }_{ 1 }, \ 0 \leq i \leq 3$
			&
			\SetCell { bg = WordAquaLighter40, fg = black, font = \bfseries \small } $\mathbf{ g }_{ i } \odot \mathbf{ g }_{ 2 }, \ 0 \leq i \leq 3$
			&
			\SetCell { bg = WordAquaLighter40, fg = black, font = \bfseries \small } $\mathbf{ g }_{ i } \odot \mathbf{ g }_{ 3 }, \ 0 \leq i \leq 3$
			\\
			\SetCell { bg = WordAquaLighter40, fg = black, font = \bfseries \small } $\mathbf{ g }_{ 0 } = 0001$
			&
			{ \footnotesize $\mathbf{ r }_{ 0 } \colon 0001000100011110$ }
			&
			{ \footnotesize $\mathbf{ r }_{ 1 } \colon 0010001000101101$ }
			&
			{ \footnotesize $\mathbf{ r }_{ 2 } \colon 0100010001001011$ }
			&
			{ \footnotesize $\mathbf{ r }_{ 3 } \colon 1000100010000111$ }
			\\
			\SetCell { bg = WordAquaLighter40, fg = black, font = \bfseries \small } $\mathbf{ g }_{ 1 } = 0010$
			&
			{ \footnotesize $\mathbf{ r }_{ 4 } \colon 0001000111100001$ }
			&
			{ \footnotesize $\mathbf{ r }_{ 5 } \colon 0010001011010010$ }
			&
			{ \footnotesize $\mathbf{ r }_{ 6 } \colon 0100010010110100$ }
			&
			{ \footnotesize $\mathbf{ r }_{ 7 } \colon 1000100001111000$ }
			\\
			\SetCell { bg = WordAquaLighter40, fg = black, font = \bfseries \small } $\mathbf{ g }_{ 2 } = 0100$
			&
			{ \footnotesize $\mathbf{ r }_{ 8 } \colon 0001111000010001$ }
			&
			{ \footnotesize $\mathbf{ r }_{ 9 } \colon 0010110100100010$ }
			&
			{ \footnotesize $\mathbf{ r }_{ 10 } \colon 0100101101000100$ }
			&
			{ \footnotesize $\mathbf{ r }_{ 11 } \colon 1000011110001000$ }
			\\
			\SetCell { bg = WordAquaLighter40, fg = black, font = \bfseries \small } $\mathbf{ g }_{ 3 } = 1000$
			&
			{ \footnotesize $\mathbf{ r }_{ 12 } \colon 1110000100010001$ }
			&
			{ \footnotesize $\mathbf{ r }_{ 13 } \colon 1101001000100010$ }
			&
			{ \footnotesize $\mathbf{ r }_{ 14 } \colon 1011010001000100$ }
			&
			{ \footnotesize $\mathbf{ r }_{ 15 } \colon 0111100010001000$ }
			\\
		\end{tblr}
	\end{table}
\end{tcolorbox}

By inspecting the above Table \ref{tbl: Pattern Basis $Q_{ 4 }$}, we may may immediately conclude that $Q_{ 4 }$ satisfies the condition set in Definition \ref{def: Pattern Basis}, and is, therefore, a pattern basis of rank $4$. Following Definition \ref{def: Function Class Realizing Pattern Basis}, we denote by $F_{ Q_{ 4 } }$ the class of Boolean functions that realizes the pattern basis $Q_{ 4 }$.

\end{example}

In a symmetric manner, we present a method from constructing new function classes using the notion of external product.

\begin{definition} {New Function Classes From Old} { New Function Classes From Old} \

	If $f_{ \mathbf{ p } } \colon \mathbb{ B }^{ n } \rightarrow \mathbb{ B }$ and $g_{ \mathbf{ q } } \colon \mathbb{ B }^{ m } \rightarrow \mathbb{ B }$ are the Boolean functions realizing the pattern bit vectors $\mathbf{ p }$ and $\mathbf{ q }$, respectively, their \emph{extended product}
	$f_{ \mathbf{ p } } \star g_{ \mathbf{ q } } \colon \mathbb{ B }^{ n + m } \rightarrow \mathbb{ B }$ is the Boolean function
	defined as
	\begin{align}
		\label{eq: Extended Product Of Functions}
		f_{ \mathbf{ p } } \star g_{ \mathbf{ q } } ( j + i 2^{ n }  )
		=
		\begin{cases}
			\ g_{ \mathbf{ q } } ( j )						& \text{if } f_{ \mathbf{ p } } ( i ) = 0 
			\\[ 1.000 ex ]
			\ \overline{ g_{ \mathbf{ q } } ( j ) }			& \text{if } f_{ \mathbf{ p } } ( i ) = 1 
		\end{cases}
		\ ,
		\quad
		0 \leq i \leq 2^{ n } - 1,
		\quad
		0 \leq j \leq 2^{ m } - 1
		\ .
	\end{align}

	If $F_{ P_{ n } }$ $=$ $( f_{ \mathbf{ p }_{ 0 } }, f_{ \mathbf{ p }_{ 1 } },$ \dots $, f_{ \mathbf{ p }_{ 2^{ n } - 1 } } )$ and $F_{ Q_{ m } }$ $=$ $( f_{ \mathbf{ q }_{ 0 } }, f_{ \mathbf{ q }_{ 1 } },$ \dots $, f_{ \mathbf{ q }_{ 2^{ m } - 1 } } )$ are the classes of Boolean functions that realize the pattern bases $P_{ n }$ and $Q_{ m }$ of rank $n$ and $m$, respectively, their \emph{extended product} $F_{ P_{ n } }$ $\star$ $F_{ Q_{ m } }$ is a class of $2^{ n + m }$ Boolean functions, which is constructed as
	\begin{align}
		\label{eq: Extended Product Of Function Classes}
		F_{ P_{ n } }
		\star
		F_{ Q_{ m } }
		\coloneq
		(
		\
		&f_{ \mathbf{ p }_{ 0 } } \star f_{ \mathbf{ q }_{ 0 } },
		f_{ \mathbf{ p }_{ 0 } } \star f_{ \mathbf{ q }_{ 1 } },
		\dots ,
		f_{ \mathbf{ p }_{ 0 } } \star f_{ \mathbf{ q }_{ 2^{ m } - 1 } },
		\nonumber
		\\
		&f_{ \mathbf{ p }_{ 1 } } \star f_{ \mathbf{ q }_{ 0 } },
		f_{ \mathbf{ p }_{ 1 } } \star f_{ \mathbf{ q }_{ 1 } },
		\dots ,
		f_{ \mathbf{ p }_{ 1 } } \star f_{ \mathbf{ q }_{ 2^{ m } - 1 } },
		\nonumber
		\\
		&\dots ,
		\nonumber
		\\
		&f_{ \mathbf{ p }_{ 2^{ n } - 1 } } \star f_{ \mathbf{ q }_{ 0 } },
		f_{ \mathbf{ p }_{ 2^{ n } - 1 } } \star f_{ \mathbf{ q }_{ 1 } },
		\dots ,
		f_{ \mathbf{ p }_{ 2^{ n } - 1 } } \star f_{ \mathbf{ q }_{ 2^{ m } - 1 } }
		\
		)
		\ .
	\end{align}
\end{definition}

In view of Definitions \ref{def: New Pattern Bases From Old} and \ref{def: New Function Classes From Old}, it is straightforward to establish the following fact.

\begin{proposition} {Extended Function Product Realizes Pattern Vector Product} { Extended Function Product Realizes Pattern Vector Product}
	If $f_{ \mathbf{ p } } \colon \mathbb{ B }^{ n } \rightarrow \mathbb{ B }$ and $g_{ \mathbf{ q } } \colon \mathbb{ B }^{ m } \rightarrow \mathbb{ B }$ are the Boolean functions realizing the pattern bit vectors $\mathbf{ p }$ and $\mathbf{ q }$, respectively, their extended product $f_{ \mathbf{ p } } \star g_{ \mathbf{ q } } \colon \mathbb{ B }^{ n + m } \rightarrow \mathbb{ B }$ is the Boolean function that realizes the pattern bit vector $\mathbf{ p } \odot \mathbf{ q }$.
\end{proposition}

Continuing this line of reasoning, we arrive at the next Theorem.

\begin{theorem}  {Extended Function Product Realizes Pattern Bases Product} { Extended Function Product Realizes Pattern Bases Product}
	If $F_{ P_{ n } }$ and $F_{ Q_{ m } }$ are the classes of Boolean functions that realize the pattern bases $P_{ n }$ and $Q_{ m }$ of rank $n$ and $m$, respectively, their extended product $F_{ P_{ n } }$ $\star$ $F_{ Q_{ m } }$ realized the pattern basis $P_{ n }$ $\odot$ $Q_{ m }$.
\end{theorem}

To ensure this investigation yields meaningful and reproducible results, it must be anchored in well-defined assumptions and a robust theoretical framework. To this end, we agree that the quantum classifiers we investigate in this work, are constructed as prescribed by the next Definition \ref{def: Quantum Classifier}, using as elementary building blocks the Hadamard transform $H$ and the classifier $C_{ 2 }$, defined by equation \eqref{eq: Unitary Transform C_2}.

\begin{definition} {Quantum Classifier} { Quantum Classifier}
	A \emph{quantum classifier} $\mathbf{ G }$ is a finite tensor product of the form
	\begin{align}
		\label{eq: General Quantum Classifier}
		\mathbf{ G }
		=
		G_{ m - 1 }
		\otimes
		\dots
		\otimes
		G_{ 0 }
		\ ,
	\end{align}
	where each $G_{ j }, 0 \leq j \leq m - 1,$ is either the Hadamard transform $H$ or the classifier $C_{ 2 }$, defined by equation \eqref{eq: Unitary Transform C_2}.
\end{definition}

Using induction on $m$, we may prove Theorem \ref{thr: Classification of Products of Pattern Bases & Function Classes}.

\begin{theorem}  {Classification of Products of Pattern Bases \& Function Classes} { Classification of Products of Pattern Bases & Function Classes}
	Consider the quantum classifier $\mathbf{ G }$ $=$ $G_{ m - 1 }$ $\otimes$ $\dots$ $\otimes$ $G_{ 0 }$, where each $G_{ j }, 0 \leq j \leq m - 1,$ is the classifier for $P_{ j }$ and $F_{ P_{ j } }$. Then $\mathbf{ G }$ classifies the class of Boolean functions $F_{ P_{ m - 1 } }$ $\star$ $\dots$ $\star$ $F_{ P_{ 0 } }$ realizing the pattern basis $P_{ m - 1 }$ $\odot$ $\dots$ $\odot$ $P_{ 0 }$.
\end{theorem}

An immediate Corollary of the Theorem \ref{thr: Classification of Products of Pattern Bases & Function Classes} is the following.

\begin{corollary}  {Correspondence among Bases, Function Classes \& Classifiers} { Correspondence Among Bases, Function Classes & Classifiers}
	Let $P$ be the product pattern bit basis
	\begin{align}
		\label{eq: General Pattern Bit Bases Product}
		P
		=
		P_{ m - 1 }
		\odot
		\dots
		\odot
		P_{ 0 }
		\ ,
	\end{align}
	where each $P_{ j }, \ 0 \leq j \leq m - 1,$ is either the basis $B_{ 1 }$ or the basis $Q_{ 2 }$, defined by equations \eqref{eq: Pattern Basis $B_{ 1 }$} and \eqref{eq: Pattern Basis $Q_{ 2 }$}, respectively.

	The class $F_{ P }$ of Boolean functions that realizes $P$ is the extended product
	\begin{align}
		\label{eq: General Function Classes Extended Product}
		F_{ P }
		=
		F_{ P_{ m - 1 } }
		\star
		\dots
		\star
		F_{ P_{ 0 } }
		\ ,
	\end{align}
	where each $F_{ P_{ j } }, \ 0 \leq j \leq m - 1,$ is $F_{ B_{ 1 } }$ if $P_{ j } = B_{ 1 }$ or $F_{ Q_{ 2 } }$ if $P_{ j } = Q_{ 2 }$.

	The quantum classifier $\mathbf{ G }$ that classifies $F_{ P }$ is the tensor product
	\begin{align}
		\label{eq: General Quantum Classifiers Tensor Product}
		\mathbf{ G }
		=
		G_{ m - 1 }
		\otimes
		\dots
		\otimes
		G_{ 0 }
		\ ,
	\end{align}
	where each $G_{ j }, \ 0 \leq j \leq m - 1,$ is $H$ if $P_{ j } = B_{ 1 }$ or $C_{ 2 }$ if $P_{ j } = Q_{ 2 }$.
\end{corollary}

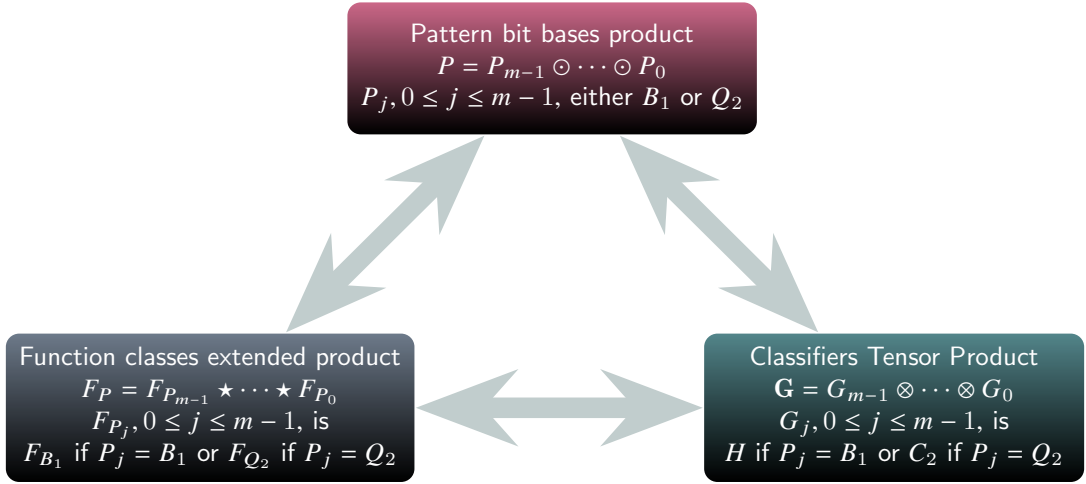
\begin{figure}[htp]
	\centering
	\begin{tikzpicture} 
		[
			scale = 0.900,
			node distance = 4.500 cm,
			concept/.style =
			{
				rectangle,							
				rounded corners = 5.000 pt,			
				minimum width = 5.000 cm,
				minimum height = 1.750 cm,
				thick,								
				inner sep = 5.000 pt,				
				align = center,						
				font = \sffamily,					
				text = white						
			}
		]
		\node [ concept, shade, top color = PaleVioletRed3, bottom color = black ] (A) at ( 0, 0 )
		{
			Pattern bit bases product \\
			$P = P_{ m - 1 } \odot \dots \odot P_{ 0 }$\\
			$P_{ j }, 0 \leq j \leq m - 1,$ either $B_{ 1 }$ or $Q_{ 2 }$
		};
		\node [ concept, shade, top color = LightSteelBlue4, bottom color = black ] (B) at ( -5.000, -5.000 )
		{
			Function classes extended product\\
			$F_{ P } = F_{ P_{ m - 1 } } \star \dots \star F_{ P_{ 0 } }$\\
			$F_{ P_{ j } }, 0 \leq j \leq m - 1,$ is\\
			$F_{ B_{ 1 } }$ if $P_{ j } = B_{ 1 }$ or $F_{ Q_{ 2 } }$ if $P_{ j } = Q_{ 2 }$
		};
		\node [ concept, shade, top color = CadetBlue4, bottom color = black ] (C) at ( 5.000, -5.000 )
		{
			Classifiers Tensor Product\\
			$\mathbf{ G } = G_{ m - 1 } \otimes \dots \otimes G_{ 0 }$\\
			$G_{ j }, 0 \leq j \leq m - 1,$ is\\
			$H$ if $P_{ j } = B_{ 1 }$ or $C_{ 2 }$ if $P_{ j } = Q_{ 2 }$
		};
		\draw [ Stealth-Stealth, line width = 7.750 pt, Azure3 ] (A) -- (B);
		\draw [ Stealth-Stealth, line width = 7.750 pt, Azure3 ] (A) -- (C);
		\draw [ Stealth-Stealth, line width = 7.750 pt, Azure3 ] (B) -- (C);
	\end{tikzpicture}
	\caption{One-to-one correspondence among pattern bases product, function classes extended product, and quantum classifiers tensor product.}
	\label{fig: One-to-one Correspondence Among Products}
\end{figure}

Corollary \ref{crl: Correspondence Among Bases, Function Classes & Classifiers} makes explicitly clear the one-to-one correspondence among pattern bases product, function classes extended product, and quantum classifiers tensor product, which is visually illustrated in Figure \ref{fig: One-to-one Correspondence Among Products}.

\begin{example} {Perfectly Classifiable Function in $F_{ Q_{ 4 } }$} { Perfectly Classifiable Function in $F_{ Q_{ 4 } }$}

Consider the pattern bit vector $\mathbf{ r }_{ 3 } = 1000 \ 1000 \ 1000 \ 0111$ $\in$ $Q_{ 4 }$, which is listed in Table \ref{tbl: Pattern Basis $Q_{ 4 }$} of Example \ref{xmp: Example Of Pattern Bases Product}. As shown there, $\mathbf{ r }_{ 3 }$ is the product $\mathbf{ g }_{ 0 } \odot \mathbf{ g }_{ 3 }$ of the pattern vectors $\mathbf{ g }_{ 0 }$ and $\mathbf{ g }_{ 3 }$, where $\mathbf{ g }_{ 0 }, \mathbf{ g }_{ 3 } \in Q_{ 2 }$. By Theorem \ref{thr: Extended Function Product Realizes Pattern Bases Product}, the Boolean function $f_{ \mathbf{ r }_{ 3 } }$ that realizes pattern bit vector $\mathbf{ r }_{ 3 }$ is extended product $f_{ \mathbf{ r }_{ 3 } }$ $=$ $f_{ \mathbf{ g }_{ 0 } } \star f_{ \mathbf\mathbf{ g }_{ 3 } }$, where $f_{ \mathbf{ g }_{ 0 } }$ and $f_{ \mathbf{ g }_{ 3 } }$ are the functions realizing $\mathbf{ g }_{ 0 }$ and $\mathbf{ g }_{ 3 }$, respectively.

From Example \ref{xmp: Classifiers for $B_{ 1 }$ and $Q_{ 2 }$} we know that $C_{ 2 }$, defined by equation \eqref{eq: Unitary Transform C_2}, is the classifier for the class of Boolean functions $F_{ Q_{ 2 } }$ that realizes the pattern basis $Q_{ 2 }$. According to Theorem \ref{thr: Classification of Products of Pattern Bases & Function Classes}, $C_{ 2 } \otimes C_{ 2 }$ is the classifier for the class $F_{ Q_{ 2 } }$ $\star$ $F_{ Q_{ 2 } }$ that realizes the pattern basis $Q_{ 2 }$ $\odot$ $Q_{ 2 }$. In this scenario, the concrete implementation of the Pattern Basis Classification Scheme outlined in Definition \ref{def: The Pattern Basis Classification Scheme} in Qiskit \cite{Qiskit2025} is shown in Figure \ref{fig: Phase4___1000100010000111___} below, where the oracle encodes $f_{ \mathbf{ r }_{ 3 } }$.

\begin{tcolorbox}
	[
		enhanced,
		breakable,
		grow to left by = 0.000 cm,
		grow to right by = 0.000 cm,
		colback = white,
		enhanced jigsaw,			
		frame hidden,
		sharp corners,
	]
	\begin{figure}[H]
		\centering
		\includegraphics [ scale = 0.500, trim = {5.750cm 1.500cm 0.000cm 2.000cm}, clip ] {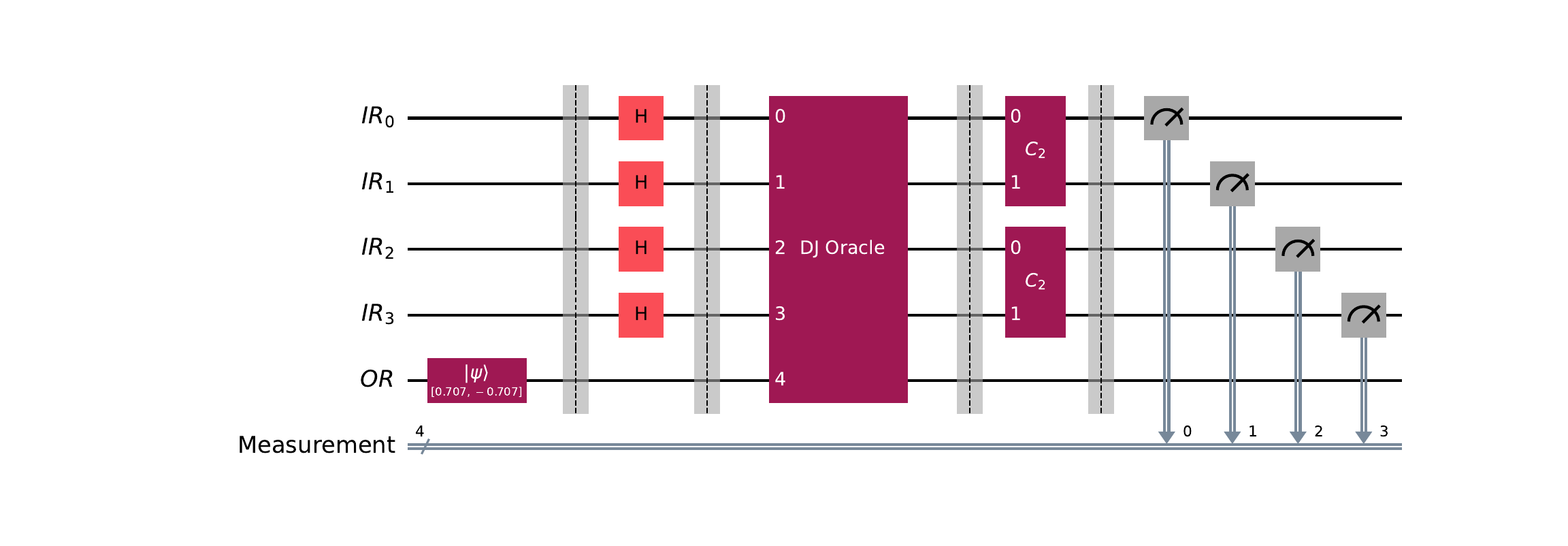}
		\caption{This figure shows the concrete implementation of the Pattern Basis Classification Scheme for the classification of the function $f_{ \mathbf{ r }_{ 3 } }$.}
		\label{fig: Phase4___1000100010000111___}
	\end{figure}
\end{tcolorbox}

\begin{tcolorbox}
	[
		enhanced,
		breakable,
		grow to left by = 0.000 cm,
		grow to right by = 0.000 cm,
		colback = white,
		enhanced jigsaw,			
		frame hidden,
		sharp corners,
	]
	\begin{minipage} [ t ] { 0.400 \textwidth }
		Before measurement, the state of the system is just $\ket{ \mathbf{ 0011 } }$. Hence, measuring the quantum circuit depicted in Figure \ref{fig: Phase4___1000100010000111___} will output the bit vector $0011$ with probability $1.0$. This is corroborated by the simulating this circuit in Qiskit for $2048$ runs, as shown in Figure \ref{fig: Phase4_Histogram_StatevectorSampler___1000100010000111___}). The binary vector $0011$ reveals the index of the function hidden in the oracle, namely $f_{ \mathbf{ r }_{ 3 } }$. This provides a conclusive demonstration of the Pattern Basis Classification Scheme.
	\end{minipage}
	\hspace{ 0.500 cm }
	\begin{minipage} [ t ] { 0.450 \textwidth }
		\begin{figure}[H]
			\centering
			\includegraphics [ scale = 0.400, trim = {0.000cm 0.000cm 0.000cm 0.000cm}, clip ] {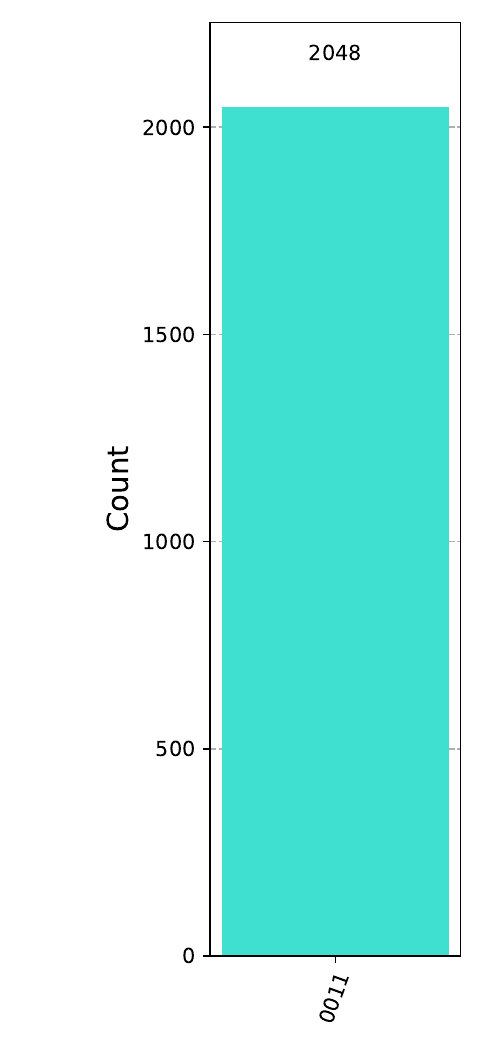}
			\caption{The measurement outcome of the quantum circuit of Figure \ref{fig: Phase4___1000100010000111___}.}
			\label{fig: Phase4_Histogram_StatevectorSampler___1000100010000111___}
		\end{figure}
	\end{minipage}
\end{tcolorbox}

\end{example}

In general, if a quantum classifier $\mathbf{ G }$ $=$ $G_{ m - 1 }$ $\otimes$ $\dots$ $\otimes$ $G_{ 1 }$ $\otimes$ $G_{ 0 }$ classifies the class of Boolean functions $F_{ P_{ n } }$, then the quantum circuit visualized in Figure \ref{fig: The BFPQC Quantum Circuit for $F_{ P_{ n } }$}, where the oracle $U_{ f }$ encodes a function $f \in F_{ P_{ n } }$, will conclusively, i.e., with probability $1.0$, classify $f$.

\begin{tcolorbox}
	[
		enhanced,
		breakable,
		grow to left by = 0.000 cm,
		grow to right by = 0.000 cm,
		colback = SkyBlue1!08,
		enhanced jigsaw,			
		frame hidden,
		sharp corners,
	]
	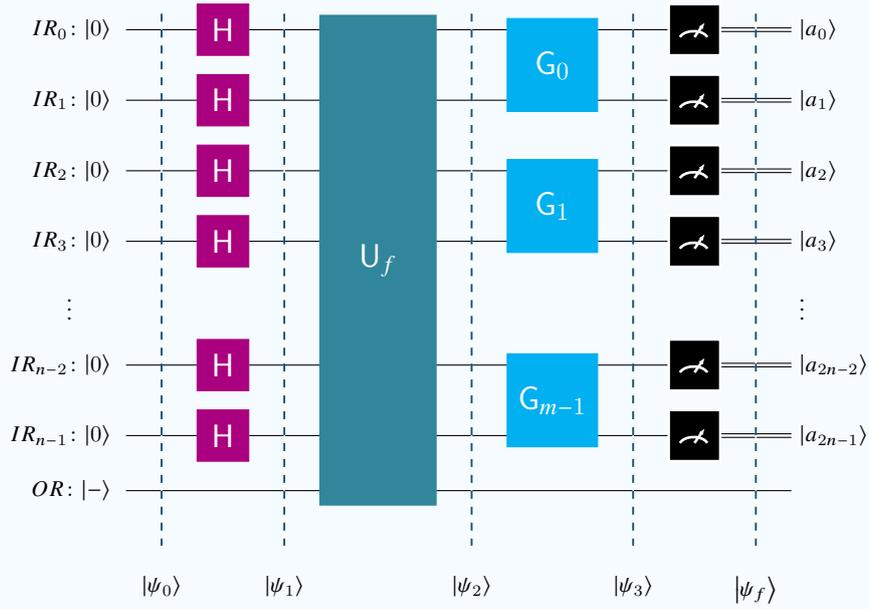
\begin{figure}[H]
		\centering
		\begin{tikzpicture} [ scale = 0.850, transform shape ]
			\begin{yquant}[ operator/separation = 3.000 mm, register/separation = 3.000 mm, every nobit output/.style = { } ]
				qubit { $IR_{ 0 } \colon \ket{ 0 }$ } IR;
				qubit { $IR_{ 1 } \colon \ket{ 0 }$ } IR [ + 1 ];
				qubit { $IR_{ 2 } \colon \ket{ 0 }$ } IR [ + 1 ];
				qubit { $IR_{ 3 } \colon \ket{ 0 }$ } IR [ + 1 ];
				qubit { $\vdots$ \hspace{ 0.475 cm } } IR [ + 1 ]; discard IR [ 4 ];
				qubit { $IR_{ n - 2 } \colon \ket{ 0 }$ } IR [ + 1 ];
				qubit { $IR_{ n - 1 } \colon \ket{ 0 }$ } IR [ + 1 ];
				qubit { $OR \colon \ket{ - }$ } OR;
				nobit AUX_1;
				[
				name = Input,
				WordBlueDarker,
				line width = 0.250 mm,
				]
				barrier ( - ) ;
				[ draw = RedPurple, fill = RedPurple, radius = 0.400 cm ] box {\color{white} \Large \sf{H}} IR [ 0 ];
				[ draw = RedPurple, fill = RedPurple, radius = 0.400 cm ] box {\color{white} \Large \sf{H}} IR [ 1 ];
				[ draw = RedPurple, fill = RedPurple, radius = 0.400 cm ] box {\color{white} \Large \sf{H}} IR [ 2 ];
				[ draw = RedPurple, fill = RedPurple, radius = 0.400 cm ] box {\color{white} \Large \sf{H}} IR [ 3 ];
				[ draw = RedPurple, fill = RedPurple, radius = 0.400 cm ] box {\color{white} \Large \sf{H}} IR [ 5 ];
				[ draw = RedPurple, fill = RedPurple, radius = 0.400 cm ] box {\color{white} \Large \sf{H}} IR [ 6 ];
				[
				name = Expansion,
				WordBlueDarker,
				line width = 0.250 mm,
				]
				barrier ( - ) ;
				[ draw = WordAquaDarker25, fill = WordAquaDarker25, x radius = 0.900 cm, y radius = 0.450 cm ] box { \color{white} \Large \sf{U}$_{ f }$} ( IR - OR );
				[
				name = Oracle,
				WordBlueDarker,
				line width = 0.250 mm,
				]
				barrier ( - ) ;
				[ draw = WordBlueVeryLight, fill = WordBlueVeryLight, x radius = 0.700 cm, y radius = 0.350 cm ] box { \color{white} \Large \sf{G}$_{ 0 }$}  ( IR [ 0 ] - IR [ 1 ] );
				[ draw = WordBlueVeryLight, fill = WordBlueVeryLight, x radius = 0.700 cm, y radius = 0.350 cm ] box { \color{white} \Large \sf{G}$_{ 1 }$}  ( IR [ 2 ] - IR [ 3 ] );
				[ draw = WordBlueVeryLight, fill = WordBlueVeryLight, x radius = 0.700 cm, y radius = 0.350 cm ] box { \color{white} \Large \sf{G}$_{ m - 1 }$}  ( IR [ 5 ] - IR [ 6 ] );
				[
				name = Classifier,
				WordBlueDarker,
				line width = 0.250 mm,
				]
				barrier ( - ) ;
				[ line width = .350 mm, draw = white, fill = black, radius = 0.400 cm ] measure IR [ 0 ];
				[ line width = .350 mm, draw = white, fill = black, radius = 0.400 cm ] measure IR [ 1 ];
				[ line width = .350 mm, draw = white, fill = black, radius = 0.400 cm ] measure IR [ 2 ];
				[ line width = .350 mm, draw = white, fill = black, radius = 0.400 cm ] measure IR [ 3 ];
				[ line width = .350 mm, draw = white, fill = black, radius = 0.400 cm ] measure IR [ 5 ];
				[ line width = .350 mm, draw = white, fill = black, radius = 0.400 cm ] measure IR [ 6 ];
				[
				name = Measurement,
				WordBlueDarker,
				line width = 0.250 mm,
				]
				barrier ( - ) ;
				output { $\ket{ a_{ 0 } }$ } IR [ 0 ];
				output { $\ket{ a_{ 1 } }$ } IR [ 1 ];
				output { $\ket{ a_{ 2 } }$ } IR [ 2 ];
				output { $\ket{ a_{ 3 } }$ } IR [ 3 ];
				output { $\vdots$ } IR [ 4 ];
				output { $\ket{ a_{ 2 n - 2 } }$ } IR [ 5 ];
				output { $\ket{ a_{ 2 n - 1 } }$ } IR [ 6 ];
				\node [ below = 4.500 cm ] at (Input) { $\ket{ \psi_{ 0 } }$ };
				\node [ below = 4.500 cm ] at (Expansion) { $\ket{ \psi_{ 1 } }$ };
				\node [ below = 4.500 cm ] at (Oracle) { $\ket{ \psi_{ 2 } }$ };
				\node [ below = 4.500 cm ] at (Classifier) { $\ket{ \psi_{ 3 } }$ };
				\node [ below = 4.500 cm ] at (Measurement) { $\ket{ \psi_{ f } }$ };
			\end{yquant}
		\end{tikzpicture}
		\caption{This figure depicts the quantum circuit that conclusively classifies all functions contained in $F_{ P_{ n } }$.}
		\label{fig: The BFPQC Quantum Circuit for $F_{ P_{ n } }$}
	\end{figure}
\end{tcolorbox}

To enhance clarity, we explain the notation used in the above figure.

\begin{itemize}
	\item	
	$IR$ is the quantum input register that contains $n$ qubits and starts its operation at state $\ket{ \mathbf{ 0 } }$.
	\item	
	$OR$ is the single-qubit output register initialized to $\ket{ - }$.
	\item	
	$H$ is the Hadamard transform.
	\item	
	$U_{ f }$ is the unitary transform corresponding to the oracle for the unknown function $f$. The latter is promised to be an element of $F_{ P_{ n } }$.
	\item	
	$G_{ j }, 0 \leq j \leq m - 1,$ are the fundamental building blocks of $\mathbf{ G }$ $=$ $G_{ m - 1 }$ $\otimes$ $\dots$ $\otimes$ $G_{ 1 }$ $\otimes$ $G_{ 0 }$ classifying the class of Boolean functions $F_{ P_{ n } }$.
\end{itemize}

We now emphasize an important point regarding the behavior of classification algorithms on general inputs. No classification algorithm is expected to produce correct results—and is unlikely to do so—for Boolean functions that fall outside the class of perfectly classifiable functions. Thus, regarding the quantum circuit of Figure \ref{fig: The BFPQC Quantum Circuit for $F_{ P_{ n } }$}, if the oracle realizes a function outside $F_{ P_{ n } }$, the final measurement, may not reveal the unknown function. In the rest of this paper, we investigate whether, and under what assumptions, it is possible to obtain some useful information when the input function is close to the class $F_{ P_{ n } }$.

\section{What about nonclassifiable Boolean functions?} \label{sec: What About Nonclassifiable Boolean Functions?}

In view of the last remark of the previous Section, we lay down the objectives of this study. Let us assume that we have fixed the class of Boolean functions $F_{ P_{ n } }$ that realizes the pattern basis $P_{ n }$, and that $\mathbf{ G }$ is the classifier for $F_{ P_{ n } }$ and $P_{ n }$. In the rest of this paper we investigate the following questions.

\begin{itemize}
	\item	
	When the input function $h$ is nonclassifiable by $\mathbf{ G }$, i.e., $h \not \in F_{ P_{ n } }$, is it possible, and under what conditions, to obtain some useful information about $h$?
	\item	
	By useful information we mean to learn as much as possible about the behavior of $h$. Within the current framework, this translates to knowing (one of) the $f \in F_{ P_{ n } }$ ``nearest'' to $h$.
	\item	
	Thus, to quantify the ``nearness'' between two Boolean functions, a notion of distance is required. In this work, we adopt one of the most general distance metric, the Hamming distance.
	\item	
	Finally, we aim to answer the question of how useful is the Hamming distance, as defined in Definition \ref{def: Hamming Distance of Boolean Functions}, for classification purposes.
\end{itemize}

Undoubtedly, any attempt to arrive at quantitative conclusions, presupposes a metric to capture the notion of ``closeness'' or distance between two Boolean functions. As mentioned above, in this study we employ one of the most conventional and most general distance metrics, namely the Hamming distance (see \cite{Bierbrauer2017,Ling2004}), and analyze its benefits and drawbacks. In \cite{Andronikos2025b} a totally different concept of distance, based on left and right clusters, was used. The classification game investigated there, with the distance restriction forcing the unknown function to be either in the left or right cluster of a perfectly classifiable function, enabled Alice to surely win. The Nearest Basis Ket Game we study in this work is totally different because it deals with the utmost general situation where Bob can pick any Boolean function with no restrictions whatsoever.

\begin{definition} {Hamming Distance of Boolean Functions} { Hamming Distance of Boolean Functions}
	Let $f$ and $h$ be two Boolean functions from $\mathbb{ B }^{ n } \to \mathbb{ B }$, and let $\mathbf{ p }_{ f }$ and $\mathbf{ p }_{ h }$ be their corresponding pattern bit vectors. We define the Hamming distance between $f$ and $h$, denoted $d_{ H } ( f, h )$, as follows:
	\begin{align}
		\label{eq: Boolean Functions Hamming Distance}
		d_{ H } ( f, h )
		\coloneq
		d_{ H } ( \mathbf{ p }_{ f }, \mathbf{ p }_{ h } )
		\ .
	\end{align}
\end{definition}

\begin{definition} {Neighborhoods of Boolean Functions} { Neighborhoods of Boolean Functions}
	Let $f$ be a Boolean function from $\mathbb{ B }^{ n } \to \mathbb{ B }$. The $r$-neighborhood of $f$, denoted by $N_{ r } ( f )$, $r \geq 0$, is the collection of all Boolean functions $h$ from $\mathbb{B}^{ n } \to \mathbb{ B }$ such that $d_{ H } ( f, h ) \leq r$.
\end{definition}

\begin{definition} {Hamming Distance from Function Class} { Hamming Distance from Function Class}
	Given the class of Boolean functions $F_{ P_{ n } }$ that realizes the pattern basis $P_{ n }$, and an arbitrary Boolean function $h \colon \mathbb{ B }^{ n } \to \mathbb{ B }$, $n \geq 1$, the Hamming distance of $h$ from $F_{ P_{ n } }$, denoted by $d_{ H } ( F_{ P_{ n } }, h )$, is defined as
	\begin{align}
		\label{eq: Hamming Distance from F}
		d_{ H } ( F_{ P_{ n } }, h ) = \min \{ d : d = d_{ H } ( f, h ), \ f \in F_{ P_{ n } } \}
		\ .
	\end{align}
\end{definition}

\begin{definition} {Nearest Neighbors in Function Class} { Nearest Neighbors in Function Class}
	Given the class of Boolean functions $F_{ P_{ n } }$ that realizes the pattern basis $P_{ n }$, and an arbitrary Boolean function $h \colon \mathbb{ B }^{ n } \to \mathbb{ B }$, $n \geq 1$, the \emph{nearest neighbors} of $h$ in $F_{ P_{ n } }$, denoted by $NN ( F_{ P_{ n } }, h )$, are the Boolean functions $f \in F_{ P_{ n } }$ such that $d_{ H } ( f, h ) = d_{ H } ( F_{ P_{ n } }, h )$.
\end{definition}

\begin{definition} {Nearest Basis Kets} { Nearest Basis Kets}
	Let $\mathbf{ G }$ be a quantum classifier for the class of Boolean functions $F_{ P_{ n } }$ realizing the pattern basis $P_{ n }$ of rank $n$. Given an arbitrary Boolean function $h \colon \mathbb{ B }^{ n } \to \mathbb{ B }$, not belonging to $F_{ P_{ n } }$, let $NN ( F_{ P_{ n } }, h ) = \{ f_{ \mathbf{ p }_{ i_{ 1 } } }, f_{ \mathbf{ p }_{ i_{ 2 } } }, \dots, f_{ \mathbf{ p }_{ i_{ r } } } \}$ be the collection of nearest neighbors of $h$ in $F_{ P_{ n } }$.

	The \emph{nearest basis kets} of $h$ in $F_{ P_{ n } }$, denoted by $NK ( F_{ P_{ n } }, h )$, are the basis kets $\{ \ket{ \mathbf{ i }_{ 1 } }, \ket{ \mathbf{ i }_{ 2 } }, \dots, \ket{ \mathbf{ i }_{ r } } \}$, where $\ket{ \mathbf{ i }_{ 1 } }, \ket{ \mathbf{ i }_{ 2 } }, \dots, \ket{ \mathbf{ i }_{ r } }$ are the classification measurements for $f_{ \mathbf{ p }_{ i_{ 1 } } }, f_{ \mathbf{ p }_{ i_{ 2 } } }, \dots, f_{ \mathbf{ p }_{ i_{ r } } }$ according to the Pattern Basis Classification Scheme outlined in Definition \ref{def: The Pattern Basis Classification Scheme}.

	The \emph{classification threshold} $\vartheta$ of $h$ is the probability that $CM ( h )$ belongs to $NK ( F_{ P_{ n } }, h )$.
\end{definition}
\begin{example} {The Hamming Distance of $g$ from $F_{ Q_{ 4 } }$} { The Hamming Distance of $g$ from $F_{ Q_{ 4 } }$}

Let us consider the pattern bit vector $\mathbf{ r }_{ 0 } = 0001 \ 0001 \ 0001 \ 1110$ $\in$ $Q_{ 4 }$, listed in Table \ref{tbl: Pattern Basis $Q_{ 4 }$} of Example \ref{xmp: Example Of Pattern Bases Product}, and the pattern vector $\mathbf{ p }_{ g } = 0000 \ 0001 \ 0001 \ 1110$ encountered in Example \ref{xmp: Functions & Patterns}. The values of the corresponding Boolean functions $f_{ \mathbf{ r }_{ 0 } }$ and $g$ are contained in Table \ref{tbl: Comparison Of The Truth Values Of $f_{ 0 }$ And $g$}. A comparison between their corresponding values shows that they only differ in the values the assume at input $\mathbf{ x } = 1100$, where $f_{ \mathbf{ r }_{ 0 } } ( \mathbf{ x } ) = 1$ whereas $g ( \mathbf{ x } ) = 0$. According to Definition \ref{def: Hamming Distance of Boolean Functions}, $d_{ H } ( f_{ \mathbf{ r }_{ 0 } }, g ) = 1$. By systematically going over all other functions of the class $F_{ Q_{ 4 } }$, we immediately verify that their distance from $g$ is $> 1$. Therefore, $NN ( F_{ Q_{ 4 } }, g )$ $=$ $\{ f_{ \mathbf{ r }_{ 0 } } \}$. Lastly, taking into account the Pattern Basis Classification Scheme outlined in Definition \ref{def: The Pattern Basis Classification Scheme} and Definition \ref{def: Nearest Basis Kets}, we conclude that the nearest basis ket of $g$ in $F_{ Q_{ 4 } }$ is $\ket{ 0000 }$, or, in other words, $NK ( F_{ Q_{ 4 } }, h )$ $=$ $\{ \ket{ 0000 } \}$.

\begin{tcolorbox}
	[
		enhanced,
		breakable,
		grow to left by = 0.500 cm,
		grow to right by = 0.000 cm,
		colback = white,
		enhanced jigsaw,			
		frame hidden,
		sharp corners,
	]
	\begin{table}[H]
		{\small
			\caption{Comparison between the truth values of the functions $g$ and $f_{ 0 }$.}
			\label{tbl: Comparison Of The Truth Values Of $f_{ 0 }$ And $g$}
			\centering
			\SetTblrInner { rowsep = 1.000 mm }
			\begin{tblr}
				{
					colspec =
					{
						Q [ c, m, 0.500 cm ]
						| [ 1.000 pt, WordAquaDarker25 ]
						| [ 1.000 pt, WordAquaDarker25 ]
						Q [ c, m, 0.450 cm ]
						| [ 0.500 pt, WordAquaDarker25 ]
						Q [ c, m, 0.450 cm ]
						| [ 0.500 pt, WordAquaDarker25 ]
						Q [ c, m, 0.450 cm ]
						| [ 0.500 pt, WordAquaDarker25 ]
						Q [ c, m, 0.450 cm ]
						| [ 0.500 pt, WordAquaDarker25 ]
						Q [ c, m, 0.450 cm ]
						| [ 0.500 pt, WordAquaDarker25 ]
						Q [ c, m, 0.450 cm ]
						| [ 0.500 pt, WordAquaDarker25 ]
						Q [ c, m, 0.450 cm ]
						| [ 0.500 pt, WordAquaDarker25 ]
						Q [ c, m, 0.450 cm ]
						| [ 0.500 pt, WordAquaDarker25 ]
						Q [ c, m, 0.450 cm ]
						| [ 0.500 pt, WordAquaDarker25 ]
						Q [ c, m, 0.450 cm ]
						| [ 0.500 pt, WordAquaDarker25 ]
						Q [ c, m, 0.450 cm ]
						| [ 0.500 pt, WordAquaDarker25 ]
						Q [ c, m, 0.450 cm ]
						| [ 0.500 pt, WordAquaDarker25 ]
						Q [ c, m, 0.450 cm ]
						| [ 0.500 pt, WordAquaDarker25 ]
						Q [ c, m, 0.450 cm ]
						| [ 0.500 pt, WordAquaDarker25 ]
						Q [ c, m, 0.450 cm ]
						| [ 0.500 pt, WordAquaDarker25 ]
						Q [ c, m, 0.450 cm ]
					},
					rowspec =
					{
						|
						[ 3.500 pt, WordAquaDarker25 ]
						|
						[ 0.750 pt, WordAquaDarker25 ]
						|
						[ 0.250 pt, white ]
						Q
						|
						[ 0.500 pt, WordAquaDarker25 ]
						Q
						|
						[ 0.500 pt, WordAquaDarker25 ]
						Q
						|
						[ 3.500 pt, WordAquaDarker50 ]
					}
				}
				&
				\SetCell { WordAquaLighter40, fg = black } $\mathbf{ 0000 }$
				&
				\SetCell { WordAquaLighter40, fg = black } $\mathbf{ 0001 }$
				&
				\SetCell { WordAquaLighter40, fg = black } $\mathbf{ 0010 }$
				&
				\SetCell { WordAquaLighter40, fg = black } $\mathbf{ 0011 }$
				&
				\SetCell { WordAquaLighter40, fg = black, font = \bfseries } $\mathbf{ 0100 }$
				&
				\SetCell { WordAquaLighter40, fg = black, font = \bfseries } $\mathbf{ 0101 }$
				&
				\SetCell { WordAquaLighter40, fg = black, font = \bfseries } $\mathbf{ 0110 }$
				&
				\SetCell { WordAquaLighter40, fg = black, font = \bfseries } $\mathbf{ 0111 }$
				&
				\SetCell { WordAquaLighter40, fg = black, font = \bfseries } $\mathbf{ 1000 }$
				&
				\SetCell { WordAquaLighter40, fg = black, font = \bfseries } $\mathbf{ 1001 }$
				&
				\SetCell { WordAquaLighter40, fg = black, font = \bfseries } $\mathbf{ 1010 }$
				&
				\SetCell { WordAquaLighter40, fg = black, font = \bfseries } $\mathbf{ 1011 }$
				&
				\SetCell { WordAquaLighter40, fg = black, font = \bfseries } $\mathbf{ 1100 }$
				&
				\SetCell { WordAquaLighter40, fg = black, font = \bfseries } $\mathbf{ 1101 }$
				&
				\SetCell { WordAquaLighter40, fg = black, font = \bfseries } $\mathbf{ 1110 }$
				&
				\SetCell { WordAquaLighter40, fg = black, font = \bfseries } $\mathbf{ 1111 }$
				\\
				$g$
				&
				$0$
				&
				$1$
				&
				$1$
				&
				$1$
				&
				$1$
				&
				$0$
				&
				$0$
				&
				$0$
				&
				$1$
				&
				$0$
				&
				$0$
				&
				$0$
				&
				$0$
				&
				$0$
				&
				$0$
				&
				$0$
				\\
				$f_{ \mathbf{ r }_{ 0 } }$
				&
				$0$
				&
				$1$
				&
				$1$
				&
				$1$
				&
				$1$
				&
				$0$
				&
				$0$
				&
				$0$
				&
				$1$
				&
				$0$
				&
				$0$
				&
				$0$
				&
				$1$
				&
				$0$
				&
				$0$
				&
				$0$
				\\
			\end{tblr}
		}
	\end{table}
\end{tcolorbox}

\end{example}

To make this investigation more intuitive and engaging, we conceptualize the classification process as a strategic interaction, formalized as a game between two players, Alice and Bob, whom we cast as our talented protagonists. This game, termed the \textbf{Nearest Basis Ket Game}, encapsulates the challenge of identifying one of the ``closest'' functions in $F_{ P_{ n } }$ to an unknown function $h$. The rules of the game are as follows:

\begin{tcolorbox}
	[
		enhanced,
		breakable,
		center title,
		fonttitle = \bfseries,
		colbacktitle = cyan4,
		coltitle = white,
		title = \textbf{Nearest Basis Ket Game},
		grow to left by = 0.000 cm,
		grow to right by = 0.000 cm,
		colframe = cyan4,
		colback = cyan9!50,
		enhanced jigsaw,			
		sharp corners,
		boxrule = 0.500 pt,
	]
	\begin{enumerate}
		[ left = 0.500 cm, labelsep = 1.000 cm, start = 1 ]
		\renewcommand \labelenumi { (\textbf{R}$_{ \theenumi }$) }
		\item
		Alice and Bob fix a classifier $\mathbf{ G }$ and the corresponding class of Boolean functions $F_{ P }$ that realizes the pattern basis $P$.
		\item
		Bob picks an arbitrary Boolean function $h$ that doesn't belong to $F_{ P }$ and reveals to Alice only the minimum Hamming distance of $h$ from $F_{ P }$.
		\item
		Bob implements the abstract classification circuit of Figure \ref{fig: The Pattern Basis Classification Scheme} using the oracle for $h$, and announces the classification measurement $CM ( h )$ to Alice.
		\item
		The challenge posed to Alice is to decide whether $CM ( h )$ belongs to $NK ( F_{ P }, h )$.
		\item	Alice wins the game if she gives the correct answer, either yes or no, to the above question, otherwise Bob is declared the winner.
	\end{enumerate}
\end{tcolorbox}

As we show in the rest of this paper, when the Hamming distance lies within certain intervals, the question whether $CM ( h ) \in NK ( F_{ P_{ n } }, h )$ can be answered affirmatively or negatively with high probability.

\section{What do experiments show?} \label{sec: What Do Experiments Show?}

\subsection{Experimental methodology} \label{subsec: Experimental Methodology}

To evaluate the usefulness of the Hamming distance as a meaningful metric to measure ``closeness'' between Boolean functions, we have conducted a series of experiments using Qiskit \cite{Qiskit2025}. This Section is devoted to the presentation and analysis of these experiments. Our experimental methodology was meticulously designed to uphold the core principles of

\begin{itemize}
	\item	
	reproducibility, and
	\item	
	interpretability,
\end{itemize}

ensuring that all procedures, parameters, and outcomes could be reliably replicated by independent researchers while providing clear, transparent explanations for the observed phenomena. To achieve this, we documented every step in detail, alongside comprehensive logging of intermediate results and computational traces.

We clarify the experimental setup that is used throughput this series of experiments. For each experiment we fix a quantum classifier $\mathbf{ G } = G_{ m - 1 } \otimes \dots \otimes G_{ 0 }$, where each factor of the tensor product is one of the elementary classifiers $H$ or $C_{ 2 }$. For the chosen $\mathbf{ G }$, the class of perfectly classifiable Boolean functions $F_{ P }$ and the pattern basis $P$ realized by this class are unambiguously specified, as prescribed by Theorem \ref{thr: Classification of Products of Pattern Bases & Function Classes} and Corollary \ref{crl: Correspondence Among Bases, Function Classes & Classifiers}. The Pattern Basis Classification Scheme is achieved via the appropriate implementation of the abstract quantum circuit depicted in Figure \ref{fig: The BFPQC Quantum Circuit for $F_{ P_{ n } }$}. Our experiments strive to shed light on the following investigation.

\begin{tcolorbox}
	[
		enhanced,
		breakable,
		center title,
		fonttitle = \bfseries,
		colbacktitle = RedPurple,
		coltitle = white,
		title = \textbf{Experimental Investigation},
		grow to left by = 0.000 cm,
		grow to right by = 0.000 cm,
		colframe = RedPurple,
		colback = RedPurple!12,
		enhanced jigsaw,			
		sharp corners,
		boxrule = 0.500 pt,
	]
	Assuming that the oracle in Figure \ref{fig: The BFPQC Quantum Circuit for $F_{ P_{ n } }$} encodes an unknown random function $h$, what is the classification threshold of $h$, that is what is probability that the classification measurement $CM ( h )$ belongs to $NK ( F_{ P }, h )$, as a function of the Hamming distance $d_{ H } ( F_{ P }, h )$ of $h$ from $F_{ P }$?
\end{tcolorbox}

Given the inherent computational constraints—such as finite memory capacity and processing time—we adopted a pragmatic and scalable approach that balanced thoroughness with feasibility. For smaller-scale analyses, we performed an exhaustive enumeration of all possible Boolean functions associated with pattern bit vectors of length up to $16$ bits. This comprehensive sweep covered the full combinatorial space of all $2^{ 16 } = 65536$ functions from $\mathbb{ B }^{ 4 } \to \mathbb{ B }$, which is quite manageable on modern hardware, allowing us to derive definitive insights without sampling biases.

For larger pattern bit vectors (lengths of $32$ bits and beyond), exhaustive enumeration becomes prohibitively expensive due to the exponential growth in complexity. For instance, we have $2^{ 32 } = 4,294,967,296$ Boolean functions $\mathbb{ B }^{ 5 } \to \mathbb{ B }$. Therefore, we shifted to a statistically robust sampling strategy. Specifically, we randomly generated a diverse set of pattern bit vectors using uniform distribution, ensuring broad coverage of the search space. For each generated vector, we systematically constructed the corresponding Boolean function and classified it under our experimental framework. This random sampling not only mitigated the risks of exhaustive computation but also enabled us to identify representative trends and generalize findings to the broader landscape of Hamming distance usability.

\subsection{The case of exhaustively enumerating all Boolean functions} \label{subsec: The Case Of Exhaustively Enumerating All Boolean Functions}

This case documents the behavior of all pattern bit vectors of length $8$ and $16$, which were fully enumerated and tested. In particular, using the elementary pattern bases $B_{ 1 }$ and $Q_{ 2 }$, defined by equations \eqref{eq: Pattern Basis $B_{ 1 }$} and \eqref{eq: Pattern Basis $Q_{ 2 }$} respectively, we constructed all possible products of pattern bases that consist of bit vectors of length $8$ and $16$. As explained in detail in Example \ref{xmp: Classifiers for $B_{ 1 }$ and $Q_{ 2 }$}, the function classes $F_{ B_{ 1 } }$ and $F_{ Q_{ 2 } }$ realize $B_{ 1 }$ and $Q_{ 2 }$. Thus, by invoking Corollary \ref{crl: Correspondence Among Bases, Function Classes & Classifiers} that makes explicit the one-to-one exact correspondence among the products of bases, function classes, and quantum classifiers (see also Figure \ref{fig: One-to-one Correspondence Among Products}), we obtain Tables \ref{tbl: Pattern Bases Of Length $8$} and \ref{tbl: Pattern Bases Of Length $16$}.

\begin{tcolorbox}
	[
		enhanced,
		breakable,
		grow to left by = 0.000 cm,
		grow to right by = 0.000 cm,
		colback = white,
		enhanced jigsaw,			
		frame hidden,
		sharp corners,
	]
	\begin{table}[H]
		\caption{This table contains the pattern bases that contain pattern bit vectors of length $8$.}
		\label{tbl: Pattern Bases Of Length $8$}
		\centering
		\SetTblrInner { rowsep = 1.200 mm }
		\begin{tblr}
			{
				colspec =
				{
					Q [ c, m, 2.000 cm ]
					| [ 0.750 pt, WordAquaDarker25 ]
					| [ 0.750 pt, WordAquaDarker25 ]
					Q [ c, m, 3.000 cm ]
					| [ 0.500 pt, WordAquaDarker25 ]
					Q [ c, m, 3.600 cm ]
					| [ 0.500 pt, WordAquaDarker25 ]
					Q [ c, m, 3.000 cm ]
				},
				rowspec =
				{
					| [ 3.500 pt, WordAquaDarker25 ]
					| [ 0.750 pt, WordAquaDarker25 ]
					| [ 0.250 pt, white ]
					Q
					| [ 0.150 pt, WordAquaDarker25 ]
					Q
					| [ 0.150 pt, WordAquaDarker25 ]
					Q
					| [ 0.150 pt, WordAquaDarker25 ]
					Q
					| [ 3.500 pt, WordAquaDarker50 ]
				}
			}
			\SetCell { bg = WordAquaDarker50, fg = white, font = \bfseries \small } Pattern Bit Vector Length
			&
			\SetCell { bg = WordAquaLighter40, fg = black, font = \bfseries \small } Pattern Basis
			&
			\SetCell { bg = WordAquaLighter40, fg = black, font = \bfseries \small } Class of Classifiable Boolean Functions
			&
			\SetCell { bg = WordAquaLighter40, fg = black, font = \bfseries \small } Quantum Classifier
			\\
			\SetCell { bg = WordAquaLighter40, fg = black, font = \bfseries \small } $8$
			&
			\SetCell { font = \small } $B_{ 1 } \odot B_{ 1 } \odot B_{ 1 }$
			&
			\SetCell { font = \small } $F_{ B_{ 1 } } \star F_{ B_{ 1 } } \star F_{ B_{ 1 } }$
			&
			\SetCell { font = \small } $H \otimes H \otimes H$
			\\
			\SetCell { bg = WordAquaLighter40, fg = black, font = \bfseries \small } $8$
			&
			\SetCell { font = \small } $B_{ 1 } \odot Q_{ 2 }$
			&
			\SetCell { font = \small } $F_{ B_{ 1 } } \star F_{ Q_{ 2 } }$
			&
			\SetCell { font = \small } $H \otimes C_{ 2 }$
			\\
			\SetCell { bg = WordAquaLighter40, fg = black, font = \bfseries \small } $8$
			&
			\SetCell { font = \small } $Q_{ 2 } \odot B_{ 1 }$
			&
			\SetCell { font = \small } $F_{ Q_{ 2 } } \star F_{ B_{ 1 } }$
			&
			\SetCell { font = \small } $C_{ 2 } \otimes H$
			\\
		\end{tblr}
	\end{table}
\end{tcolorbox}

We begin the presentation of the experimental results with the pattern bases that contain bit vectors of length $8$. The experimental results for the three classes of Boolean functions contained in Table \ref{tbl: Pattern Bases Of Length $8$} are virtually identical and this conclusion is summarized in the following text box. Hence, in Figures \ref{fig: Probability Histogram For $F_{ B_{ 1 } } F_{ Q_{ 2 } }$} and \ref{fig: Probability Plot For $F_{ B_{ 1 } } F_{ Q_{ 2 } }$}, we have chosen to display only the results for the class $F_{ B_{ 1 } } \star F_{ Q_{ 2 } }$.

\begin{tcolorbox}
	[
		enhanced,
		breakable,
		center title,
		fonttitle = \bfseries,
		colbacktitle = RedPurple,
		coltitle = white,
		title = \textbf{Conclusion for the pattern bit vectors of length $\mathbf{ 8 }$},
		grow to left by = 0.000 cm,
		grow to right by = 0.000 cm,
		colframe = RedPurple,
		colback = white,
		enhanced jigsaw,			
		sharp corners,
		boxrule = 0.500 pt,
	]
	The probability distribution of the classification threshold is virtually identical for the function classes $F_{ B_{ 1 } } \star F_{ B_{ 1 } } \star F_{ B_{ 1 } }$, $F_{ B_{ 1 } } \star F_{ Q_{ 2 } }$, and $F_{ Q_{ 2 } } \star F_{ B_{ 1 } }$.
\end{tcolorbox}

In this particular scenario, we have the class of Boolean functions $F_{ B_{ 1 } } \star F_{ Q_{ 2 } }$ that realizes the pattern basis $B_{ 1 } \odot Q_{ 2 }$. In Figure \ref{fig: Probability Histogram For $F_{ B_{ 1 } } F_{ Q_{ 2 } }$} the results are shown as histograms with the red bar depicting the classification threshold of a random $h$ as a function of the Hamming distance $d_{ H } ( F_{ B_{ 1 } } \star F_{ Q_{ 2 } }, h )$ of $h$ from $F_{ B_{ 1 } } \star F_{ Q_{ 2 } }$. The green bar shows the number of Boolean Functions for which the Hamming distance from $F_{ B_{ 1 } } \star F_{ Q_{ 2 } }$ takes a specific value. Likewise, in Figure \ref{fig: Probability Plot For $F_{ B_{ 1 } } F_{ Q_{ 2 } }$} the red line shows the probability and the green line shows the number of functions per distance. As we have emphasized, the results are identical for all three function classes and are summarized in Table \ref{tbl: Probability vs. Hamming Distance Table For Pattern Bit Vectors of Length $8$}.

\begin{tcolorbox}
	[
		enhanced,
		breakable,
		grow to left by = 0.000 cm,
		grow to right by = 0.000 cm,
		colback = white,
		enhanced jigsaw,			
		frame hidden,
		sharp corners,
	]
	\begin{table}[H]
		\caption{The probability distribution of the classification threshold as a function of the Hamming distance for the classes $F_{ B_{ 1 } } \star F_{ B_{ 1 } } \star F_{ B_{ 1 } }$, $F_{ B_{ 1 } } \star F_{ Q_{ 2 } }$, and $F_{ Q_{ 2 } } \star F_{ B_{ 1 } }$ in tabular form.}
		\label{tbl: Probability vs. Hamming Distance Table For Pattern Bit Vectors of Length $8$}
		\centering
		\SetTblrInner { rowsep = 1.200 mm }
		\begin{tblr}
			{
				colspec =
				{
					Q [ c, m, 3.000 cm ]
					| [ 0.750 pt, GreenLighter2!50 ]
					| [ 0.750 pt, GreenLighter2!50 ]
					Q [ c, m, 9.000 cm ]
				},
				rowspec =
				{
					| [ 3.500 pt, GreenLighter2!75 ]
					| [ 0.750 pt, GreenLighter2!75 ]
					| [ 0.250 pt, white ]
					Q
					| [ 0.250 pt, white ]
					Q
					| [ 0.150 pt, GreenLighter2 ]
					Q
					| [ 0.150 pt, GreenLighter2 ]
					Q
					| [ 0.150 pt, GreenLighter2 ]
					Q
					| [ 0.150 pt, GreenLighter2 ]
					Q
					| [ 3.500 pt, GreenLighter2 ]
				}
			}
			&
			\SetCell { bg = GreenLighter2, fg = white, font = \small \bfseries } Classification Threshold $\boldsymbol{ \vartheta }$
			\\
			\SetCell { bg = GreenLighter2, fg = white, font = \small \bfseries }
			Hamming Distance
			&
			\SetCell { bg = GreenLighter2, fg = white, font = \small \bfseries } $F_{ B_{ 1 } } \star F_{ B_{ 1 } } \star F_{ B_{ 1 } }$, $F_{ B_{ 1 } } \star F_{ Q_{ 2 } }$, and $F_{ Q_{ 2 } } \star F_{ B_{ 1 } }$
			\\
			$1$
			&
			$0.56$
			\\
			$2$
			&
			$0.54$
			\\
			$3$
			&
			$0.22$
			\\
			$4, 5, 6, 7, 8$
			&
			$0.00$
			\\
		\end{tblr}
	\end{table}
\end{tcolorbox}

\begin{figure}[H]
		\centering
		\begin{subfigure} { 0.440 \textwidth }
			\centering
			\includegraphics [ angle = 90, width = 0.950 \textwidth, trim = {0.000cm 0.250cm 0.000cm 0.250cm}, clip ] {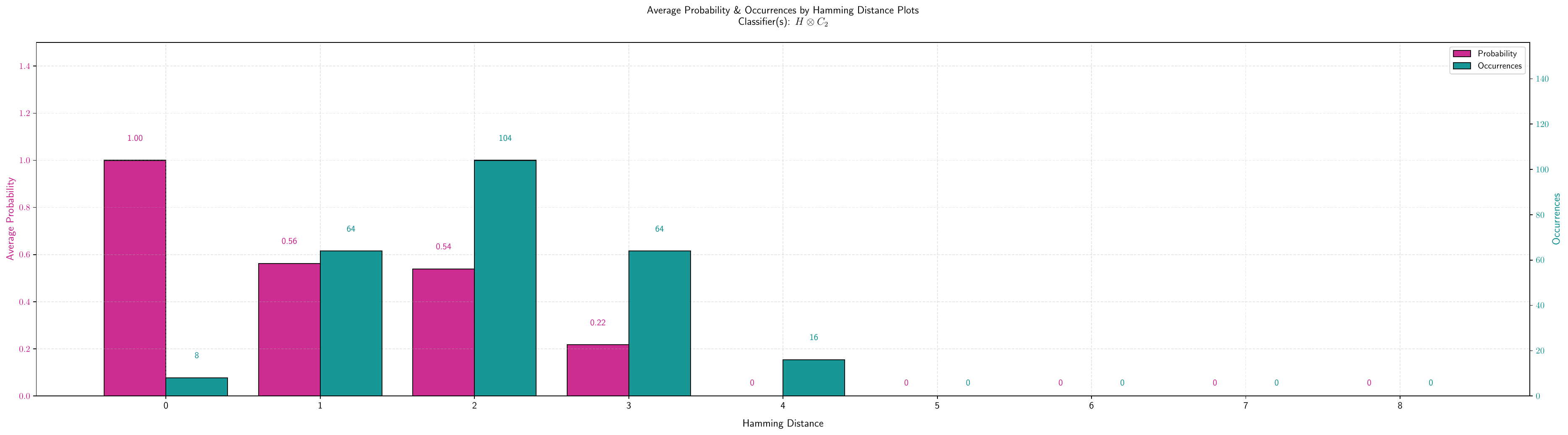}
			\caption{The probability histogram for $F_{ B_{ 1 } } \star F_{ Q_{ 2 } }$.}
			\label{fig: Probability Histogram For $F_{ B_{ 1 } } F_{ Q_{ 2 } }$}
		\end{subfigure}
		\hfill
		\begin{subfigure} { 0.440 \textwidth }
			\centering
			\includegraphics [ angle = 90, width = 0.950 \textwidth, trim = {0.000cm 0.250cm 0.000cm 0.250cm}, clip ] {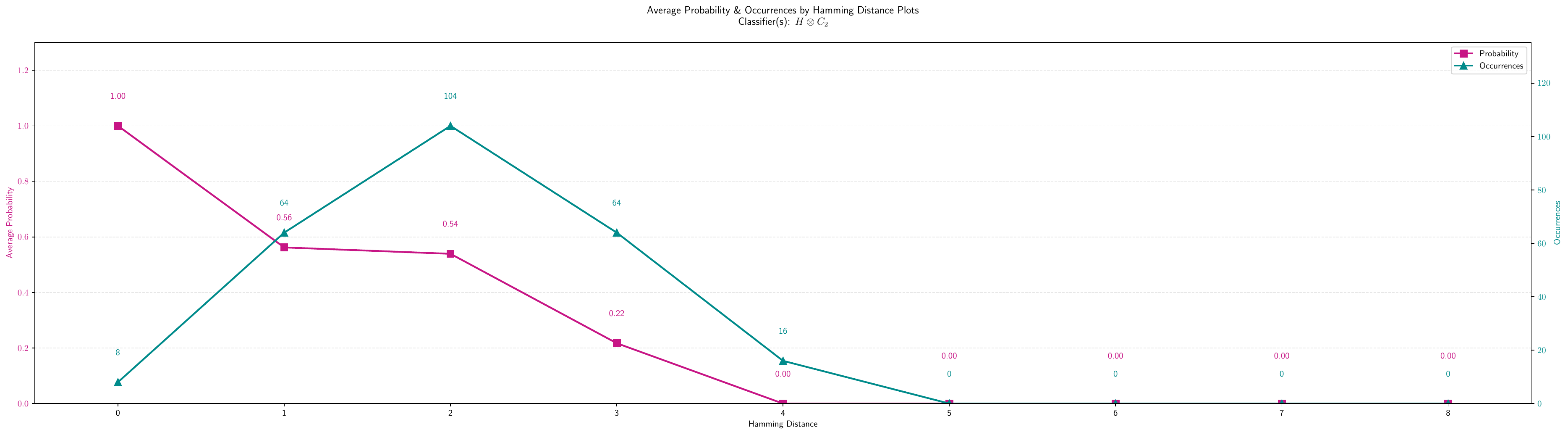}
			\caption{The probability distribution plot for $F_{ B_{ 1 } } \star F_{ Q_{ 2 } }$.}
			\label{fig: Probability Plot For $F_{ B_{ 1 } } F_{ Q_{ 2 } }$}
		\end{subfigure}
		\caption{The probability distribution for $F_{ B_{ 1 } } \star F_{ Q_{ 2 } }$ as a histogram and as a plot.}
		\label{fig: Probability Histogram & Plot For $F_{ B_{ 1 } } F_{ Q_{ 2 } }$}
\end{figure}

We continue with the experimental results for the pattern bases with bit vectors of length $16$ contained in Table \ref{tbl: Pattern Bases Of Length $16$}. Although, for bit vectors of length $8$ the results are virtually identical, this time we observe a very interesting variation. This confirms our intuition that the characteristic traits of the specific pattern vectors comprising the basis affect in a most interesting way the probability distribution. The conclusion for this scenario is summarized in the next text box.

\begin{tcolorbox}
	[
		enhanced,
		breakable,
		grow to left by = 0.000 cm,
		grow to right by = 0.000 cm,
		colback = white,
		enhanced jigsaw,			
		frame hidden,
		sharp corners,
	]
	\begin{table}[H]
		\caption{This table contains the pattern bases that contain pattern bit vectors of length $16$.}
		\label{tbl: Pattern Bases Of Length $16$}
		\centering
		\SetTblrInner { rowsep = 1.200 mm }
		\begin{tblr}
			{
				colspec =
				{
					Q [ c, m, 2.000 cm ]
					| [ 0.750 pt, WordAquaDarker25 ]
					| [ 0.750 pt, WordAquaDarker25 ]
					Q [ c, m, 3.000 cm ]
					| [ 0.500 pt, WordAquaDarker25 ]
					Q [ c, m, 3.600 cm ]
					| [ 0.500 pt, WordAquaDarker25 ]
					Q [ c, m, 3.000 cm ]
				},
				rowspec =
				{
					| [ 3.500 pt, WordAquaDarker25 ]
					| [ 0.750 pt, WordAquaDarker25 ]
					| [ 0.250 pt, white ]
					Q
					| [ 0.150 pt, WordAquaDarker25 ]
					Q
					| [ 0.150 pt, WordAquaDarker25 ]
					Q
					| [ 0.150 pt, WordAquaDarker25 ]
					Q
					| [ 0.150 pt, WordAquaDarker25 ]
					Q
					| [ 0.150 pt, WordAquaDarker25 ]
					Q
					| [ 3.500 pt, WordAquaDarker50 ]
				}
			}
			\SetCell { bg = WordAquaDarker50, fg = white, font = \bfseries \small } Pattern Bit Vector Length
			&
			\SetCell { bg = WordAquaLighter40, fg = black, font = \bfseries \small } Pattern Basis
			&
			\SetCell { bg = WordAquaLighter40, fg = black, font = \bfseries \small } Class of Classifiable Boolean Functions
			&
			\SetCell { bg = WordAquaLighter40, fg = black, font = \bfseries \small } Quantum Classifier
			\\
			\SetCell { bg = WordAquaLighter40, fg = black, font = \bfseries \small } $16$
			&
			\SetCell { font = \small } $B_{ 1 } \odot B_{ 1 } \odot B_{ 1 } \odot B_{ 1 }$
			&
			\SetCell { font = \small } $F_{ B_{ 1 } } \star F_{ B_{ 1 } } \star F_{ B_{ 1 } } \star F_{ B_{ 1 } }$
			&
			\SetCell { font = \small } $H \otimes H \otimes H \otimes H$
			\\
			\SetCell { bg = WordAquaLighter40, fg = black, font = \bfseries \small } $16$
			&
			\SetCell { font = \small } $B_{ 1 } \odot B_{ 1 } \odot Q_{ 2 }$
			&
			\SetCell { font = \small } $F_{ B_{ 1 } } \star F_{ B_{ 1 } } \star F_{ Q_{ 2 } }$
			&
			\SetCell { font = \small } $H \otimes H \otimes C_{ 2 }$
			\\
			\SetCell { bg = WordAquaLighter40, fg = black, font = \bfseries \small } $16$
			&
			\SetCell { font = \small } $B_{ 1 } \odot Q_{ 2 } \odot B_{ 1 }$
			&
			\SetCell { font = \small } $F_{ B_{ 1 } } \star F_{ Q_{ 2 } } \star F_{ B_{ 1 } }$
			&
			\SetCell { font = \small } $H \otimes C_{ 2 } \otimes H$
			\\
			\SetCell { bg = WordAquaLighter40, fg = black, font = \bfseries \small } $16$
			&
			\SetCell { font = \small } $Q_{ 2 } \odot B_{ 1 } \odot B_{ 1 }$
			&
			\SetCell { font = \small } $F_{ Q_{ 2 } } \star F_{ B_{ 1 } } \star F_{ B_{ 1 } }$
			&
			\SetCell { font = \small } $C_{ 2 } \otimes H \otimes H$
			\\
			\SetCell { bg = WordAquaLighter40, fg = black, font = \bfseries \small } $16$
			&
			\SetCell { font = \small } $Q_{ 2 } \odot Q_{ 2 }$
			&
			\SetCell { font = \small } $F_{ Q_{ 2 } } \star F_{ Q_{ 2 } }$
			&
			\SetCell { font = \small } $C_{ 2 } \otimes C_{ 2 }$
			\\
		\end{tblr}
	\end{table}
\end{tcolorbox}

\begin{tcolorbox}
	[
		enhanced,
		breakable,
		center title,
		fonttitle = \bfseries,
		colbacktitle = RedPurple,
		coltitle = white,
		title = \textbf{Conclusion for the pattern bit vectors of length $\mathbf{ 16 }$},
		grow to left by = 0.000 cm,
		grow to right by = 0.000 cm,
		colframe = RedPurple,
		colback = white,
		enhanced jigsaw,			
		sharp corners,
		boxrule = 0.500 pt,
	]
	\begin{itemize}
		\item	
		The probability distribution of the classification threshold $\vartheta$ is virtually identical for the function classes $F_{ B_{ 1 } } \star F_{ B_{ 1 } } \star F_{ B_{ 1 } } \star F_{ B_{ 1 } }$, $F_{ B_{ 1 } } \star F_{ B_{ 1 } } \star F_{ Q_{ 2 } }$, $F_{ B_{ 1 } } \star F_{ Q_{ 2 } } \star F_{ B_{ 1 } }$, and $F_{ Q_{ 2 } } \star F_{ B_{ 1 } } \star F_{ B_{ 1 } }$ and closely mirrors the shape encountered in the case of bit vectors of length $8$.
		\item	
		For the class $F_{ Q_{ 2 } } \star F_{ Q_{ 2 } }$ the probability distribution of the classification threshold $\vartheta$ exhibits a different probability distribution with nonzero probability values for Hamming distances $9$ and $10$.
	\end{itemize}
\end{tcolorbox}

Figures \ref{fig: Probability Histogram For $F_{ B_{ 1 } } F_{ Q_{ 2 } } F_{ B_{ 1 } }$} and \ref{fig: Probability Plot For $F_{ B_{ 1 } } F_{ Q_{ 2 } } F_{ B_{ 1 } }$}, display the results for the class $F_{ B_{ 1 } } \star F_{ Q_{ 2 } } \star F_{ B_{ 1 } }$, as representative of the first category above, while Figures \ref{fig: Probability Histogram For $F_{ Q_{ 2 } } F_{ Q_{ 2 } }$} and \ref{fig: Probability Plot For $F_{ Q_{ 2 } } F_{ Q_{ 2 } }$}, display the results for the class $F_{ Q_{ 2 } } \star F_{ Q_{ 2 } }$. According to our convention, in all Figures the red bars and the red line display probability values, whereas the green bars and the green line show the number of Boolean functions per distance. The results are contained in Table \ref{tbl: Probability vs. Hamming Distance Table For Pattern Bit Vectors of Length $16$}. The class $F_{ Q_{ 2 } } \star F_{ Q_{ 2 } }$ exhibits a distinctly different classification threshold compared to the other four classes $F_{ B_{ 1 } } \star F_{ B_{ 1 } } \star F_{ B_{ 1 } }$, $F_{ B_{ 1 } } \star F_{ Q_{ 2 } }$, and $F_{ Q_{ 2 } } \star F_{ B_{ 1 } }$ because of the nonzero values at distances $9$ and $10$. It is worth noting that for distance $10$, in particular, the classification threshold becomes $1.00$. The reason for this behavior will be explained in subsection \ref{subsec: Interpretation Of The Experimental Results}.

\begin{tcolorbox}
	[
		enhanced,
		breakable,
		grow to left by = 0.000 cm,
		grow to right by = 0.000 cm,
		colback = white,
		enhanced jigsaw,			
		frame hidden,
		sharp corners,
	]
	\begin{table}[H]
		\caption{The probability distribution of the classification threshold as a function of the Hamming distance for the classes $F_{ B_{ 1 } } \star F_{ B_{ 1 } } \star F_{ B_{ 1 } } \star F_{ B_{ 1 } }$, $F_{ B_{ 1 } } \star F_{ B_{ 1 } } \star F_{ Q_{ 2 } }$, $F_{ B_{ 1 } } \star F_{ Q_{ 2 } } \star F_{ B_{ 1 } }$, $F_{ Q_{ 2 } } \star F_{ B_{ 1 } } \star F_{ B_{ 1 } }$, and $F_{ Q_{ 2 } } \star F_{ Q_{ 2 } }$ in tabular form.}
		\label{tbl: Probability vs. Hamming Distance Table For Pattern Bit Vectors of Length $16$}
		\centering
		\SetTblrInner { rowsep = 1.200 mm }
		\begin{tblr}
			{
				colspec =
				{
					Q [ c, m, 3.000 cm ]
					| [ 0.750 pt, GreenLighter2!50 ]
					| [ 0.750 pt, GreenLighter2!50 ]
					Q [ c, m, 5.000 cm ]
					| [ 0.750 pt, GreenLighter2!50 ]
					Q [ c, m, 3.000 cm ]
				},
				rowspec =
				{
					| [ 3.500 pt, GreenLighter2!75 ]
					| [ 0.750 pt, GreenLighter2!75 ]
					| [ 0.250 pt, white ]
					Q
					| [ 0.250 pt, white ]
					Q
					| [ 0.150 pt, GreenLighter2 ]
					Q
					| [ 0.150 pt, GreenLighter2 ]
					Q
					| [ 0.150 pt, GreenLighter2 ]
					Q
					| [ 0.150 pt, GreenLighter2 ]
					Q
					| [ 0.150 pt, GreenLighter2 ]
					Q
					| [ 0.150 pt, GreenLighter2 ]
					Q
					| [ 0.150 pt, GreenLighter2 ]
					Q
					| [ 0.150 pt, GreenLighter2 ]
					Q
					| [ 0.150 pt, GreenLighter2 ]
					Q
					| [ 0.150 pt, GreenLighter2 ]
					Q
					| [ 0.150 pt, GreenLighter2 ]
					Q
					| [ 3.500 pt, GreenLighter2 ]
				}
			}
			&
			\SetCell [ c = 2 ] { bg = GreenLighter2, fg = white, font = \small \bfseries } Classification Threshold $\boldsymbol{ \vartheta }$
			\\
			\SetCell { bg = GreenLighter2, fg = white, font = \small \bfseries }
			Hamming Distance
			&
			\SetCell { bg = GreenLighter2, fg = white, font = \small \bfseries } $F_{ B_{ 1 } } \star F_{ B_{ 1 } } \star F_{ B_{ 1 } } \star F_{ B_{ 1 } }$, $F_{ B_{ 1 } } \star F_{ B_{ 1 } } \star F_{ Q_{ 2 } }$, $F_{ B_{ 1 } } \star F_{ Q_{ 2 } } \star F_{ B_{ 1 } }$, and $F_{ Q_{ 2 } } \star F_{ B_{ 1 } } \star F_{ B_{ 1 } }$
			&
			\SetCell { bg = black!50!WordBlueVeryLight, fg = white, font = \small \bfseries } $F_{ Q_{ 2 } } \star F_{ Q_{ 2 } }$
			\\
			$1$
			&
			$0.76$
			&
			$0.77$
			\\
			$2$
			&
			$0.56$
			&
			$0.56$
			\\
			$3$
			&
			$0.39$
			&
			$0.39$
			\\
			$4$
			&
			$0.34$
			&
			$0.34$
			\\
			$5$
			&
			$0.36$
			&
			$0.36$
			\\
			$6$
			&
			$0.27$
			&
			$0.27$
			\\
			$7$
			&
			$0.10$
			&
			$0.09$
			\\
			$8$
			&
			$0.00$
			&
			$0.00$
			\\
			$9$
			&
			$0.00$
			&
			\SetCell { bg = black!50!WordBlueVeryLight, fg = white, font = \small } $0.16$
			\\
			$10$
			&
			$0.00$
			&
			\SetCell { bg = black!50!WordBlueVeryLight, fg = white, font = \small } $1.00$
			\\
			$11, 12, \dots, 16$
			&
			$0.00$
			&
			$0.00$
			\\
		\end{tblr}
	\end{table}
\end{tcolorbox}

\begin{figure}[H]
		\centering
		\begin{subfigure} { 0.440 \textwidth }
			\centering
			\includegraphics [ angle = 90, width = 0.950 \textwidth, trim = {0.000cm 0.250cm 0.000cm 0.250cm}, clip ] {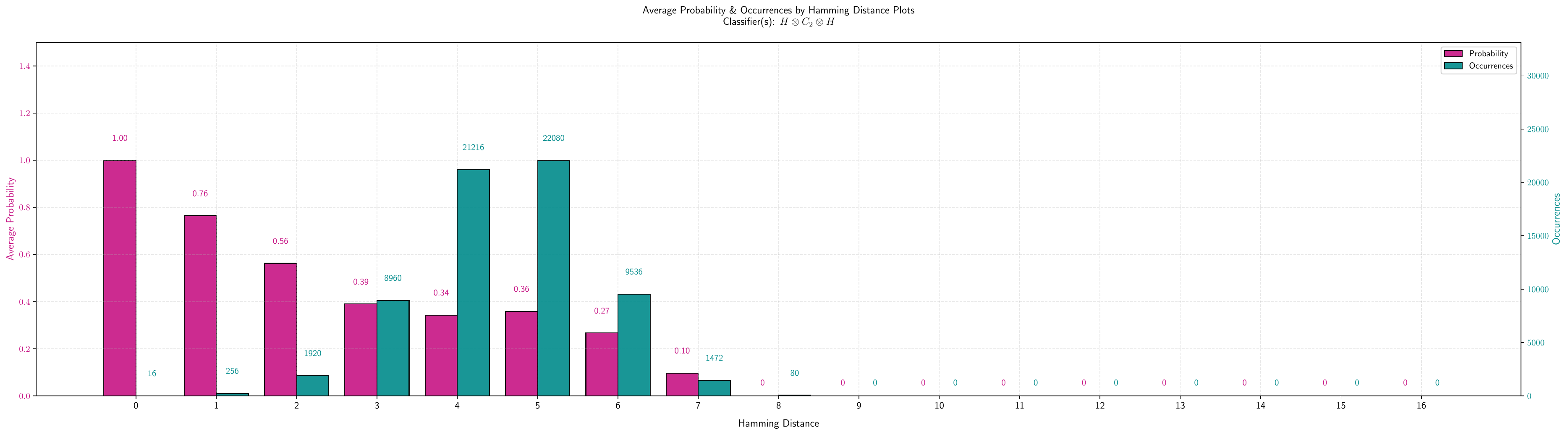}
			\caption{The probability histogram for $F_{ B_{ 1 } } \star F_{ Q_{ 2 } } \star F_{ B_{ 1 } }$.}
			\label{fig: Probability Histogram For $F_{ B_{ 1 } } F_{ Q_{ 2 } } F_{ B_{ 1 } }$}
		\end{subfigure}
		\hfill
		\begin{subfigure} { 0.440 \textwidth }
			\centering
			\includegraphics [ angle = 90, width = 0.950 \textwidth, trim = {0.000cm 0.250cm 0.000cm 0.250cm}, clip ] {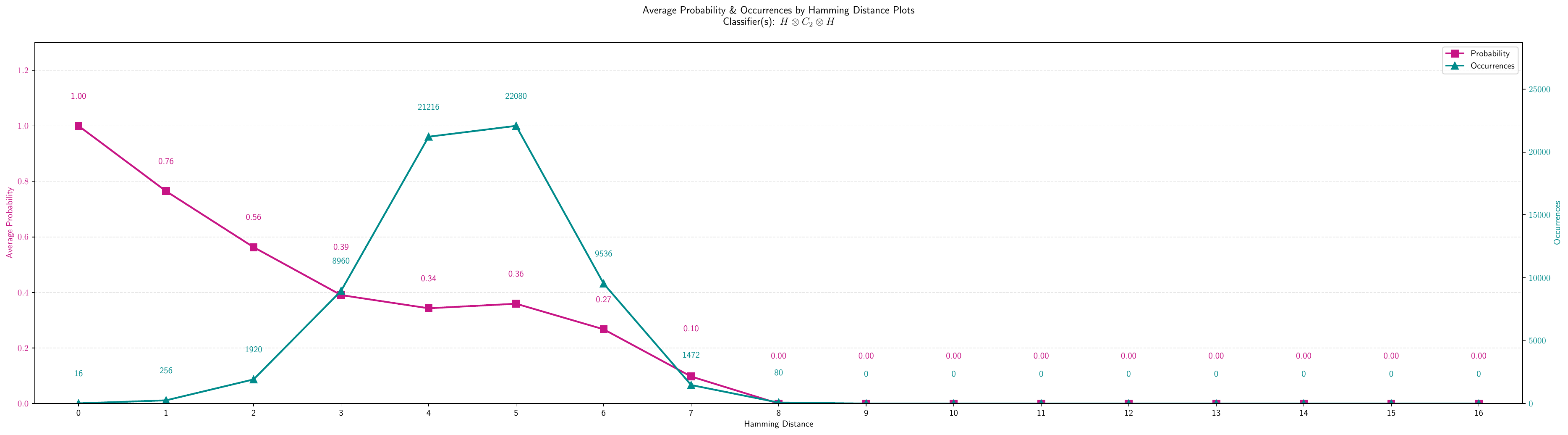}
			\caption{The probability plot for $F_{ B_{ 1 } } \star F_{ Q_{ 2 } } \star F_{ B_{ 1 } }$.}
			\label{fig: Probability Plot For $F_{ B_{ 1 } } F_{ Q_{ 2 } } F_{ B_{ 1 } }$}
		\end{subfigure}
		\caption{The probability distribution for $F_{ B_{ 1 } } \star F_{ Q_{ 2 } } \star F_{ B_{ 1 } }$ as a histogram and as a plot.}
		\label{fig: Probability Histogram & Plot For $F_{ B_{ 1 } } F_{ Q_{ 2 } } F_{ B_{ 1 } }$}
\end{figure}

\begin{figure}[H]
		\centering
		\begin{subfigure} { 0.440 \textwidth }
			\centering
			\includegraphics [ angle = 90, width = 0.950 \textwidth, trim = {0.000cm 0.250cm 0.000cm 0.250cm}, clip ] {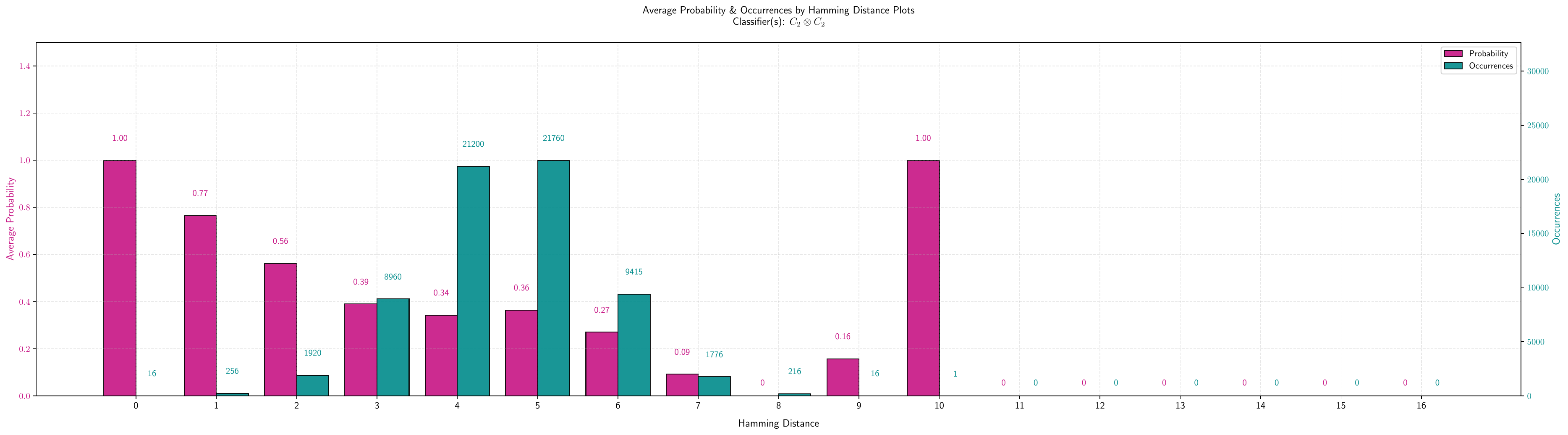}
			\caption{The probability histogram for $F_{ Q_{ 2 } } \star F_{ Q_{ 2 } }$.}
			\label{fig: Probability Histogram For $F_{ Q_{ 2 } } F_{ Q_{ 2 } }$}
		\end{subfigure}
		\hfill
		\begin{subfigure} { 0.440 \textwidth }
			\centering
			\includegraphics [ angle = 90, width = 0.950 \textwidth, trim = {0.000cm 0.250cm 0.000cm 0.250cm}, clip ] {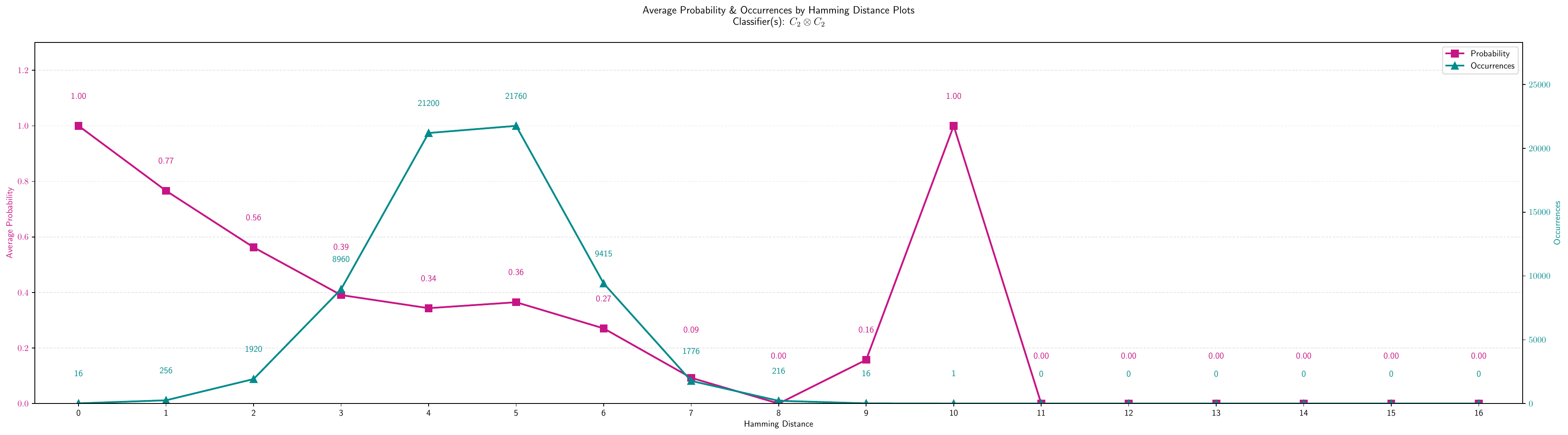}
			\caption{The probability distribution plot for $F_{ Q_{ 2 } } \star F_{ Q_{ 2 } }$.}
			\label{fig: Probability Plot For $F_{ Q_{ 2 } } F_{ Q_{ 2 } }$}
		\end{subfigure}
		\caption{The probability distribution for $F_{ Q_{ 2 } } \star F_{ Q_{ 2 } }$ as a histogram and as a plot.}
		\label{fig: Probability Histogram & Plot For $F_{ Q_{ 2 } } F_{ Q_{ 2 } }$}
\end{figure}
\subsection{The case of sampling Boolean functions} \label{subsec: The Case Of Sampling Boolean Functions}

This case investigates the behavior of sampled pattern bit vectors of length $32$ and $64$. For such bit vectors, exhaustive enumeration becomes inefficient and impractical due to the exponential growth in complexity. Thus, we employed sampling by randomly generating a set of pattern bit vectors. For each generated vector, we constructed the corresponding Boolean function and oracle in order to classify via the Pattern Basis Classification Scheme outlined in Definition \ref{def: The Pattern Basis Classification Scheme}. This random sampling approach allowed us to generalize our findings regarding Hamming distance usability. As before, the elementary pattern bases $B_{ 1 }$ and $Q_{ 2 }$, defined by equations \eqref{eq: Pattern Basis $B_{ 1 }$} and \eqref{eq: Pattern Basis $Q_{ 2 }$} respectively, are the fundamental building blocks for the construction of all possible products of pattern bases that comprise bit vectors of length $32$ and $64$. Corollary \ref{crl: Correspondence Among Bases, Function Classes & Classifiers} elucidates the one-to-one exact correspondence among the products of bases, function classes, and quantum classifiers (see also Figure \ref{fig: One-to-one Correspondence Among Products}), leading to Tables \ref{tbl: Pattern Bases Of Length $32$} and \ref{tbl: Pattern Bases Of Length $64$}.

\begin{tcolorbox}
	[
		enhanced,
		breakable,
		grow to left by = 0.500 cm,
		grow to right by = 0.000 cm,
		colback = white,
		enhanced jigsaw,			
		frame hidden,
		sharp corners,
	]
	\begin{table}[H]
		\caption{This table contains the pattern bases that contain pattern bit vectors of length $32$.}
		\label{tbl: Pattern Bases Of Length $32$}
		\centering
		\SetTblrInner { rowsep = 1.200 mm }
		\begin{tblr}
			{
				colspec =
				{
					Q [ c, m, 2.000 cm ]
					| [ 0.750 pt, WordAquaDarker25 ]
					| [ 0.750 pt, WordAquaDarker25 ]
					Q [ c, m, 3.000 cm ]
					| [ 0.500 pt, WordAquaDarker25 ]
					Q [ c, m, 3.600 cm ]
					| [ 0.500 pt, WordAquaDarker25 ]
					Q [ c, m, 3.000 cm ]
				},
				rowspec =
				{
					| [ 3.500 pt, WordAquaDarker25 ]
					| [ 0.750 pt, WordAquaDarker25 ]
					| [ 0.250 pt, white ]
					Q
					| [ 0.150 pt, WordAquaDarker25 ]
					Q
					| [ 0.150 pt, WordAquaDarker25 ]
					Q
					| [ 0.150 pt, WordAquaDarker25 ]
					Q
					| [ 0.150 pt, WordAquaDarker25 ]
					Q
					| [ 0.150 pt, WordAquaDarker25 ]
					Q
					| [ 0.150 pt, WordAquaDarker25 ]
					Q
					| [ 0.150 pt, WordAquaDarker25 ]
					Q
					| [ 0.150 pt, WordAquaDarker25 ]
					Q
					| [ 3.500 pt, WordAquaDarker50 ]
				}
			}
			\SetCell { bg = WordAquaDarker50, fg = white, font = \bfseries \small } Pattern Bit Vector Length
			&
			\SetCell { bg = WordAquaLighter40, fg = black, font = \bfseries \small } Pattern Basis
			&
			\SetCell { bg = WordAquaLighter40, fg = black, font = \bfseries \small } Class of Classifiable Boolean Functions
			&
			\SetCell { bg = WordAquaLighter40, fg = black, font = \bfseries \small } Quantum Classifier
			\\
			\SetCell { bg = WordAquaLighter40, fg = black, font = \bfseries \small } $32$
			&
			\SetCell { font = \small } $B_{ 1 } \odot B_{ 1 } \odot B_{ 1 } \odot B_{ 1 } \odot B_{ 1 }$
			&
			\SetCell { font = \small } $F_{ B_{ 1 } } \star F_{ B_{ 1 } } \star F_{ B_{ 1 } } \star F_{ B_{ 1 } } \star F_{ B_{ 1 } }$
			&
			\SetCell { font = \small } $H \otimes H \otimes H \otimes H \otimes H$
			\\
			\SetCell { bg = WordAquaLighter40, fg = black, font = \bfseries \small } $32$
			&
			\SetCell { font = \small } $B_{ 1 } \odot B_{ 1 } \odot B_{ 1 } \odot Q_{ 2 }$
			&
			\SetCell { font = \small } $F_{ B_{ 1 } } \star F_{ B_{ 1 } } \star F_{ B_{ 1 } } \star F_{ Q_{ 2 } }$
			&
			\SetCell { font = \small } $H \otimes H \otimes H \otimes C_{ 2 }$
			\\
			\SetCell { bg = WordAquaLighter40, fg = black, font = \bfseries \small } $32$
			&
			\SetCell { font = \small } $B_{ 1 } \odot B_{ 1 } \odot Q_{ 2 } \odot B_{ 1 }$
			&
			\SetCell { font = \small } $F_{ B_{ 1 } } \star F_{ B_{ 1 } } \star F_{ Q_{ 2 } } \star F_{ B_{ 1 } }$
			&
			\SetCell { font = \small } $H \otimes H \otimes C_{ 2 } \otimes H$
			\\
			\SetCell { bg = WordAquaLighter40, fg = black, font = \bfseries \small } $32$
			&
			\SetCell { font = \small } $B_{ 1 } \odot Q_{ 2 } \odot B_{ 1 } \odot B_{ 1 }$
			&
			\SetCell { font = \small } $F_{ B_{ 1 } } \star F_{ Q_{ 2 } } \star F_{ B_{ 1 } } \star F_{ B_{ 1 } }$
			&
			\SetCell { font = \small } $H \otimes C_{ 2 } \otimes H \otimes H$
			\\
			\SetCell { bg = WordAquaLighter40, fg = black, font = \bfseries \small } $32$
			&
			\SetCell { font = \small } $Q_{ 2 } \odot B_{ 1 } \odot B_{ 1 } \odot B_{ 1 }$
			&
			\SetCell { font = \small } $F_{ Q_{ 2 } } \star F_{ B_{ 1 } } \star F_{ B_{ 1 } } \star F_{ B_{ 1 } }$
			&
			\SetCell { font = \small } $C_{ 2 } \otimes H \otimes H \otimes H$
			\\
			\SetCell { bg = WordAquaLighter40, fg = black, font = \bfseries \small } $32$
			&
			\SetCell { font = \small } $B_{ 1 } \odot Q_{ 2 } \odot Q_{ 2 }$
			&
			\SetCell { font = \small } $F_{ B_{ 1 } } \star F_{ Q_{ 2 } } \star F_{ Q_{ 2 } }$
			&
			\SetCell { font = \small } $H \otimes C_{ 2 } \otimes C_{ 2 }$
			\\
			\SetCell { bg = WordAquaLighter40, fg = black, font = \bfseries \small } $32$
			&
			\SetCell { font = \small } $Q_{ 2 } \odot B_{ 1 } \odot Q_{ 2 }$
			&
			\SetCell { font = \small } $F_{ Q_{ 2 } } \star F_{ B_{ 1 } } \star F_{ Q_{ 2 } }$
			&
			\SetCell { font = \small } $C_{ 2 } \otimes H \otimes C_{ 2 }$
			\\
			\SetCell { bg = WordAquaLighter40, fg = black, font = \bfseries \small } $32$
			&
			\SetCell { font = \small } $Q_{ 2 } \odot Q_{ 2 } \odot B_{ 1 }$
			&
			\SetCell { font = \small } $F_{ Q_{ 2 } } \star F_{ Q_{ 2 } } \star F_{ B_{ 1 } }$
			&
			\SetCell { font = \small } $C_{ 2 } \otimes C_{ 2 } \otimes H$
			\\
		\end{tblr}
	\end{table}
\end{tcolorbox}

The experimental results corroborate that all the above function classes have a virtually identical probability distribution of the classification threshold. Hence, this series of experiments confirms the pattern already observed, i.e., that the classification threshold is a decreasing function of the Hamming distance.

\begin{tcolorbox}
	[
		enhanced,
		breakable,
		center title,
		fonttitle = \bfseries,
		colbacktitle = RedPurple,
		coltitle = white,
		title = \textbf{Conclusion for the pattern bit vectors of length $\mathbf{ 32 }$},
		grow to left by = 0.000 cm,
		grow to right by = 0.000 cm,
		colframe = RedPurple,
		colback = white,
		enhanced jigsaw,			
		sharp corners,
		boxrule = 0.500 pt,
	]
	The probability distribution for the classification threshold is the same for all the function classes $F_{ B_{ 1 } } \star F_{ B_{ 1 } } \star F_{ B_{ 1 } } \star F_{ B_{ 1 } } \star F_{ B_{ 1 } }$, $F_{ B_{ 1 } } \star F_{ B_{ 1 } } \star F_{ B_{ 1 } } \star F_{ Q_{ 2 } }$, $F_{ B_{ 1 } } \star F_{ B_{ 1 } } \star F_{ Q_{ 2 } } \star F_{ B_{ 1 } }$, $F_{ B_{ 1 } } \star F_{ Q_{ 2 } } \star F_{ B_{ 1 } } \star F_{ B_{ 1 } }$, $F_{ Q_{ 2 } } \star F_{ B_{ 1 } } \star F_{ B_{ 1 } } \star F_{ B_{ 1 } }$, $F_{ B_{ 1 } } \star F_{ Q_{ 2 } } \star F_{ Q_{ 2 } }$, $F_{ Q_{ 2 } } \star F_{ B_{ 1 } } \star F_{ Q_{ 2 } }$, and $F_{ Q_{ 2 } } \star F_{ Q_{ 2 } } \star F_{ B_{ 1 } }$.
\end{tcolorbox}

For this reason we illustrate only the results for the class $F_{ Q_{ 2 } } \star F_{ Q_{ 2 } } \star F_{ B_{ 1 } }$ in Figures \ref{fig: Probability Histogram For $F_{ Q_{ 2 } } F_{ Q_{ 2 } } F_{ B_{ 1 } }$} and \ref{fig: Probability Plot For $F_{ Q_{ 2 } } F_{ Q_{ 2 } } F_{ B_{ 1 } }$}. For enhanced readability, these results are repeated in tabular form in Table \ref{tbl: Probability vs. Hamming Distance Table For Pattern Bit Vectors of Length $32$}.

\begin{figure}[H]
		\centering
		\begin{subfigure} { 0.440 \textwidth }
			\centering
			\includegraphics [ angle = 90, width = 0.950 \textwidth, trim = {0.000cm 0.250cm 0.000cm 0.250cm}, clip ] {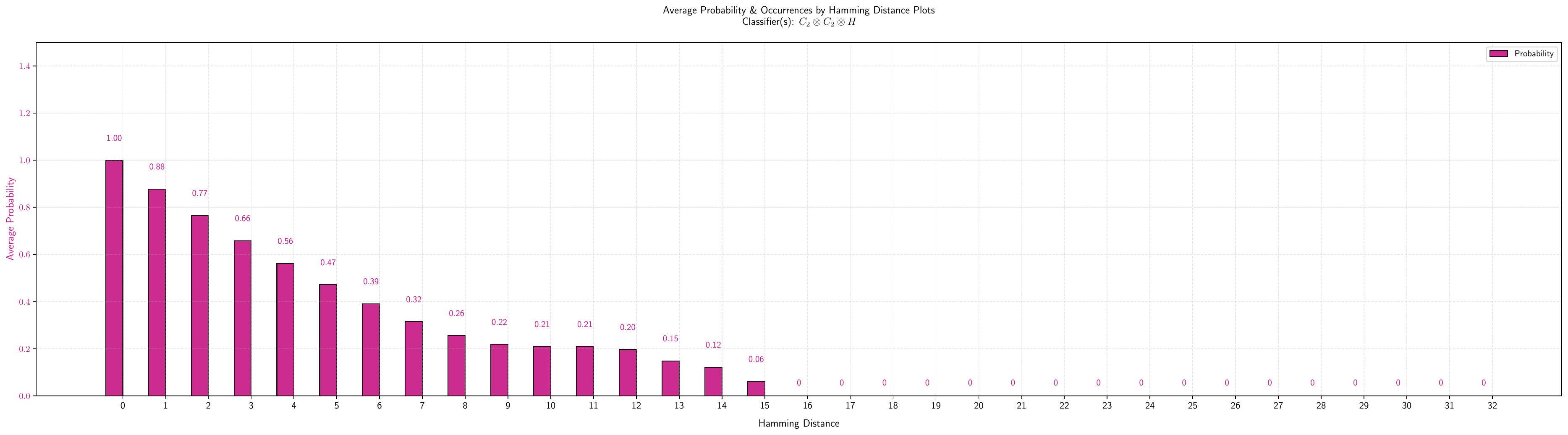}
			\caption{The probability histogram for $F_{ Q_{ 2 } } \star F_{ Q_{ 2 } } \star F_{ B_{ 1 } }$.}
			\label{fig: Probability Histogram For $F_{ Q_{ 2 } } F_{ Q_{ 2 } } F_{ B_{ 1 } }$}
		\end{subfigure}
		\hfill
		\begin{subfigure} { 0.440 \textwidth }
			\centering
			\includegraphics [ angle = 90, width = 0.950 \textwidth, trim = {0.000cm 0.250cm 0.000cm 0.250cm}, clip ] {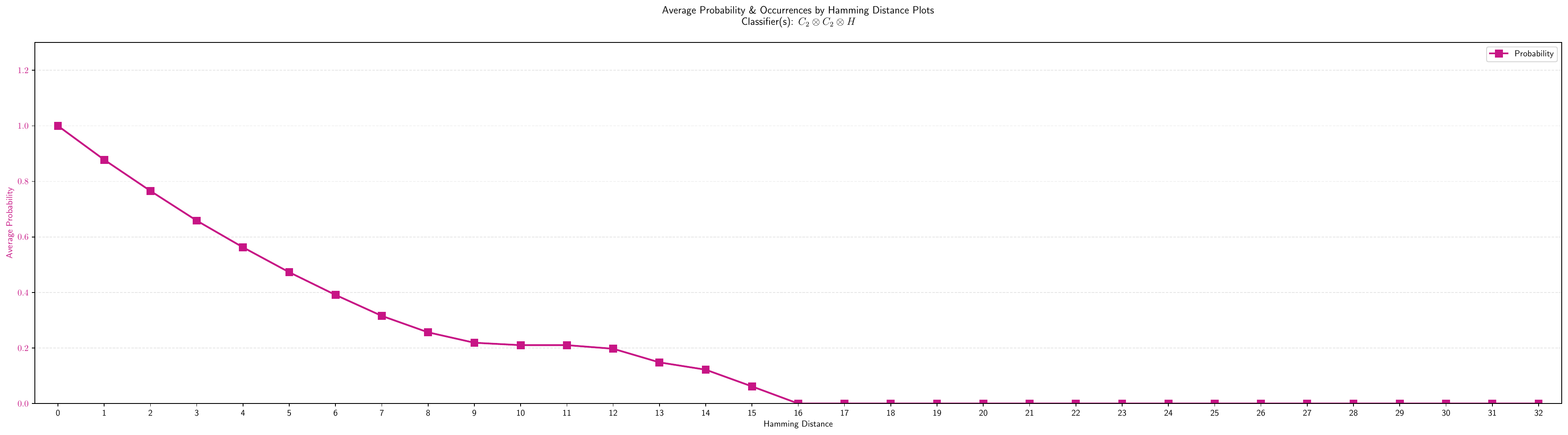}
			\caption{The probability plot for $F_{ Q_{ 2 } } \star F_{ Q_{ 2 } } \star F_{ B_{ 1 } }$.}
			\label{fig: Probability Plot For $F_{ Q_{ 2 } } F_{ Q_{ 2 } } F_{ B_{ 1 } }$}
		\end{subfigure}
		\caption{The probability distribution for $F_{ Q_{ 2 } } \star F_{ Q_{ 2 } } \star F_{ B_{ 1 } }$ as a histogram and as a plot.}
		\label{fig: Probability Histogram & Plot For $F_{ Q_{ 2 } } F_{ Q_{ 2 } } F_{ B_{ 1 } }$}
\end{figure}

\begin{tcolorbox}
	[
		enhanced,
		breakable,
		grow to left by = 0.000 cm,
		grow to right by = 0.000 cm,
		colback = white,
		enhanced jigsaw,			
		frame hidden,
		sharp corners,
	]
	\begin{table}[H]
		\caption{The probability distribution of the classification threshold as a function of the Hamming distance for the classes $F_{ B_{ 1 } } \star F_{ B_{ 1 } } \star F_{ B_{ 1 } } \star F_{ B_{ 1 } } \star F_{ B_{ 1 } }$, $F_{ B_{ 1 } } \star F_{ B_{ 1 } } \star F_{ B_{ 1 } } \star F_{ Q_{ 2 } }$, $F_{ B_{ 1 } } \star F_{ B_{ 1 } } \star F_{ Q_{ 2 } } \star F_{ B_{ 1 } }$, $F_{ B_{ 1 } } \star F_{ Q_{ 2 } } \star F_{ B_{ 1 } } \star F_{ B_{ 1 } }$, $F_{ Q_{ 2 } } \star F_{ B_{ 1 } } \star F_{ B_{ 1 } } \star F_{ B_{ 1 } }$, $F_{ B_{ 1 } } \star F_{ Q_{ 2 } } \star F_{ Q_{ 2 } }$, $F_{ Q_{ 2 } } \star F_{ B_{ 1 } } \star F_{ Q_{ 2 } }$, and $F_{ Q_{ 2 } } \star F_{ Q_{ 2 } } \star F_{ B_{ 1 } }$ in tabular form.}
		\label{tbl: Probability vs. Hamming Distance Table For Pattern Bit Vectors of Length $32$}
		\centering
		\SetTblrInner { rowsep = 1.200 mm }
		\begin{tblr}
			{
				colspec =
				{
					Q [ c, m, 3.000 cm ]
					| [ 0.750 pt, GreenLighter2!50 ]
					| [ 0.750 pt, GreenLighter2!50 ]
					Q [ c, m, 9.000 cm ]
				},
				rowspec =
				{
					| [ 3.500 pt, GreenLighter2!75 ]
					| [ 0.750 pt, GreenLighter2!75 ]
					| [ 0.250 pt, white ]
					Q
					| [ 0.250 pt, white ]
					Q
					| [ 0.150 pt, GreenLighter2 ]
					Q
					| [ 0.150 pt, GreenLighter2 ]
					Q
					| [ 0.150 pt, GreenLighter2 ]
					Q
					| [ 3.500 pt, GreenLighter2 ]
				}
			}
			&
			\SetCell { bg = GreenLighter2, fg = white, font = \small \bfseries } Classification Threshold $\boldsymbol{ \vartheta }$
			\\
			\SetCell { bg = GreenLighter2, fg = white, font = \small \bfseries }
			Hamming Distance
			&
			\SetCell { bg = GreenLighter2, fg = white, font = \small \bfseries } $F_{ B_{ 1 } } \star F_{ B_{ 1 } } \star F_{ B_{ 1 } } \star F_{ B_{ 1 } } \star F_{ B_{ 1 } }$, $F_{ B_{ 1 } } \star F_{ B_{ 1 } } \star F_{ B_{ 1 } } \star F_{ Q_{ 2 } }$, $F_{ B_{ 1 } } \star F_{ B_{ 1 } } \star F_{ Q_{ 2 } } \star F_{ B_{ 1 } }$, $F_{ B_{ 1 } } \star F_{ Q_{ 2 } } \star F_{ B_{ 1 } } \star F_{ B_{ 1 } }$, $F_{ Q_{ 2 } } \star F_{ B_{ 1 } } \star F_{ B_{ 1 } } \star F_{ B_{ 1 } }$, $F_{ B_{ 1 } } \star F_{ Q_{ 2 } } \star F_{ Q_{ 2 } }$, $F_{ Q_{ 2 } } \star F_{ B_{ 1 } } \star F_{ Q_{ 2 } }$, and $F_{ Q_{ 2 } } \star F_{ Q_{ 2 } } \star F_{ B_{ 1 } }$
			\\
			$1, \dots, 4$
			&
			$0.50 < \vartheta < 1.00$
			\\
			$5, \dots, 15$
			&
			$0.00 < \vartheta < 0.50$
			\\
			$16, \dots, 32$
			&
			$0.00$
			\\
		\end{tblr}
	\end{table}
\end{tcolorbox}

The fact that the classification of random functions with respect to their nearest neighbors in all the above function classes follows the same pattern, is a strong indication of the robustness of our results. A small deviation from this dominating behavior is observed when the function class is $F_{ Q_{ 2 } } \star F_{ Q_{ 2 } }$. To test whether this is a statistical artifact or something inherent in classes that are obtained as the external product of the same factor $F_{ Q_{ 2 } }$, we have experimented with the class $F_{ Q_{ 2 } } \star F_{ Q_{ 2 } } \star F_{ Q_{ 2 } }$, as shown in Table \ref{tbl: Pattern Bases Of Length $64$}. The results for the class $F_{ Q_{ 2 } } \star F_{ Q_{ 2 } } \star F_{ Q_{ 2 } }$ are presented in Figures \ref{fig: Probability Histogram For $F_{ Q_{ 2 } } F_{ Q_{ 2 } } F_{ Q_{ 2 } }$} and \ref{fig: Probability Plot For $F_{ Q_{ 2 } } F_{ Q_{ 2 } } F_{ Q_{ 2 } }$}. The conclusion we have arrived at is given below. For clarity, the results are repeated in tabular form in Table \ref{tbl: Probability vs. Hamming Distance Table For Pattern Bit Vectors of Length $64$}.

\begin{tcolorbox}
	[
		enhanced,
		breakable,
		grow to left by = 0.500 cm,
		grow to right by = 0.000 cm,
		colback = white,
		enhanced jigsaw,			
		frame hidden,
		sharp corners,
	]
	\begin{table}[H]
		\caption{This table shows the class $F_{ Q_{ 2 } } \star F_{ Q_{ 2 } } \star F_{ Q_{ 2 } }$ that realizes the pattern basis $Q_{ 2 } \odot Q_{ 2 } \odot Q_{ 2 }$ consisting of pattern bit vectors of length $64$.}
		\label{tbl: Pattern Bases Of Length $64$}
		\centering
		\SetTblrInner { rowsep = 1.200 mm }
		\begin{tblr}
			{
				colspec =
				{
					Q [ c, m, 2.000 cm ]
					| [ 0.750 pt, WordAquaDarker25 ]
					| [ 0.750 pt, WordAquaDarker25 ]
					Q [ c, m, 3.000 cm ]
					| [ 0.500 pt, WordAquaDarker25 ]
					Q [ c, m, 3.600 cm ]
					| [ 0.500 pt, WordAquaDarker25 ]
					Q [ c, m, 3.000 cm ]
				},
				rowspec =
				{
					| [ 3.500 pt, WordAquaDarker25 ]
					| [ 0.750 pt, WordAquaDarker25 ]
					| [ 0.250 pt, white ]
					Q
					| [ 0.150 pt, WordAquaDarker25 ]
					Q
					| [ 3.500 pt, WordAquaDarker50 ]
				}
			}
			\SetCell { bg = WordAquaDarker50, fg = white, font = \bfseries \small } Pattern Bit Vector Length
			&
			\SetCell { bg = WordAquaLighter40, fg = black, font = \bfseries \small } Pattern Basis
			&
			\SetCell { bg = WordAquaLighter40, fg = black, font = \bfseries \small } Class of Classifiable Boolean Functions
			&
			\SetCell { bg = WordAquaLighter40, fg = black, font = \bfseries \small } Quantum Classifier
			\\
			\SetCell { bg = WordAquaLighter40, fg = black, font = \bfseries \small } $64$
			&
			\SetCell { font = \small } $Q_{ 2 } \odot Q_{ 2 } \odot Q_{ 2 }$
			&
			\SetCell { font = \small } $F_{ Q_{ 2 } } \star F_{ Q_{ 2 } } \star F_{ Q_{ 2 } }$
			&
			\SetCell { font = \small } $C_{ 2 } \otimes C_{ 2 } \otimes C_{ 2 }$
			\\
		\end{tblr}
	\end{table}
\end{tcolorbox}

\begin{tcolorbox}
	[
		enhanced,
		breakable,
		center title,
		fonttitle = \bfseries,
		colbacktitle = RedPurple,
		coltitle = white,
		title = \textbf{Conclusion for the pattern bit vectors of length $\mathbf{ 8 }$},
		grow to left by = 0.000 cm,
		grow to right by = 0.000 cm,
		colframe = RedPurple,
		colback = white,
		enhanced jigsaw,			
		sharp corners,
		boxrule = 0.500 pt,
	]
	For the most part, the probability distribution of the classification threshold of a random function $h$ with respect to the class $F_{ Q_{ 2 } } \star F_{ Q_{ 2 } } \star F_{ Q_{ 2 } }$ is similar to that of other class, i.e., a decreasing function of the Hamming distance. However, there is a noteworthy difference at Hamming distance $36$, where, instead of being $0.$, it spikes to $1.0$.
\end{tcolorbox}

\begin{figure}[H]
		\centering
		\begin{subfigure} { 0.440 \textwidth }
			\centering
			\includegraphics [ angle = 90, width = 0.950 \textwidth, trim = {0.000cm 0.250cm 0.000cm 0.250cm}, clip ] {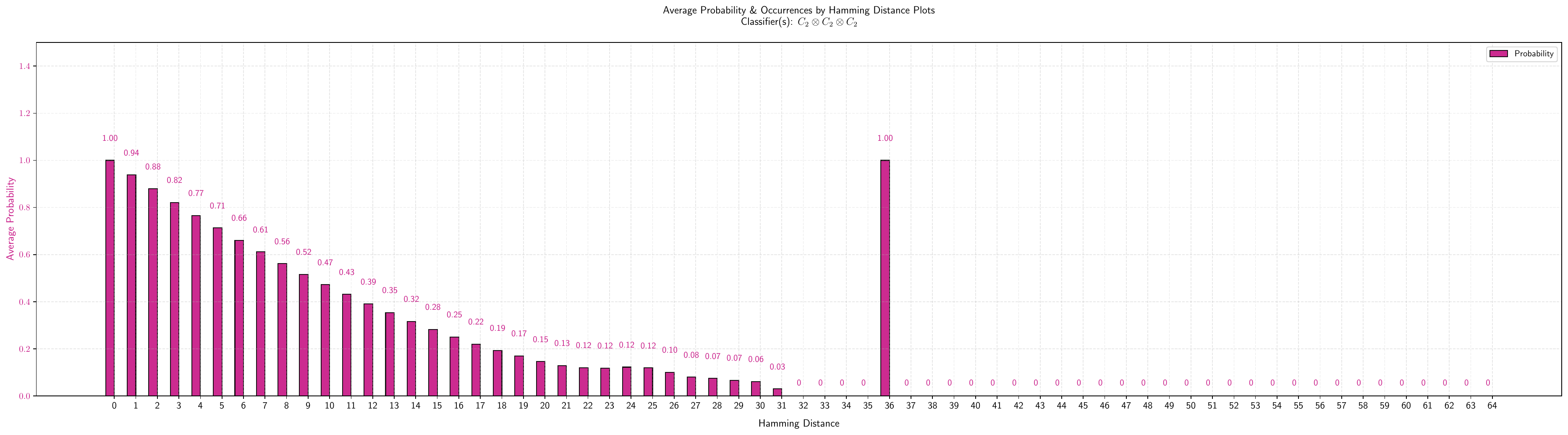}
			\caption{The probability histogram for $F_{ Q_{ 2 } } \star F_{ Q_{ 2 } } \star F_{ Q_{ 2 } }$.}
			\label{fig: Probability Histogram For $F_{ Q_{ 2 } } F_{ Q_{ 2 } } F_{ Q_{ 2 } }$}
		\end{subfigure}
		\hfill
		\begin{subfigure} { 0.440 \textwidth }
			\centering
			\includegraphics [ angle = 90, width = 0.950 \textwidth, trim = {0.000cm 0.250cm 0.000cm 0.250cm}, clip ] {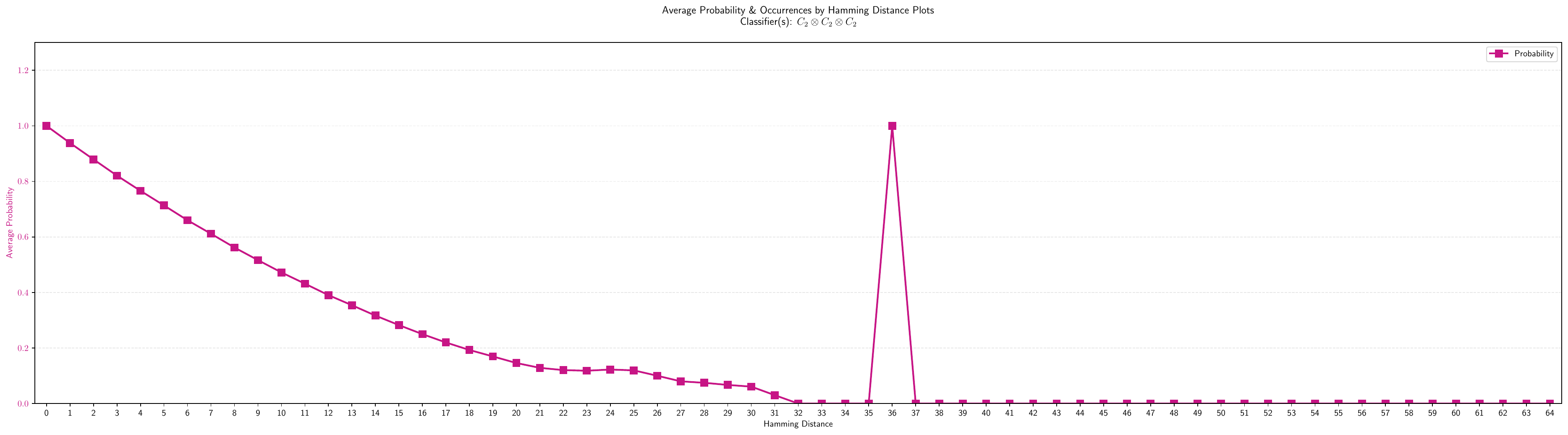}
			\caption{The probability plot for $F_{ Q_{ 2 } } \star F_{ Q_{ 2 } } \star F_{ Q_{ 2 } }$.}
			\label{fig: Probability Plot For $F_{ Q_{ 2 } } F_{ Q_{ 2 } } F_{ Q_{ 2 } }$}
		\end{subfigure}
		\caption{The probability distribution for $F_{ Q_{ 2 } } \star F_{ Q_{ 2 } } \star F_{ Q_{ 2 } }$ as a histogram and as a plot.}
		\label{fig: Probability Histogram & Plot For $F_{ Q_{ 2 } } F_{ Q_{ 2 } } F_{ Q_{ 2 } }$}
\end{figure}

\begin{tcolorbox}
	[
		enhanced,
		breakable,
		grow to left by = 0.000 cm,
		grow to right by = 0.000 cm,
		colback = white,
		enhanced jigsaw,			
		frame hidden,
		sharp corners,
	]
	\begin{table}[H]
		\caption{The probability distribution of the classification threshold as a function of the Hamming distance for the class $F_{ Q_{ 2 } } \star F_{ Q_{ 2 } } \star F_{ Q_{ 2 } }$ in tabular format.}
		\label{tbl: Probability vs. Hamming Distance Table For Pattern Bit Vectors of Length $64$}
		\centering
		\SetTblrInner { rowsep = 1.200 mm }
		\begin{tblr}
			{
				colspec =
				{
					Q [ c, m, 3.000 cm ]
					| [ 0.750 pt, GreenLighter2!50 ]
					| [ 0.750 pt, GreenLighter2!50 ]
					Q [ c, m, 9.000 cm ]
				},
				rowspec =
				{
					| [ 3.500 pt, GreenLighter2!75 ]
					| [ 0.750 pt, GreenLighter2!75 ]
					| [ 0.250 pt, white ]
					Q
					| [ 0.250 pt, white ]
					Q
					| [ 0.150 pt, GreenLighter2 ]
					Q
					| [ 0.150 pt, GreenLighter2 ]
					Q
					| [ 0.150 pt, GreenLighter2 ]
					Q
					| [ 0.150 pt, GreenLighter2 ]
					Q
					| [ 0.150 pt, GreenLighter2 ]
					Q
					| [ 3.500 pt, GreenLighter2 ]
				}
			}
			&
			\SetCell { bg = GreenLighter2, fg = white, font = \small \bfseries } Classification Threshold $\boldsymbol{ \vartheta }$
			\\
			\SetCell { bg = GreenLighter2, fg = white, font = \small \bfseries }
			Hamming Distance
			&
			\SetCell { bg = GreenLighter2, fg = white, font = \small \bfseries } $F_{ Q_{ 2 } } \star F_{ Q_{ 2 } } \star F_{ Q_{ 2 } }$
			\\
			$1, \dots, 9$
			&
			$0.50 < \vartheta < 1.00$
			\\
			$9, \dots, 31$
			&
			$0.00 < \vartheta < 0.50$
			\\
			$32, \dots, 35$
			&
			$0.00$
			\\
			$36$
			&
			$1.00$
			\\
			$37, \dots, 64$
			&
			$0.00$
			\\
		\end{tblr}
	\end{table}
\end{tcolorbox}

In the coming Section, we interpret the results and provide an explanation regarding the probability spike observed when the product function class is constructed entirely by factors of the class $F_{ Q_{ 2 } }$.

\subsection{Interpretation of the experimental results} \label{subsec: Interpretation Of The Experimental Results}

By synthesizing the comprehensive experimental outcomes, a definitive pattern emerges: within an arbitrary function class, the classification threshold manifests as a monotonically decreasing function with respect to the Hamming distance. This trend, while robust across diverse scenarios, is complicated by predictable anomalies in certain specialized configurations, allowing us to articulate a foundational ``rule'' tempered by precisely delineated exceptions. To formalize this insight, we introduce the parameter $n$, representing the arity of the Boolean functions mapping $\mathbb{ B }^{ n } \to \mathbb{ B }$ within a given class $F_{ P }$, alongside the vector length $2^{ n }$ of the pattern bit vectors that constitute the associated pattern basis $P$. This framework enables the specification of subintervals that systematically partition the full spectrum from $0$ to $2^{ n }$ into three regions, as illustrated in Table \ref{tbl: Probability vs. Hamming Distance Table For ArbitraryPattern Bit Vectors of Length $2^{ n }$}. Notably, when the function class is an extended product of the form $F_{ Q_{ 2 } }^{ \otimes m }$, for some $m \geq 1$, a pronounced irregularity arises in the third region, manifesting as a sharp peak where the probability surges to precisely $1.00$—a departure that underscores the nuanced interplay between structural properties and probabilistic outcomes.

This distinctive anomaly can be traced to an intrinsic hallmark of the functions populating $F_{ Q_{ 2 } }^{ \otimes m }$, $m \geq 1$. As established in \cite{Andronikos2025a}, every such function maintains an identical proportion of $1$ values across the $2^{ n }$ possible outputs, or equivalently, exhibits a fixed count of both $0$ and $1$ values. Let $\varrho$ denote the count of $0$ values; a straightforward computation reveals that for $F_{ Q_{ 2 } }$, $F_{ Q_{ 2 } }^{ \otimes 2 }$, and $F_{ Q_{ 2 } }^{ \otimes 3 }$, this quantity assumes values of $3$, $10$, and $36$, respectively. The general formula for this ratio in these special classes is derived in \cite{Andronikos2025a}. By stark contrast, the remaining classes investigated in this study encompass functions with heterogeneous distributions of $0$s and $1$s, introducing variability that dilutes such uniform behaviors. Therefore, for the family of classes $F_{ Q_{ 2 } }^{ \otimes m }$, $m \geq 1$, the constant Boolean function $h \colon \mathbb{ B }^{ n } \to \mathbb{ B }$ defined by $h ( \mathbf{ x } ) = 1$ for all $\mathbf{ x } \in \mathbb{ B }^{ n }$ maintains a uniform Hamming distance of $\varrho$ relative to \textbf{all} basis kets. This invariance ensures that, irrespective of the specific classification measurement obtained, $h$ will invariably reside among the nearest basis kets, thereby securing a classification probability of exactly $1.0$—an outcome that not only validates the metric's sensitivity but also highlights its potential for deterministic predictions in balanced function spaces.

At its core, this body of results affirms that the classification threshold serves as a reliably decreasing function of the Hamming distance, albeit with the caveat of this singular irregularity in special classes. Moreover, the versatile parameter $n$, which encapsulates the scale and complexity of the function class, proves instrumental in demarcating three fundamental subintervals along the Hamming distance axis. The first interval, encompassing distances from $1$ through $\frac { 2^{ n } } { 8 }$ inclusive, is characterized by a classification threshold that remains firmly above $0.50$, signaling robust confidence in nearest-neighbor predictions. Transitioning to the second interval, from $\frac { 2^{ n } } { 8 } + 1$ to $\frac { 2^{ n } } { 2 } - 1$, the threshold, while persistently positive, undergoes a steady attenuation, asymptotically approaching $0.00$ as distances increase, reflecting a progressive erosion of classification certainty. Finally, the third interval, stretching from $\frac { 2^{ n } } { 2 }$ to $2^{ n }$, reveals a uniform threshold of $0.00$, indicative of non-resemblance. Armed with knowledge of the subinterval containing the Hamming distance of an unknown function $h$, we can prognosticate the classification verdict with appreciable precision: alignment with a nearest basis ket portends the identification of a class member most akin to $h$, what amounts to a ``positive classification'', or, conversely, a ``negative classification'' that confidently precludes membership, all underpinned by elevated probabilistic guarantees. This duality not only substantiates the Hamming distance's efficacy as a versatile metric but also elevates its role in enabling both affirmative results and refutations, thereby broadening its applicability in a classification framework.

The ensuing Table \ref{tbl: Probability vs. Hamming Distance Table For ArbitraryPattern Bit Vectors of Length $2^{ n }$} encapsulates these findings in a concise overview, juxtaposing the dynamics for a generic class $F_{ P }$ against the idiosyncrasies of the special $F_{ Q_{ 2 } }^{ \otimes m }$, $m \geq 1$, to facilitate a comparative analysis.

\begin{tcolorbox}
	[
	enhanced,
	breakable,
	grow to left by = 0.000 cm,
	grow to right by = 0.000 cm,
	colback = white,
	enhanced jigsaw,			
	frame hidden,
	sharp corners,
	]
	\begin{table}[H]
		\caption{This table gives the probability distribution of the classification threshold as a function of the Hamming distance for a general class $F_{ P }$, and for a special class $F_{ Q_{ 2 } }^{ \otimes m }$. The results are parameterized by $n$, the arity of the Boolean functions in the class.}
		\label{tbl: Probability vs. Hamming Distance Table For ArbitraryPattern Bit Vectors of Length $2^{ n }$}
		\centering
		\SetTblrInner { rowsep = 1.500 mm }
		\begin{tblr}
			{
				colspec =
				{
					Q [ c, m, 3.000 cm ]
					| [ 0.750 pt, GreenLighter2!50 ]
					| [ 0.750 pt, GreenLighter2!50 ]
					Q [ c, m, 5.000 cm ]
					| [ 0.750 pt, GreenLighter2!50 ]
					Q [ c, m, 3.000 cm ]
				},
				rowspec =
				{
					| [ 3.500 pt, GreenLighter2!75 ]
					| [ 0.750 pt, GreenLighter2!75 ]
					| [ 0.250 pt, white ]
					Q
					| [ 0.250 pt, white ]
					Q
					| [ 0.150 pt, GreenLighter2 ]
					Q
					| [ 0.150 pt, GreenLighter2 ]
					Q
					| [ 0.150 pt, GreenLighter2 ]
					Q
					| [ 0.150 pt, GreenLighter2 ]
					Q
					| [ 0.150 pt, GreenLighter2 ]
					Q
					| [ 3.500 pt, GreenLighter2 ]
				}
			}
			&
			\SetCell [ c = 2 ] { bg = GreenLighter2, fg = white, font = \small \bfseries } Classification Threshold $\boldsymbol{ \vartheta }$
			\\
			\SetCell { bg = GreenLighter2, fg = white, font = \small \bfseries }
			Hamming Distance
			&
			\SetCell { bg = GreenLighter2, fg = white, font = \small \bfseries } $F_{ P }$
			&
			\SetCell { bg = black!50!WordBlueVeryLight, fg = white, font = \small \bfseries } $F_{ Q_{ 2 } }^{ \otimes m }$
			\\
			$1, \dots, \frac { 2^{ n } } { 8 }$
			&
			$0.50 < \vartheta < 1.00$
			&
			$0.50 < \vartheta < 1.00$
			\\
			$\frac { 2^{ n } } { 8 } + 1, \dots, \frac { 2^{ n } } { 2 } - 1$
			&
			$0.00 < \vartheta < 0.50$
			&
			$0.00 < \vartheta < 0.50$
			\\
			$\frac { 2^{ n } } { 2 }, \dots, \varrho - 1$
			&
			$0.00$
			&
			$0.00$
			\\
			\SetCell { bg = black!50!WordBlueVeryLight, fg = white, font = \small } $\varrho$
			&
			$0.00$
			&
			\SetCell { bg = black!50!WordBlueVeryLight, fg = white, font = \small } $1.00$
			\\
			$\varrho + 1, \ldots, 2^{ n }$
			&
			$0.00$
			&
			$0.00$
			\\
		\end{tblr}
	\end{table}
\end{tcolorbox}

\section{Discussion and conclusions} \label{sec: Discussion and Conclusions}

Quantum classification represents a vibrant and rapidly evolving area of research, especially when considered alongside the groundbreaking developments in quantum machine learning and quantum artificial intelligence. Quantum classification algorithms demonstrate exceptional proficiency in categorizing functions that fall within their intended ``promised'' class, that is the specific set of functions for which they have been engineered. The literature is replete with sophisticated investigations into the classification of Boolean functions exhibiting particular attributes, such as those adhering to well-defined structural or computational properties that place them into distinct categories. However, to the best of our knowledge, there remains a notable scarcity of research exploring the insights that might emerge when the input Boolean function deviates from these expected categories, failing to align with the algorithm's presupposed framework.

This paper introduces a novel investigative lens, illustrating that substantial and meaningful insights can nonetheless be extracted under suitably engineered conditions. We posit that deriving useful information from functions lying beyond an algorithm's core competency holds profound practical value, as it bridges the gap between idealized theoretical models and the messy realities of diverse computational inputs. In particular, we contend that, for a quantum algorithm tailored to conclusively classify functions within a designated class, denoted as $F$, and confronted with an unknown function $h$ that resides outside $F$, the most desirable resolution would involve identifying one of the nearest neighbors of $h$ within $F$. Such a neighbor would embody the function in $F$ that most faithfully mirrors the behavioral pattern of $h$, thereby offering a suitable approximation for understanding and approximating the outlier.

To advance this line of inquiry, it becomes imperative to establish a rigorous metric that encapsulates the intuitive concept of ``nearness'' among Boolean functions, enabling quantitative assessments of similarity. The Hamming distance stands out as one of the most established and versatile measures for this purpose, owing to its simplicity, computational efficiency, and interpretability in terms of bit-level discrepancies. That said, alternative distance metrics between Boolean functions are conceivable and could enrich the analytical toolkit—for instance, the innovative ``cluster'' concept delineated in \cite{Andronikos2025b}. The present work undertakes a comprehensive examination of the Hamming distance's efficacy as a proximity metric, specifically in the context of mapping unknown Boolean functions to their nearest counterparts within the set of perfectly classifiable functions. Through this focused approach, we probe not only the metric's robustness but also its implications for enhancing the adaptability and diagnostic power of quantum classifiers. In the ensuing text box, we encapsulate our principal discoveries and critically appraise their broader ramifications for the field.

\begin{tcolorbox}
	[
	enhanced,
	breakable,
	center title,
	fonttitle = \bfseries,
	colbacktitle = cyan4,
	coltitle = white,
	title = \textbf{Synopsis and evaluation of the derived results},
	grow to left by = 0.000 cm,
	grow to right by = 0.000 cm,
	colframe = cyan4,
	colback = cyan9!50,
	enhanced jigsaw,			
	sharp corners,
	boxrule = 0.500 pt,
	]
	\begin{itemize}
		\item	
		The comprehensive suite of experimental results unequivocally validates the utility of the Hamming distance as a pivotal metric, adeptly suited for the critical task of affirming or disqualifying prospective nearest neighbor candidates in the classification process.
		\item	
		As anticipated, our investigations have solidified the intuitive yet fundamental observation that the classification threshold is a monotonically decreasing function with respect to the Hamming distance, subject to the intriguing exceptions encountered in the specialized classes analyzed in subsection \ref{subsec: Interpretation Of The Experimental Results}. The importance of this revelation cannot be overstated; it compellingly illustrates how individual functions within a given class can precipitate meaningful divergences from the prevailing pattern, thereby enriching our understanding of intra-class heterogeneity and its implications for algorithmic robustness.
		\item	
		Moreover, even amidst these special scenarios where the class deviates from conventional behavior, our analysis unveils the pivotal revelation that such irregularities are not unpredictable but rather amenable to rigorous quantification and foresightful prediction, thereby transforming potential pitfalls into manageable phenomena.
		\item	
		To the best of our knowledge, this work constitutes the first contribution that delineates precise intervals for Hamming distances, thereby quantifying and demarcating the spectrum of values that the classification threshold may assume, offering a foundational tool for probabilistic assessments in quantum classification paradigms.
		\item	
		This fact is of cardinal significance, as it equips practitioners with a sophisticated framework: one that facilitates the endorsement of a classification result with amplified certainty when it plausibly unveils one of the nearest neighbors to the unknown input function, or conversely, leads to the resolute dismissal of such a result when probabilistic suggests that its alignment with a true nearest neighbor to be inconsequentially remote.
	\end{itemize}
\end{tcolorbox}

\subsection{The Nearest Basis Ket Game} \label{subsec: The Nearest Basis Ket Game}

Drawing upon the aforementioned conclusions, we are now well-equipped to furnish an evidence-based response to a logical question: Does the Nearest Basis Ket Game, as delineated in Section \ref{sec: What About Nonclassifiable Boolean Functions?}, possess substantive merit within the broader landscape of quantum classification paradigms? We assert with unwavering conviction an affirmative response to this question. A perusal of Table \ref{tbl: Probability vs. Hamming Distance Table For ArbitraryPattern Bit Vectors of Length $2^{ n }$} illuminates the game's intrinsic value, revealing nuanced strategic dynamics between the protagonists, Alice and Bob. Specifically, should Bob, perhaps through inadvertence or suboptimal deliberation, select an unknown function $h$ that exhibits a modest Hamming distance from the target class, spanning from $1$ to $\frac { 2^{ n } } { 8 }$, Alice is positioned to issue a resolute ``Yes'' verdict with substantial confidence. In such scenarios, the probabilistic tilt heavily favors Alice, affording her a greater likelihood of prevailing in the contest.

In a mirror-image fashion, provided the game unfolds outside the purview of a special class, an analogous advantage accrues to Alice when Bob's selection of $h$ yields a substantial Hamming distance from the class, falling from $\frac { 2^{ n } } { 2 }$ to $2^{ n }$. Here, Alice's assured proclamation of ``No'' aligns seamlessly with the empirical probabilities, propelling her toward victory with near-certainty. This bilateral symmetry not only highlights the game's robustness as a diagnostic instrument for proximity assessments but also accentuates its pedagogical utility in elucidating the thresholds where quantum classifiers transition from reliable affirmation to confident refutation. Furthermore, for a special class of the form $F_{ Q_{ 2 } }^{ \otimes m }$, with $m \geq 1$, a rational Bob, ever mindful of game-theoretic imperatives, would assiduously eschew any $h$ whose distance manifests as $\varrho$, for such a selection would certainly lead Alice to surely win, devoid of any probabilistic ambiguity.

Yet, this revelation does not render Bob impotent; on the contrary, it unveils a deliberate and efficacious stratagem whereby he can increase his odds of success to approximate $0.50$. This entails a methodical selection of the secret function $h$ such that its Hamming distance from the class aligns precisely, or in intimate proximity, to $\frac { 2^{ n } } { 8 }$, the turning point at which the success probability for Alice declines, dipping below the $0.50$ threshold. By wielding this insight judiciously, Bob transforms the Nearest Basis Ket Game into an equitable analogue of the venerable coin toss, the iconic classical game of fairness and  impartiality. This infuses quantum adversarial interactions with a veneer of fairness and strategic depth that could inspire extensions to more intricate multi-player or adaptive variants in quantum game theory.

\bibliographystyle{ieeetr}
\bibliography{PLQCPBFHD}

\end{document}